\documentclass[floatfix,
 aps, prc,     
 showpacs, 
 showkeys, 
 superscriptaddress, 
 reprint, 
 nofootinbib 
]{revtex4-1}

\usepackage[english]{babel}
\usepackage[T1]{fontenc}
\usepackage[utf8]{inputenc} 
\usepackage{amsmath} 
\usepackage{amssymb, bbm} 
\usepackage{amstext} 
\usepackage{mathrsfs} 
\usepackage{booktabs} 
\usepackage{array} 
\usepackage{graphicx} \graphicspath{ {./files/} } 
\usepackage{xcolor} \definecolor{lblue}{RGB}{70,126,185}
\usepackage[pdftex,pdfpagelabels]{hyperref} 
\hypersetup{
   colorlinks=true, linktocpage=true, urlcolor=lblue, linkcolor=lblue, citecolor=lblue,
   pdftitle={Hyperon-nuclear interactions from SU(3) chiral effective field theory},
   pdfauthor={Stefan Petschauer, Johann Haidenbauer, Norbert Kaiser, Ulf-G. Mei\ss{}ner, Wolfram Weise}
}
\usepackage{overpic}
\usepackage{slashed} 
\usepackage{xspace} 

\usepackage{subcaption}
\newcommand{\Ps}{P^{(\sigma)}}
\newcommand{\Pt}{P^{(\tau)}}
\newcommand{\ir}[1]{\mathbf{#1}}
\newcommand{\extfield}{(\nabla^i \phi)}
\protected\def\lc{C}
\protected\def\ld{D}
\protected\def\lC{H}
\newcommand{\trace}[1]{\ensuremath{\langle #1\rangle}}
\newcommand\numberthis{\addtocounter{equation}{1}\tag{\theequation}}
\makeatletter
\protected\def\fig{\@ifstar\@fig\@@fig} \def\@fig#1{\ref{#1}} \def\@@fig#1{Fig.~\ref{#1}}
\protected\def\tab{\@ifstar\@tab\@@tab} \def\@tab#1{\ref{#1}} \def\@@tab#1{Tab.~\ref{#1}}
\protected\def\eq{\@ifstar\@eq\@@eq} \def\@eq#1{\eqref{#1}} \def\@@eq#1{Eq.~\eqref{#1}}
\protected\def\ch{\@ifstar\@ch\@@ch} \def\@ch#1{\ref{#1}} \def\@@ch#1{Ch.~\ref{#1}}
\protected\def\sect{\@ifstar\@sect\@@sect} \def\@sect#1{\ref{#1}} \def\@@sect#1{Sec.~\ref{#1}}
\protected\def\ssect{\@ifstar\@ssect\@@ssect} \def\@ssect#1{\ref{#1}} \def\@@ssect#1{Subsec.~\ref{#1}}
\protected\def\app{\@ifstar\@app\@@app} \def\@app#1{\cite{#1}} \def\@@app#1{App.~\ref{#1}}
\protected\def\ct{\@ifstar\@ct\@@ct} \def\@ct#1{\cite{#1}} \def\@@ct#1{Ref.~\cite{#1}}
\protected\def\figs{\@ifstar\@figs\@@figs} \def\@figs#1{\ref{#1}} \def\@@figs#1{Figs.~\ref{#1}}
\protected\def\tabs{\@ifstar\@tabs\@@tabs} \def\@tabs#1{\ref{#1}} \def\@@tabs#1{Tabs.~\ref{#1}}
\protected\def\eqs{\@ifstar\@eqs\@@eqs} \def\@eqs#1{\eqref{#1}} \def\@@eqs#1{Eqs.~\eqref{#1}}
\protected\def\chs{\@ifstar\@chs\@@chs} \def\@chs#1{\ref{#1}} \def\@@chs#1{Chs.~\ref{#1}}
\protected\def\sects{\@ifstar\@sects\@@sects} \def\@sects#1{\ref{#1}} \def\@@sects#1{Secs.~\ref{#1}}
\protected\def\ssects{\@ifstar\@ssects\@@ssects} \def\@ssects#1{\ref{#1}} \def\@@ssects#1{Subsecs.~\ref{#1}}
\protected\def\apps{\@ifstar\@apps\@@apps} \def\@apps#1{\cite{#1}} \def\@@apps#1{Apps.~\ref{#1}}
\protected\def\cts{\@ifstar\@cts\@@cts} \def\@cts#1{\cite{#1}} \def\@@cts#1{Refs.~\cite{#1}}
\makeatother
\newcommand{\ie}{i.e., }

\newcommand{\eg}{e.g., }

\newcommand{\cf}{cf.\@\xspace}

\newcommand{\cheft}{\(\chi\)EFT\xspace}
\newcommand{\Cheft}{Chiral EFT\xspace}

\newlength{\usualunit} 
\unitlength = \usualunit
\DeclareMathOperator{\tr}{tr}
\DeclareMathOperator{\diag}{diag}
\renewcommand\Re{\operatorname{Re}}

\newcommand{\defeq}{\ensuremath{:=}}
\newcommand{\apr}{\ensuremath{\approx}}
\newcommand{\figscale}{1.0}

\begin{document}

\title{Hyperon-nuclear interactions from SU(3) chiral effective field theory}

\author{Stefan~Petschauer}
   \email{stefan.petschauer@tngtech.com}
   \affiliation{TNG Technology Consulting GmbH, D-85774 Unterföhring, Germany}
   \affiliation{Physik Department, Technische Universit\"at M\"unchen, D-85747 Garching, Germany}
\author{Johann~Haidenbauer}
   \affiliation{Institute for Advanced Simulation and J\"ulich Center for Hadron Physics,\\
                Institut f\"ur Kernphysik, Forschungszentrum J\"ulich, D-52425 J\"ulich, Germany}
\author{Norbert~Kaiser}
   \affiliation{Physik Department, Technische Universit\"at M\"unchen, D-85747 Garching, Germany}
\author{Ulf-G.~Mei\ss{}ner}
   \affiliation{Helmholtz-Institut f\"ur Strahlen- und Kernphysik and Bethe Center\\
                for Theoretical Physics, Universit\"at Bonn, D-53115 Bonn, Germany}
   \affiliation{Institute for Advanced Simulation and J\"ulich Center for Hadron Physics,\\
                Institut f\"ur Kernphysik, Forschungszentrum J\"ulich, D-52425 J\"ulich, Germany}
   \affiliation{Tbilisi State University, 0186 Tbilisi, Georgia}
\author{Wolfram~Weise}
   \affiliation{Physik Department, Technische Universit\"at M\"unchen, D-85747 Garching, Germany}

\date{\today}

\begin{abstract}
The interaction between hyperons and nucleons has a wide range of applications in strangeness nuclear physics and is a topic of continuing great interest.
These interactions are not only important for hyperon-nucleon scattering but also essential as basic input to studies of hyperon-nuclear few- and many-body systems including hypernuclei and neutron star matter.
We review the systematic derivation and construction of such baryonic forces from the symmetries of quantum chromodynamics within non-relativistic SU(3) chiral effective field theory.
Several applications of the resulting potentials are presented for topics of current interest in strangeness nuclear physics.
\end{abstract}


\maketitle
\tableofcontents

\section{Introduction} \label{sec:intro}

Strangeness nuclear physics is an important topic of ongoing research, addressing for example scattering of baryons including strangeness, properties of hypernuclei or strangeness in infinite nuclear matter and in neutron star matter.
The theoretical foundation for such investigations are interaction potentials between nucleons and strange baryons such as the \(\Lambda\) hyperon.

Nuclear many-body systems are (mainly) governed by the strong interaction, described at the fundamental level by \emph{quantum chromodynamics} (QCD).
The elementary degrees of freedom of QCD are quarks and gluons.
However, in the low-energy regime of QCD quarks and gluons are confined into colorless hadrons.
This is the region where (hyper-)nuclear systems are formed.
In this region QCD cannot be solved in a perturbative way.
Lattice QCD is approaching this problem via large-scale numerical simulations:
the (Euclidean) space-time is discretized and QCD is solved on a finite grid~\cite{Aoki2012,Beane2011a,Beane2012,Nemura2018}.
Since the seminal work of Weinberg~\cite{Weinberg1968,Weinberg1979} \emph{chiral effective field theory} (\cheft) has become a powerful tool for calculating systematically the strong interaction dynamics for low-energy hadronic processes~\cite{Gasser1984,Gasser1988,Bernard1995}.
\Cheft employs the same symmetries and symmetry breaking patterns at low-energies as QCD, but it uses the proper degrees of freedom, namely hadrons instead of quarks and gluons.
In combination with an appropriate expansion in small external momenta, the results can be improved systematically, by going to higher order in the power counting, and at the same time theoretical errors can be estimated.
Furthermore, two- and three-baryon forces can be constructed in a consistent fashion.
The unresolved short-distance dynamics is encoded in \cheft in contact terms, with a priori unknown low-energy constants (LECs).

The \(NN\) interaction is empirically known to very high precision.
Corresponding two-nucleon potentials have been derived to high accuracy in phenomenological approaches \ct*{Stoks1994,Wiringa1995,Machleidt2000}.
Nowadays the systematic theory to construct nuclear forces is \cheft \ct*{Epelbaum2009,Machleidt2011}.
(Note however that there are still debates about the Weinberg power counting schemes and how it is employed in practice~\cite{Kaplan1998b,Nogga2005,Epelbaum:2006pt}.)
In contrast, the \(YN\) interaction is presently not known in such detail.
The scarce experimental data (about 35 data points for low-energy total cross sections) do not allow for a unique determination of the hyperon-nucleon interaction.
The limited accuracy of the \(YN\) scattering data does not permit a unique phase shift analysis.
However, at experimental facilities such as J-PARC in Japan or later at FAIR in Germany, a significant amount of beam time will be devoted to strangeness nuclear physics.
Various phenomenological approaches have been employed to describe the \(YN\) interaction, in particular boson-exchange 
models~\cite{Holzenkamp1989,Reuber1994a,Rijken1998,Haidenbauer2005,Rijken2010,Nagels2019}
or quark models~\cite{Kohno1999,Fujiwara2006,Garcilazo2007}.
However, given the poor experimental data base, these interactions differ considerably from each other.
Obviously there is a need for a more systematic investigation based on the underlying theory of the strong interaction, QCD\@.
Some aspects of \(YN\) scattering and hyperon mass shifts in nuclear matter using EFT methods have been covered in \cts{Savage1996,Korpa2001}.
The \(YN\) interaction has been investigated at leading order (LO) in SU(3) \cheft~\cite{Polinder2006,Polinder2007,Haidenbauer2010a} by extending the very successful \cheft framework for the nucleonic sector~\cite{Epelbaum2009,Machleidt2011} to the strangeness sector.
This work has been extended to next-to-leading order (NLO) in \cts{Petschauer2013a,Haidenbauer2013a,Haidenbauer2019b} where an excellent description of the strangeness \(-1\) sector has been achieved, comparable to most advanced phenomenological hyperon-nucleon interaction models.
An extension to systems with more strangeness has been done in \cts{Haidenbauer2015, Haidenbauer2015c,Haidenbauer2019}.
Systems including decuplet baryons have been investigated in \ct{Haidenbauer2017c} at leading order in non-relativistic \cheft.
Recently calculations within leading order covariant \cheft have been performed for \(YN\) interactions in the strangeness sector~\cite{Li2016,Ren2018,Li2018,Song2018,Li:2018tbt} with comparable results, see also \ct{Ren:2019qow}.
It is worth to briefly discuss the differences between the covariant and the heavy-baryon approach.
In the latter, due to the expansion in the inverse of the baryon masses, some terms are relegated  to higher orders.
Also, it can happen that  the analytic structure is distorted in the strict heavy-baryon limit.
This can easily be remedied by including the kinetic energy  term in the baryon propagator \cite{Bernard:1993ry}.
In what follows, we will present results based on the heavy-baryon approach.

Numerous advanced few- and many-body techniques have been developed to employ such phenomenological or chiral interactions, in order to calculate the properties of nuclear systems with and without strangeness.
For example, systems with three or four particles can be reliably treated by Faddeev-Yakubovsky theory \cite{Miyagawa1993,Miyagawa1995,Nogga2002,Nogga2014a},
somewhat heavier (hyper)nuclei with approaches like 
the no-core-shell model \cite{Bogner2009,Wirth2014,Wirth2016,Wirth2018,Gazda:2015qyt,Gazda:2016qva}.
In the nucleonic sector many-body approaches such as Quantum Monte Carlo calculations \cts*{Gandolfi2007,Gandolfi2009,Lonardoni2013c}, or nuclear lattice simulations \cts*{Borasoy2007,Epelbaum2011,Lahde:2019npb} have been successfully applied and can be extended to the strangeness sector.
Furthermore, nuclear matter is well described by many-body perturbation theory with chiral low-momentum interactions \cts*{Holt2013,Coraggio2014a,Sammarruca2015}.
Concerning \(\Lambda\) and \(\Sigma\) hyperons in nuclear matter, specific long-range processes related to two-pion exchange between hyperons and nucleons in the nuclear medium have been studied in \cts{Kaiser:2004fe,Kaiser2005}.
Conventional Brueckner theory \cts*{Brueckner1954,Brueckner1955,Day1967} at first order in the hole-line expansion, the so-called Bruecker-Hartree-Fock approximation, has been widely applied to calculations of
hypernuclear matter \cts*{Rijken1998,Kohno1999,Schulze1998,Vidana1999} employing phenomenological two-body potentials.
This approach is also used in investigations of neutron star matter \cts*{Baldo1999,Schulze2006,Schulze2011}.
Recently, corresponding calculations of the properties of hyperons in nuclear matter have been also performed with chiral \(YN\) interaction potentials~\cite{Haidenbauer2015a,Petschauer2015,Haidenbauer2019}.

Employing the high precision \(NN\) interactions described above, even "simple" nuclear systems such as triton cannot be described satisfactorily with two-body interactions alone.
The introduction of three-nucleon forces (3NF) substantially improves this situation \ct*{Pieper2001a,Epelbaum2002,KalantarNayestanaki:2011wz,Hammer:2012id} and also in the context of infinite nuclear matter 3NF are essential to achieve saturation of nuclear matter.
These 3NF are introduced either phenomenologically, such as the families of Tuscon-Melbourne \ct*{McKellar1968,Coon1975}, Brazilian \ct*{Coelho1983} or Urbana-Illinois \ct*{Pudliner1997,Pieper2001} 3NF, or constructed according to the basic principles of \cheft{} \ct*{Epelbaum2002,Weinberg1990,Weinberg1991,Weinberg1992,VanKolck1994,Ishikawa2007,Bernard2008,Bernard2011,Krebs2012,Krebs2013a}.
Within an EFT approach, 3NF arise naturally and consistently together with two-nucleon forces.
Chiral three-nucleon forces are important in order to get saturation of nuclear matter from chiral low-momentum two-body interactions treated in many-body perturbation theory~\cite{Coraggio2014a}.
In the strangeness sectors the situation is similar:
Three-baryon forces (3BF), especially the \(\Lambda NN\) interaction, seem to be important for a satisfactorily description of hypernuclei and hypernuclear matter
\ct*{Lonardoni2013c,Bhaduri1967,Bhaduri1967a,Gal1971,Gal1972,Gal1978,Bodmer1988,Usmani1995,Lonardoni2013,Logoteta:2019utx}.
Especially in the context of neutron stars, 3BF are frequently discussed.
The observation of two-solar-mass neutron stars~\cite{Demorest2010,Antoniadis2013} sets strong constraints on the stiffness of the equation-of-state (EoS) of dense baryonic matter \ct*{Hebeler2010b, Hell2014,Steiner2015,Vidana2015,Vidana2018}. The analysis of recently observed gravitational wave signals from a two merging neutron stars \cite{TheLIGOScientific:2017qsa, GBM:2017lvd} provides further conditions, by constraining the tidal deformability of neutron star matter.

A naive introduction of \(\Lambda\)-hyperons as an additional baryonic degree of freedom would soften the EoS such that it is not possible to stabilize a two-solar-mass neutron star against gravitational collapse~\cite{Djapo2010}.
To solve this so-called \emph{hyperon puzzle}, several ad-hoc mechanisms have so far been invoked, \eg through vector meson exchange~\cite{Weissenborn2012,Weissenborn2012a}, multi-Pomeron exchange~\cite{Yamamoto2014} or a suitably adjusted repulsive \(\Lambda NN\) three-body interaction~\cite{Takatsuka2008,Vidana2010,Lonardoni2015}.
Clearly, a more systematic approach to the three-baryon interaction within \cheft is needed, to estimate whether the 3BF can provide the necessary repulsion and thus keep the equation-of-state sufficiently stiff.
A first step in this direction was done in \ct{Petschauer2016}, where the leading 3BFs have been derived within SU(3) \cheft.
The corresponding low-energy constants have been estimated by decuplet saturation in \ct{Petschauer2017a}.
The effect of these estimated 3BF has been investigated in \cts{Petschauer2017a,Kohno2018}.

In this review article we present, on a basic level, the emergence of nuclear interactions in the strangeness sector from the perspective of (heavy-baryon) chiral effective field theory.
After a brief introduction to SU(3) \cheft in~\sect{sec:ChEFT}, we present how the interaction between hyperons and nucleons is derived at NLO from these basic principles for two-baryon interactions (\sect{sec:BB}) and for three-baryon interactions (\sect{sec:BBB}).
In~\sect{sec:application} applications of these potentials are briefly reviewed for \(YN\) scattering, infinite nuclear matter, hypernuclei and neutron star matter.

\section{SU(3) chiral effective field theory}  \label{sec:ChEFT}

An \emph{effective field theory} (EFT) is a low-energy approximation to a more fundamental theory.
Physical quantities can be calculated in terms of a low-energy expansion in powers of small energies and momenta over some characteristic large scale.
The basic idea of an EFT is to include the relevant degrees of freedom explicitly, while heavier (frozen) degrees of freedom are integrated out.
An effective Lagrangian is obtained by constructing the most general Lagrangian including the active degrees of freedom, that is consistent with the symmetries of the underlying fundamental theory \ct*{Weinberg1979}.
At a given order in the expansion, the theory is characterized by a finite number of coupling constants, called \emph{low-energy constants} (LECs).
The LECs encode the unresolved short-distance dynamics and furthermore allow for an order-by-order renormalization of the theory.
These constants are a priori unknown, but once determined from one experiment or from the underlying theory, predictions for physical observables can be made.
However, due to the low-energy expansion and the truncation of degrees of freedom, an EFT has only a limited range of validity.

The underlying theory of \emph{chiral effective field theory} is quantum chromodynamics.
QCD is characterized by two important properties.
For high energies the (running) coupling strength of QCD becomes weak, hence a perturbative approach in the high-energy regime of QCD is possible.
This famous feature is called \emph{asymptotic freedom} of QCD and originates from the non-Abelian structure of QCD\@.
However, at low energies and momenta the coupling strength of QCD is of order one, and a perturbative approach is no longer possible.
This is the region of \emph{non-perturbative QCD}, in which we are interested in.
Several strategies to approach this regime have been developed, such as lattice simulations, Dyson-Schwinger equations, QCD sum rules or chiral perturbation theory.
The second important feature of QCD is  the so-called \emph{color confinement}:
isolated quarks and gluons are not observed in nature, but only color-singlet objects.
These color-neutral particles, the hadrons, are the active degrees of freedom in \cheft.

But already before QCD was established, the ideas of an effective field theory were used in the context of the strong interaction.
In the sixties the Ward identities related to spontaneously broken chiral symmetry were explored by using current algebra methods, \eg by Adler and Dashen \ct*{Adler1968}.
The group-theoretical foundations for constructing phenomenological Lagrangians in the presence of spontaneous symmetry breaking have been developed by Weinberg \cts*{Weinberg1968} and Callan, Coleman, Wess and Zumino \cts*{Coleman1969,Callan1969}.
With Weinberg's seminal paper \ct*{Weinberg1979} it became clear how to systematically construct an EFT and generate loop corrections to tree level results.
This method was improved later by Gasser and Leutwyler \cts*{Gasser1984,Gasser1985}.
A systematic introduction of nucleons as degrees of freedom was done by Gasser, Sainio and Svarc \ct*{Gasser1988}.
They showed that a fully relativistic treatment of nucleons is problematic, as the nucleon mass does not vanish in the chiral limit and thus adds an extra scale.
A solution for this problem was proposed by Jenkins and Manohar \ct*{Jenkins1991} by considering baryons as heavy static sources.
This approach was further developed using a systematic path-integral framework in \ct{Bernard:1992qa}.
The nucleon-nucleon interaction and related topics were considered by Weinberg in \ct{Weinberg1990}.
Nowadays \cheft is used as a powerful tool for calculating systematically the strong interaction dynamics of hadronic processes, such as the accurate description of nuclear forces \cts*{Epelbaum2009,Machleidt2011}.

In this section, we give a short introduction to the underlying symmetries of QCD and their breaking pattern.
The basic concepts of \cheft are explained, especially the explicit degrees of freedom and the connection to the symmetries of QCD\@.
We state in more detail how the chiral Lagrangian can be constructed from basic principles.
However, it is beyond the scope of this work to give a detailed introduction to \cheft and QCD\@.
Rather we will introduce only the concepts necessary for the derivation of hyperon-nuclear forces.
We follow \cts{Bernard1995,Epelbaum2009,Machleidt2011,Weise2001,Scherer2012,Petschauer2015a} and refer the reader for more details to these references (and references therein).

\subsection{Low-energy quantum chromodynamics} \label{subsec:qcd}

Let us start the discussion with the QCD Lagrangian
\begin{equation} \label{eq:qcd}
 \mathscr{L}_{\mathrm{QCD}} = \sum_{f=u,d,s,c,b,t} \bar q_f \left(\mathrm i \slashed{D} - m_f\right) q_f - \frac14 G_{\mu\nu,a} G^{\mu\nu}_a \,,
\end{equation}
with the six quark flavors \(f\) and the gluonic field-strength tensor \(G_{\mu\nu,a}(x)\).
The gauge covariant derivative is defined by \(D_\mu = \mathbbm{1} \partial_\mu - \mathrm ig A^a_\mu \frac{\lambda_a}2\), where \(A^a_\mu(x)\) are the gluon fields and \(\lambda_a\) the Gell-Mann matrices.
The  QCD Lagrangian is symmetric under the local color gauge symmetry, under global Lorentz transformations, and the discrete symmetries parity, charge conjugation and time reversal.
In the following we will introduce the so-called \emph{chiral symmetry}, an approximate global continuous symmetry of the QCD Lagrangian.
The chiral symmetry is essential for chiral effective field theory.
In view of the application to low energies, we divide the quarks into three light quarks \(u,d,s\) and three heavy quarks \(c,b,t\), since the quark masses fulfill a hierarchical ordering:
\begin{equation}
m_u,m_d,m_s \ll 1~\mathrm{GeV} \leq m_c,m_b,m_t\,.
\end{equation}
At energies and momenta well below \(1~\mathrm{GeV}\), the heavy quarks can be treated effectively as static.
Therefore, the light quarks are the only active degrees of freedom of QCD for the low-energy region we are interested in.
In the following we approximate the QCD Lagrangian by using only the three light quarks.
Compared to characteristic hadronic scales, such as the nucleon mass (\(M_N\approx939~\mathrm{MeV}\)), the light quark masses are small.
Therefore, a good starting point for our discussion of low-energy QCD are massless quarks \(m_u=m_d=m_s=0\), which is referred to as the \emph{chiral limit}.
The QCD Lagrangian becomes in the chiral limit
\begin{equation} \label{eq:L0QCD}
 \mathscr{L}^0_{\mathrm{QCD}} = \sum_{f=u,d,s} \bar q_f \mathrm i \slashed{D}q_f - \frac14 G_{\mu\nu,a} G^{\mu\nu}_a\,.
\end{equation}
Now each quark field \(q_f(x)\) is decomposed into its \emph{chiral components}
\begin{equation}
q_{f,\mathrm L}=P_\mathrm L\, q_f\,,\qquad q_{f,\mathrm R}=P_\mathrm R\, q_f\,.
\end{equation}
using the left- and right-handed projection operators
\begin{equation}
P_\mathrm L=\frac12 \left(1-\gamma_5\right)\,,\qquad
P_\mathrm R=\frac12 \left(1+\gamma_5\right)\,,
\end{equation}
with the chirality matrix \(\gamma_5\).
These projectors are called left- and right-handed since in the chiral limit they project the free quark fields on helicity eigenstates, \( \hat h\, q_{L,R} = \pm\, q_{L,R}\), with \(\hat h=\vec\sigma \cdot \vec p\, / \left\lvert\vec p\,\right\rvert\).
For massless free fermions helicity is equal to chirality.

Collecting the three quark-flavor fields \(q=(q_u, q_d, q_s)\) and equivalently for the left and right handed components, we can express the QCD Lagrangian in the chiral limit as
\begin{equation}
 \mathscr{L}^0_{\mathrm{QCD}} = \bar q_{\mathrm R} \mathrm i \slashed{D}q_{\mathrm R} + \bar q_{\mathrm L} \mathrm i \slashed{D}q_{\mathrm L} - \frac14 G_{\mu\nu,a} G^{\mu\nu}_a\,.
\end{equation}
Obviously the right- and left-handed components of the massless quarks are separated.
The Lagrangian is invariant under a global transformation
\begin{equation} \label{eq:chtrans}
 q_\mathrm L \rightarrow L\, q_\mathrm L\,,\quad q_\mathrm R \rightarrow R\, q_\mathrm R\,,
\end{equation}
with \emph{independent} unitary \(3\times3\) matrices \(L\) and \(R\) acting in flavor space.
This means that \(\mathscr{L}_{\mathrm{QCD}}^0\) possesses (at the classical, unquantized level) a global \(\mathrm U(3)_\mathrm L\times\mathrm U(3)_\mathrm R\) symmetry, isomorphic to a global \(\mathrm{SU}(3)_\mathrm L\times \mathrm U(1)_\mathrm L\times\mathrm{SU}(3)_\mathrm R\times \mathrm U(1)_\mathrm R\) symmetry.
\(U(1)_\mathrm L\times \mathrm U(1)_\mathrm R\) are often rewritten into a vector and an axial-vector part \(U(1)_\mathrm V\times \mathrm U(1)_\mathrm A\), named after the transformation behavior of the corresponding conserved currents under parity transformation.
The flavor-singlet vector current originates from rotations of the left- and right-handed quark fields with the same phase (``\(V=L+R\,\)'') and the corresponding conserved charge is the \emph{baryon number}.
After quantization, the conservation of the flavor-singlet axial vector current, with transformations of left- an right-handed quark fields with opposite phase (``\(A=L-R\,\)''), gets broken due to the so-called Adler-Bell-Jackiw anomaly~\cite{Adler1969,Bell1969}.
The symmetry group \(\mathrm{SU}(3)_\mathrm L\times\mathrm{SU}(3)_\mathrm R\) refers to the \emph{chiral symmetry}.
Similarly the conserved currents can be rewritten into flavor-octet vector and flavor-octet axial-vector currents, where the vector currents correspond to the diagonal subgroup \(\mathrm{SU}(3)_\mathrm V\) of \(\mathrm{SU}(3)_\mathrm L\times\mathrm{SU}(3)_\mathrm R\) with \(L=R\).

After the introduction of small \emph{non-vanishing quark masses}, the quark mass term of the QCD Lagrangian \eqref{eq:qcd} can be expressed as
\begin{equation}
 \mathscr{L}_M=-\bar q M q = -\left(\bar q_\mathrm R M q_\mathrm L + \bar q_\mathrm L M q_\mathrm R \right)\,,
\end{equation}
with the diagonal quark mass matrix \( M = \diag(m_u,m_d,m_s)\).
Left- and right-handed quark fields are mixed in \(\mathscr{L}_M\) and the chiral symmetry is explicitly broken.
The baryon number is still conserved, but the flavor-octet vector and axial-vector currents are no longer conserved.
The axial-vector current is not conserved for any small quark masses.
However, the flavor-octet vector current remains conserved, if the quark masses are equal, \(m_u=m_d=m_s\), referred to as the \emph{(flavor) SU(3) limit}.

Another crucial aspect of QCD is the so-called \emph{spontaneous chiral symmetry breaking}.
The chiral symmetry of the Lagrangian is not a symmetry of the ground state of the system, the QCD vacuum.
The structure of the hadron spectrum allows to conclude that the chiral symmetry \(\mathrm{SU}(3)_\mathrm L\times\mathrm{SU}(3)_\mathrm R\) is spontaneously broken to its vectorial subgroup \(\mathrm{SU}(3)_\mathrm V\), the so-called \emph{Nambu-Goldstone realization} of the chiral symmetry.
The spontaneous breaking of chiral symmetry can be characterized by a non-vanishing \emph{chiral quark condensate} \(\langle \bar q q \rangle \neq 0\), \ie the vacuum involves strong correlations of scalar quark-antiquark pairs.

The eight Goldstone bosons corresponding to the spontaneous symmetry breaking of the chiral symmetry are identified with the eight lightest hadrons, the pseudoscalar mesons (\(\pi^\pm,\pi^0,K^\pm,K^0,\bar K^0,\eta\)).
They are pseudoscalar particles, due to the parity transformation behavior of the flavor-octet axial-vector currents.
The explicit chiral symmetry breaking due to non-vanishing quark masses leads to non-zero masses of the pseudoscalar mesons.
However, there is a substantial mass gap, between the masses of the pseudoscalar mesons and the lightest hadrons of the remaining hadronic spectrum.
For non-vanishing but equal quark masses, \(\mathrm{SU}(3)_\mathrm V\) remains a symmetry of the ground state.
In this context \(\mathrm{SU}(3)_\mathrm V\) is often called the flavor group \(\mathrm{SU}(3)\), which provides the basis for the classification of low-lying hadrons in multiplets.
In the following we will consider the so-called \emph{isospin symmetric limit}, with \(m_u=m_d\neq m_s\).
The remaining symmetry is the \(\mathrm{SU}(2)\) isospin symmetry.
An essential feature of low-energy QCD is, that the pseudoscalar mesons \emph{interact weakly} at low energies.
This is a direct consequence of their Goldstone-boson nature.
This feature allows for the construction of a low-energy effective field theory enabling a systematic expansion in small momenta and quark masses.

Let us introduce one more tool for the systematic development of \cheft called the \emph{external-field method}.
The chiral symmetry gives rise to so-called \emph{chiral Ward identities}:
relations between the divergence of Green functions that include a symmetry current (vector or axial-vector currents) to linear combinations of Green functions.
Even if the symmetry is explicitly broken, Ward identities related to the symmetry breaking term exist.
The chiral Ward identities do not rely on perturbation theory, but are also valid in the non-perturbative region of QCD\@.
The external-field method is an elegant way to formally combine all chiral Ward identities in terms of invariance properties of a generating functional.
Following the procedure of Gasser and Leutwyler \cts*{Gasser1984,Gasser1985} we introduce (color neutral) \emph{external fields}, \(s(x)\), \(p(x)\), \(v_\mu(x)\), \(a_\mu(x)\),  of the form of Hermitian \(3\times3\) matrices that couple to scalar, pseudoscalar, vector and axial-vector currents of quarks:
\begin{align*} \label{eq:LagrangianExt}
 \mathscr L &= \mathscr{L}^0_{\mathrm{QCD}} + \mathscr{L}_{\mathrm{ext}} \\
 &= \mathscr L^0_\mathrm{QCD} + \bar q \gamma^\mu(v_\mu+\gamma_5 a_\mu)q - \bar q(s-\mathrm i \gamma_5 p)q \,. \numberthis
\end{align*}
All chiral Ward identities are encoded in the corresponding generating functional, if the global chiral symmetry  \(\mathrm{SU}(3)_\mathrm L \times \mathrm{SU}(3)_\mathrm R\) of \(\mathscr{L}^0_{\mathrm{QCD}}\) is promoted to a \emph{local gauge symmetry} of \(\mathscr L\) \ct*{Leutwyler1994}.
Since \(\mathscr{L}^0_{\mathrm{QCD}}\) is only invariant under the global chiral symmetry, the external fields have to fulfill a suitable transformation behavior:
\begin{align*}
 v_\mu+a_\mu &\rightarrow R (v_\mu+a_\mu) R^\dagger + \mathrm i\, R\partial_\mu R^\dagger\,, \\
 v_\mu-a_\mu &\rightarrow L (v_\mu-a_\mu) L^\dagger + \mathrm i\, L\partial_\mu L^\dagger\,, \\
 s+\mathrm i\,p &\rightarrow R \left(s+\mathrm i\,p\right) L^\dagger\,, \\
 s-\mathrm i\,p &\rightarrow L \left(s-\mathrm i\,p\right) R^\dagger\,, \numberthis \label{eq:extrans}
\end{align*}
where \(L(x)\) and \(R(x)\) are (independent) space-time-dependent elements of \(\mathrm{SU}(3)_\mathrm{L}\) and \(\mathrm{SU}(3)_\mathrm{R}\).

Furthermore, we still require the full Lagrangian \(\mathscr L\) to be invariant under \(P\), \(C\) and \(T\).
As the transformation properties of the quarks are well-known, the transformation behavior of the external fields can be determined and is displayed in \tab{tab:transpropext}.
Time reversal symmetry is not considered explicitly, since it is automatically fulfilled due to the \(CPT\) theorem.

\begin{table}
 \centering
\setlength{\tabcolsep}{12pt}
 \begin{tabular}{>{\(}c<{\)}>{\(}c<{\)}>{\(}c<{\)}>{\(}c<{\)}>{\(}c<{\)}}
 \toprule
 & v^\mu & a^\mu & s & p \\
 \cmidrule(lr){1-1}
 \cmidrule(lr){2-5}
 P & {P^\mu}_\nu v^\nu & - {P^\mu}_\nu a^\nu & s & -p \\
 C & -{v^\mu}^\top & {a^\mu}^\top & s^\top & p^\top \\
 \bottomrule
 \end{tabular}
\setlength{\tabcolsep}{6pt}
 \caption{Transformation properties of the external fields under parity and charge conjugation.
 For \(P\) a change of the spatial arguments \(\left(t,\vec x\,\right) \rightarrow \left(t,-\vec x\,\right)\) is implied and we defined the matrix \({P^\mu}_\nu=\diag(+1,-1,-1,-1)\).} \label{tab:transpropext}
\end{table}

Another central aspect of the external-field method is the addition of terms to the three-flavor QCD Lagrangian in the chiral limit, \(\mathscr{L}^0_{\mathrm{QCD}}\).
Non-vanishing current quark masses and therefore the \emph{explicit breaking} of chiral symmetry can be included by setting the scalar field equal to the quark mass matrix, \(s(x)=M=\diag\left(m_u,m_d,m_s\right)\).
Similarly \emph{electroweak interactions} can be introduced through appropriate external vector and axial vector fields.
This feature is important, to systematically include explicit chiral symmetry breaking or couplings to electroweak gauge fields into the chiral effective Lagrangian.

\subsection{Explicit degrees of freedom} \label{subsec:dof}

In the low-energy regime of QCD, hadrons are the observable states.
The active degrees of freedom of \cheft are identified as the pseudoscalar Goldstone-boson octet.
The soft scale of the low-energy expansion is given by the small external momenta and the small masses of the pseudo-Goldstone bosons, while the large scale is a typical hadronic scale of about \(1\ \textrm{GeV}\).
The effective Lagrangian has to fulfill the same symmetry properties as QCD: invariance under Lorentz and parity transformations, charge conjugation and time reversal symmetry.
Especially the chiral symmetry and its spontaneous symmetry breaking has to be incorporated.
Using the external-field method, the \emph{same} external fields \(v,a,s,p\) as in \eq{eq:LagrangianExt}, with the same transformation behavior, are included in the effective Lagrangian.

\begin{figure}
\centering
{\small
\setlength{\unitlength}{0.00135\textwidth}
\begin{picture}(170,150)(-90,-80) \thicklines \linethickness{0.6pt}
    \put(-70,-62){\vector(0,1){124}}
    \put(-70,-62){\vector(1,0){140}}
    \put(-80,62){$S$}
    \put(-72,42){\line(1,0){4}}
    \put(-90,39){$+1$}
    \put(-72,0){\line(1,0){4}}
    \put(-84,-3){$0$}
    \put(-72,-42){\line(1,0){4}}
    \put(-90,-45){$-1$}
    \put(70,-72){$I_3$}
    \put(-50,-64){\line(0,1){4}}
    \put(-57,-77){$-1$}
    \put(-25,-64){\line(0,1){4}}
    \put(-32,-77){$-\frac12$}
    \put(0,-64){\line(0,1){4}}
    \put(-2,-77){$0$}
    \put(25,-64){\line(0,1){4}}
    \put(22,-77){$\frac12$}
    \put(50,-64){\line(0,1){4}}
    \put(48,-77){$1$}
    \put(-25,42){\color{lblue}\line(-3,-5){25}}
    \put(-25,42){\color{lblue}\line(1, 0){50}}
    \put(25,42){\color{lblue}\line(3,-5){25}}
    \put(-25,-42){\color{lblue}\line(1,0){50}}
    \put(-25,-42){\color{lblue}\line(-3, 5){25}}
    \put(25,-42){\color{lblue}\line(3, 5){25}}
    \put(-25,42){\circle*{3}}
    \put(-25,-42){\circle*{3}}
    \put(25,42){\circle*{3}}
    \put(25,-42){\circle*{3}}
    \put(-50,0){\circle*{3}}
    \put(50,0){\circle*{3}}
    \put(0,0){\circle*{3}}
    \put(0,0){\circle{6}}
    \put(-28,47){$K^0$}
    \put(28,47){$K^+$}
    \put(-28,-55){$K^{-}$}
    \put(28,-55){$\bar K^{0}$}
    \put(55,0){$\pi^{+}$}
    \put(-65,0){$\pi^{-}$}
    \put(0, 8){$\pi^{0}$}
    \put(0, -13){$\eta$}
\end{picture}
\setlength{\unitlength}{0.00135\textwidth}
\begin{picture}(170,150)(-90,-80) \thicklines \linethickness{0.6pt}
    \put(-70,-62){\vector(0,1){124}}
    \put(-70,-62){\vector(1,0){140}}
    \put(-80,62){$S$}
    \put(-72,42){\line(1,0){4}}
    \put(-84,39){$0$}
    \put(-72,0){\line(1,0){4}}
    \put(-90,-3){$-1$}
    \put(-72,-42){\line(1,0){4}}
    \put(-90,-45){$-2$}
    \put(70,-72){$I_3$}
    \put(-50,-64){\line(0,1){4}}
    \put(-57,-77){$-1$}
    \put(-25,-64){\line(0,1){4}}
    \put(-32,-77){$-\frac12$}
    \put(0,-64){\line(0,1){4}}
    \put(-2,-77){$0$}
    \put(25,-64){\line(0,1){4}}
    \put(22,-77){$\frac12$}
    \put(50,-64){\line(0,1){4}}
    \put(48,-77){$1$}
    \put(-25,42){\color{lblue}\line(-3,-5){25}}
    \put(-25,42){\color{lblue}\line(1, 0){50}}
    \put(25,42){\color{lblue}\line(3,-5){25}}
    \put(-25,-42){\color{lblue}\line(1,0){50}}
    \put(-25,-42){\color{lblue}\line(-3, 5){25}}
    \put(25,-42){\color{lblue}\line(3, 5){25}}
    \put(-25,42){\circle*{3}}
    \put(-25,-42){\circle*{3}}
    \put(25,42){\circle*{3}}
    \put(25,-42){\circle*{3}}
    \put(-50,0){\circle*{3}}
    \put(50,0){\circle*{3}}
    \put(0,0){\circle*{3}}
    \put(0,0){\circle{6}}
    \put(-28, 47){$n$}
    \put(28, 47){$p$}
    \put(-28, -55){$\Xi^{-}$}
    \put(28, -55){$\Xi^{0}$}
    \put(55, 0){$\Sigma^{+}$}
    \put(-65, 0){$\Sigma^{-}$}
    \put(0, 8){$\Sigma^{0}$}
    \put(0, -13){$\Lambda$}
\end{picture} \\[2ex]
\setlength{\unitlength}{0.00115\textwidth}
\begin{picture}(220,200)(-115,-124) \thicklines \linethickness{0.6pt}
    \put(-95,-100){\vector(0,1){166}}
    \put(-95,-100){\vector(1,0){190}}
    \put(-110,66){$S$}
    \put(-98,42){\line(1,0){6}}
    \put(-113,38){$0$}
    \put(-98,0){\line(1,0){6}}
    \put(-123,-4){$-1$}
    \put(-98,-42){\line(1,0){6}}
    \put(-123,-46){$-2$}
    \put(-98,-84){\line(1,0){6}}
    \put(-123,-88){$-3$}
    \put(95,-115){$I_3$}
    \put(-75,-103){\line(0,1){6}}
    \put(-85,-120){$-\frac32$}
    \put(-50,-103){\line(0,1){6}}
    \put(-60,-120){$-1$}
    \put(-25,-103){\line(0,1){6}}
    \put(-35,-120){$-\frac12$}
    \put(0,-103){\line(0,1){6}}
    \put(-3,-120){$0$}
    \put(25,-103){\line(0,1){6}}
    \put(22,-120){$\frac12$}
    \put(50,-103){\line(0,1){6}}
    \put(47,-120){$1$}
    \put(75,-103){\line(0,1){6}}
    \put(72,-120){$\frac32$}
    \put(-75,42){\color{lblue}\line(3,-5){75}}
    \put(-75, 42){\color{lblue}\line(1, 0){150}}
    \put(75, 42){\color{lblue}\line(-3,-5){75}}
    \put(-25, 42){\circle*{4}}
    \put(-25, -42){\circle*{4}}
    \put(25, 42){\circle*{4}}
    \put(25, -42){\circle*{4}}
    \put(-50, 0){\circle*{4}}
    \put(50, 0){\circle*{4}}
    \put(0, 0){\circle*{4}}
    \put(0, -84){\circle*{4}}
    \put(-75, 42){\circle*{4}}
    \put(75, 42){\circle*{4}}
    \put(-78, 47){$\Delta^{-}$}
    \put(-28, 47){$\Delta^{0}$}
    \put(28, 47){$\Delta^{+}$}
    \put(78, 47){$\Delta^{++}$}
    \put(-20, -45){$\Xi^{*-}$}
    \put(33, -45){$\Xi^{*0}$}
    \put(59, 0){$\Sigma ^{*+}$}
    \put(-45, 0){$\Sigma^{*-}$}
    \put(5, 0){$\Sigma^{*0}$}
    \put(7, -86){$\Omega^{-}$}
\end{picture}
\setlength{\unitlength}{\usualunit}
}
\caption{Pseudoscalar meson octet (\(J^P = 0^-\)), baryon octet (\(J^P=1/2^+\)) and baryon decuplet (\(J^P=3/2^+\)).} \label{fig:multiplets}
\end{figure}
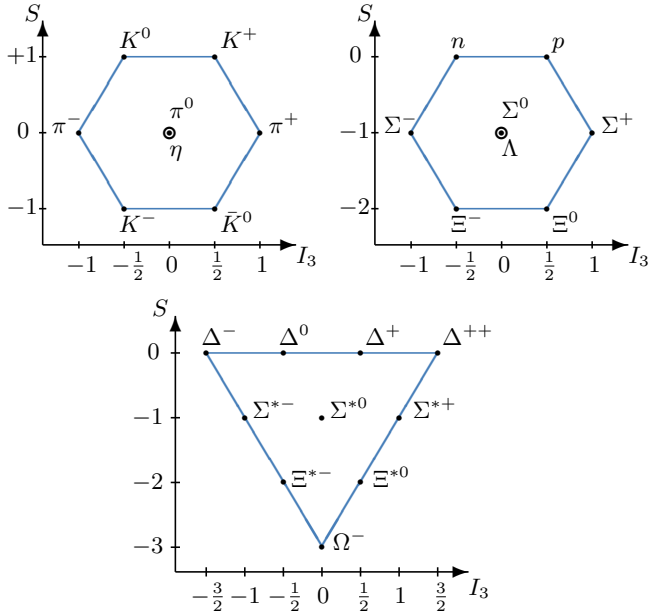

As the QCD vacuum is approximately invariant under the flavor symmetry group \(\mathrm{SU}(3)\), one expects the hadrons to organize themselves in multiplets of irreducible representations of \(\mathrm{SU}(3)\).
The pseudoscalar mesons form an octet, \cf \fig{fig:multiplets}.
The members of the octet are characterized by the strangeness quantum number \(S\) and the third component \(I_3\) of the isospin.
The symbol \(\eta\) stands for the octet component (\(\eta_8\)).
As an approximation we identify \(\eta_8\) with the physical \(\eta\), ignoring possible mixing with the singlet state \(\eta_1\).
For the lowest-lying baryons one finds an octet and a decuplet, see also \fig{fig:multiplets}.
In the following we summarize how these explicit degrees of freedom are included in the chiral Lagrangian in the standard non-linear realization of chiral symmetry~\ct*{Coleman1969,Callan1969}.

The chiral symmetry group \(\mathrm{SU}(3)_\mathrm L\times\mathrm{SU}(3)_\mathrm R\) is spontaneously broken to its diagonal subgroup \(\mathrm{SU}(3)_\mathrm V\).
Therefore the \emph{Goldstone-boson octet} should transform under \(\mathrm{SU}(3)_\mathrm L\times\mathrm{SU}(3)_\mathrm R\) such that an irreducible \(\mathbf 8\)-representation results for \(\mathrm{SU}(3)_\mathrm V\).
A convenient choice to describe the pseudoscalar mesons under these conditions is a unitary \(3\times3\) matrix \(U(x)\) in flavor space, which fulfills
\begin{equation}
 U^\dagger U = 1\,,\qquad \det U = 1 \,.
\end{equation}
The transformation behavior under chiral symmetry reads
\begin{equation} \label{eq:mestrans}
U \to R U L^\dagger \,,
\end{equation}
where \(L(x)\), \(R(x)\) are elements of \(\mathrm{SU}(3)_{L,R}\).
An explicit parametrization of \(U(x)\) in terms of the pseudoscalar mesons is given by
\begin{equation} \label{eq:mesonU}
 U(x) = \exp\left[\mathrm i\,\phi(x)/f_0\right]\,,
\end{equation}
with the traceless Hermitian matrix
\begin{align*} \label{eq:defphi}
 \phi(x)&=\sum_{a=1}^8 \phi_a(x) \lambda_a \\
 &=
 \begin{pmatrix}
  \pi^0 + \frac{1}{\sqrt 3}\eta & \sqrt 2 \pi^+ & \sqrt 2 K^+ \\
  \sqrt 2 \pi^- & -\pi^0 + \frac{1}{\sqrt 3}\eta & \sqrt 2 K^0 \\
  \sqrt 2 K^- & \sqrt 2 \bar K^0 & -\frac{2}{\sqrt 3}\eta
 \end{pmatrix}\,. \numberthis
\end{align*}
The constant \(f_0\) is the \emph{decay constant} of the pseudoscalar Goldstone bosons in the chiral limit.
For a transformation of the subgroup \(\mathrm{SU}(3)_\mathrm V\) with \(L=R=V\), the meson matrix \(U\) transforms as
\begin{equation}
U \to V U V^\dagger \,,
\end{equation}
\ie the mesons \(\phi_a(x)\) transform in the adjoint (irreducible) \(\mathbf 8\)-representation of \(\mathrm{SU}(3)\).
The parity transformation behavior of the pseudoscalar mesons is \(\phi_a(t,\vec x\,)\stackrel{P}{\to} -\phi_a(t,-\vec x\,)\) or, equivalently, \(U(t,\vec x\,)\stackrel{P}{\to} U^\dagger(t,-\vec x\,)\).
Under charge conjugation the particle fields are mapped to antiparticle fields, leading to \(U\stackrel{C}{\to} U^\top\).

The \emph{octet baryons} are described by Dirac spinor fields and represented in a traceless \(3\times3\) matrix \(B(x)\) in flavor space,
\begin{align*} \label{eq:defbarmatrix}
 B&=\sum_{a=1}^8 \frac{B_a \lambda_a}{\sqrt 2} \\
 &=
 \begin{pmatrix}
  \frac{1}{\sqrt 2}\Sigma^0 + \frac{1}{\sqrt 6}\Lambda & \Sigma^+ & p \\
  \Sigma^- & -\frac{1}{\sqrt 2}\Sigma^0 + \frac{1}{\sqrt 6}\Lambda & n \\
  \Xi^- & \Xi^0 & -\frac{2}{\sqrt 6}\Lambda
 \end{pmatrix} \,. \numberthis
\end{align*}
We use the convenient \ct*{Georgi1984} non-linear realization of chiral symmetry for the baryons, which lifts the well-known flavor transformations to the chiral symmetry group.
The matrix \(B(x)\) transforms under the chiral symmetry group \(\mathrm{SU(3)}_\mathrm L\times \mathrm{SU(3)}_\mathrm R\) as
\begin{equation} \label{eq:bartrans}
B \rightarrow K B K^\dagger\,,
\end{equation}
with the SU(3)-valued compensator field
\begin{equation} \label{eq:comp}
K\left(L,R,U\right) = \sqrt{LU^\dagger R^\dagger} R \sqrt U \,.
\end{equation}
 Note that \(K\left(L,R,U\right)\) also depends on the meson matrix \(U\).
The square root of the meson matrix,%
\begin{equation}
u=\sqrt U\,,
\end{equation}
transforms as \(u \rightarrow \sqrt{RUL^\dagger} = R u K^\dagger = K u L^\dagger\).
For transformations under the subgroup \(\mathrm{SU}(3)_\mathrm V\) the baryons transform as an octet, \ie the adjoint representation of \(\mathrm{SU}(3)\):
\begin{equation} \label{eq:barrepr}
 B\rightarrow V B V^\dagger\,.
\end{equation}
The octet-baryon fields transform under parity and charge conjugation as
\(B_a\left(t,\vec x\,\right) \stackrel{P}{\to} \gamma^0 B_a\left(t,-\vec x\,\right)\) and
\(B_{\alpha,a} \stackrel{C}{\to} C_{\alpha\beta} \bar B_{\beta,a}\)
with the Dirac-spinor indices \(\alpha,\beta\) and with \(C=\mathrm i\gamma^2 \gamma^0\).

A natural choice to represent the \emph{decuplet baryons} is a totally symmetric three-index tensor \(T\).
It transforms under the chiral symmetry \(\mathrm{SU}(3)_\mathrm L \times \mathrm{SU}(3)_\mathrm R\) as
\begin{equation}
 T_{abc} \rightarrow K_{ad}K_{be}K_{cf}T_{def} \,,
\end{equation}
with the compensator field \(K(L,R,U)\) of Eq.~\eqref{eq:comp}.
For an \(\mathrm{SU}(3)_\mathrm V\) transformation the decuplet fields transform as an irreducible representation of \(\mathrm{SU}(3)\):%
\begin{equation}
 T_{abc} \rightarrow V_{ad}V_{be}V_{cf}T_{def} \,.
\end{equation}
The physical fields are assigned to the following components of the totally antisymmetric tensor:
\begin{align*} \label{eq:Tfields}
  &T^{111}=\Delta^{++}\,, T^{112}=\tfrac{1}{\sqrt{3}}\Delta^{+}\,, T^{122}=\tfrac{1}{\sqrt{3}}\Delta^{0}\,, T^{222}=\Delta^{-}\,, \\
  &T^{113}=\tfrac{1}{\sqrt{3}}\Sigma^{*+}\,,\ T^{123}=\tfrac{1}{\sqrt{6}}\Sigma^{*0}\,,\ T^{223}=\tfrac{1}{\sqrt{3}}\Sigma^{*-}\,, \\
  &T^{133}=\tfrac{1}{\sqrt{3}}\Xi^{*0}\,,\ T^{233}=\tfrac{1}{\sqrt{3}}\Xi^{*-}\,, \\
  &T^{333}=\Omega^-\,. \numberthis
\end{align*}
Since decuplet baryons are spin-3/2 particles, each component is expressed through \emph{Rarita-Schwinger fields}.
Within the scope of this article, decuplet baryons are only used for estimating LECs via decuplet resonance saturation.
In that case it is sufficient to treat them in their non-relativistic form, where no complications with the Rarita-Schwinger formalism arise.

Now the representation of the explicit degrees of freedom and their transformation behavior are established.
Together with the external fields the construction of the chiral effective Lagrangian is straightforward.

\subsection{Construction of the chiral Lagrangian} \label{subsec:constchlag}

The chiral Lagrangian can be ordered according to the number of baryon fields:
\begin{equation} \label{eq:barLagrExpand}
\mathscr{L}_\mathrm{eff} = \mathscr{L}_\phi + \mathscr{L}_{B} + \mathscr{L}_{BB} + \mathscr{L}_{BBB} + \dots \,,
\end{equation}
where \(\mathscr{L}_\phi\) denotes the purely mesonic part of the Lagrangian.
Each part is organized in the number of small momenta (\ie derivatives) or small meson masses, \eg
\begin{equation}
\mathscr{L}_\phi = \mathscr{L}^{(2)}_\phi + \mathscr{L}^{(4)}_\phi + \mathscr{L}^{(6)}_\phi + \dots \,.
\end{equation}
\(\mathscr{L}_\phi\) has been constructed to \(\mathcal{O}(q^6)\) in \cts{Fearing1996,Bijnens1999}.
The chiral Lagrangian for the baryon-number-one sector has been investigated in various works.
The chiral effective pion-nucleon Lagrangian of order \(\mathcal O (q^4)\) has been constructed in \ct{Fettes2000}.
The three-flavor Lorentz invariant chiral meson-baryon Lagrangians \(\mathscr{L}_{B}\) at order \(\mathcal O (q^2)\) and \(\mathcal O (q^3)\) have been first formulated in \ct{Krause1990} and were later completed in \cts{Oller2006,Frink2006}.
Concerning the nucleon-nucleon contact terms, the relativistically invariant contact Lagrangian at order \(\mathcal O (q^2)\) for two flavors (without any external fields) has been constructed in \ct{Girlanda2010}.
The baryon-baryon interaction Lagrangian  \(\mathscr{L}_{BB}\) has been considered up to NLO in \cts{Savage1996,Polinder2006,Petschauer2013a}.
Furthermore the leading three-baryon contact interaction Lagrangian \(\mathscr{L}_{BBB}\) has been derived in \ct{Petschauer2016}.

We follow closely \ct{Petschauer2013a} to summarize the basic procedure for constructing systematically the three-flavor chiral effective Lagrangian \ct*{Coleman1969,Callan1969} with the inclusion of external fields \ct*{Gasser1984,Gasser1985}.
The effective chiral Lagrangian has to fulfill all discrete and continuous symmetries of the strong interaction.
Therefore it has to be invariant under parity (\(P\)), charge conjugation (\(C\)), Hermitian conjugation (\(H\)) and the proper, orthochronous Lorentz transformations.
Time reversal symmetry is then automatically fulfilled via the \(CPT\) theorem.
Especially \emph{local} chiral symmetry has to be fulfilled.
A common way to construct the chiral Lagrangian is to define so-called \emph{building blocks}, from which the effective Lagrangian can be determined as an invariant polynomial.
Considering the chiral transformation properties, a convenient choice for the building blocks is
\begin{align*} \label{eq:buildingblocks}
 u_\mu &= \mathrm i \left[ u^\dagger\left(\partial_\mu-\mathrm i\, r_\mu \right) u - u\left(\partial_\mu - \mathrm i\, l_\mu\right)u^\dagger\right]\,,\\
 \chi_\pm &= u^\dagger\chi u^\dagger \pm u\chi^\dagger u\,,\\
 f_{\mu\nu}^\pm &= uf_{\mu\nu}^\mathrm L u^\dagger \pm u^\dagger f_{\mu\nu}^\mathrm R u\,, \numberthis
\end{align*}
with the combination
\begin{equation}
\chi= 2B_0\left(s+\mathrm i\,p\right)\,,
\end{equation}
containing the new parameter \(B_0\) and the external scalar and pseudoscalar fields.
One defines external field strength tensors by
\begin{align*}
 f^\mathrm R_{\mu\nu}&= \partial_\mu r_\nu - \partial_\nu r_\mu - \mathrm i\left[r_\mu,r_\nu\right]\,, \\
 f^\mathrm L_{\mu\nu}&= \partial_\mu l_\nu - \partial_\nu l_\mu - \mathrm i\left[l_\mu,l_\nu\right]\,, \numberthis
\end{align*}
where the fields
\begin{equation}
 r_\mu=v_\mu+a_\mu\,,\quad l_\mu=v_\mu-a_\mu\,,
\end{equation}
describe right handed and left handed external vector fields.
In the absence of flavor singlet couplings one can assume \(\trace{a_\mu}=\trace{v_\mu}=0\), where \(\trace{\dots}\) denotes the flavor trace.
Therefore, the fields \(u_\mu\) and \(f_{\mu\nu}^\pm\) in \eq{eq:buildingblocks} are all traceless.

Using the \emph{transformation behavior} of the pseudoscalar mesons and octet baryons in \eq{eq:mestrans} and \eq{eq:bartrans}, and the transformation properties of the external fields in \eq{eq:extrans}, one can determine the transformation behavior of the building blocks.
All building blocks \(A\), and therefore all products of these, transform according to the adjoint (octet) representation of SU(3), \ie \(A\rightarrow KAK^\dagger\).
Note that traces of products of such building blocks are invariant under local chiral symmetry, since \(K^\dagger K=\mathbbm1\).
The chiral covariant derivative of such a building block \(A\) is given by
\begin{equation} \label{eq:covder}
 D_\mu A = \partial_\mu A + \left[\Gamma_\mu,A\right]\,,
\end{equation}
with the chiral connection
\begin{equation}
 \Gamma_\mu = \frac12 \left[ u^\dagger \left(\partial_\mu -\mathrm i\, r_\mu\right) u + u \left(\partial_\mu -\mathrm i\, l_\mu\right) u^\dagger \right]\,.
\end{equation}
The covariant derivative transforms homogeneously under the chiral group as \(D_\mu A \rightarrow K \left(D_\mu A\right) K^\dagger\).
The chiral covariant derivative of the baryon field \(B\) is given by \eq{eq:covder} as well.

A Lorentz-covariant \emph{power counting scheme} has been introduced by Krause in \ct{Krause1990}.
Due  to the large baryon mass \(M_0\) in the chiral limit, a time-derivative acting on a baryon field \(B\) cannot be counted as small.
Only baryon three-momenta are small on typical chiral scales.
This leads to the following counting rules for baryon fields and their covariant derivatives,
\begin{equation}
 B\,,\ \bar B\,,\ D_\mu B \sim \mathcal{O}\big(q^0\big)\,,\qquad \left(\mathrm i \slashed{D} - M_0\right) B \sim \mathcal{O}\left(q\right)\,.
\end{equation}
The chiral dimension of the chiral building blocks and baryon bilinears \(\bar B\Gamma B\) are given in \tab{tab:blocksbar}.
A covariant derivative acting on a building block (but not on \(B\)) raises the chiral dimension by one.

A building block \(A\) transforms under parity, charge conjugation and Hermitian conjugation as
\begin{equation} \label{eq:buildtransnum}
 A^P = (-1)^{p} A\,,\quad A^C = (-1)^{c} A^\top\,,\quad A^\dagger = (-1)^{h} A\,,
\end{equation}
with the exponents (modulo two) \(p,c,h\in\{0,1\}\) given in \tab{subtab:blockbar}, and \(\top\) denotes the transpose of a (flavor) matrix.
A sign change of the spatial argument, \(\left(t,\vec x\right) \rightarrow \left(t,-\vec x\right)\), is implied in the fields in case of parity transformation \(P\).
Lorentz indices transform with the matrix \({P^\mu}_\nu=\diag(+1,-1,-1,-1)\) under parity transformation, \eg \((u^\mu)^P=(-1)^p{P^\mu}_\nu u^\nu\).
The transformation behavior of commutators and anticommutators of two building blocks \(A_1,\ A_2\) is the same as for building block and should therefore be used instead of simple products, \eg
\begin{align*}
 [A_1,A_2]_\pm^C &= (-1)^{c_1+c_2}(A_1^\top A_2^\top \pm A_2^\top A_1^\top) \\
 &= \pm (-1)^{c_1+c_2}[A_1,A_2]_\pm^\top\,. \numberthis
\end{align*}
The behavior under Hermitian conjugation is the same.
\begin{table}
\setlength{\tabcolsep}{11pt}
\centering
\begin{subtable}[t]{.43\textwidth}
 \centering
 \begin{tabular}{>{\(}c<{\)}>{\(}c<{\)}>{\(}c<{\)}>{\(}c<{\)}>{\(}c<{\)}}
 \toprule
 & p & c & h & \mathcal{O} \\
 \cmidrule(lr){1-1}\cmidrule(lr){2-5}
 u_\mu & 1 & 0 & 0 & \mathcal{O}\left(q^1\right)\\
 f_{\mu\nu}^+ & 0 & 1 & 0 & \mathcal{O}\left(q^2\right)\\
 f_{\mu\nu}^- & 1 & 0 & 0 & \mathcal{O}\left(q^2\right)\\
 \chi_+ & 0 & 0 & 0 & \mathcal{O}\left(q^2\right)\\
 \chi_- & 1 & 0 & 1 & \mathcal{O}\left(q^2\right)\\
 \bottomrule
 \end{tabular}
 \caption{Chiral building blocks}\label{subtab:blockbar}
\end{subtable}\\[2ex]
\begin{subtable}[t]{.43\textwidth}
 \centering
 \begin{tabular}{>{\(}c<{\)}>{\(}c<{\)}>{\(}c<{\)}>{\(}c<{\)}>{\(}c<{\)}}
 \toprule
 \Gamma & p & c & h & \mathcal{O} \\
 \cmidrule(lr){1-1}\cmidrule(lr){2-5}
 \mathbbm{1} & 0 & 0 & 0 & \mathcal{O}\left(q^0\right)\\
 \gamma_5 & 1 & 0 & 1 & \mathcal{O}\left(q^1\right)\\
 \gamma_\mu & 0 & 1 & 0 & \mathcal{O}\left(q^0\right)\\
 \gamma_5\gamma_\mu & 1 & 0 & 0 & \mathcal{O}\left(q^0\right)\\
 \sigma_{\mu\nu} & 0 & 1 & 0 & \mathcal{O}\left(q^0\right)\\
 \bottomrule
 \end{tabular}
 \caption{Baryon bilinears \(\bar B \Gamma B\)}\label{subtab:clifford}
\end{subtable}
\caption{Behavior under parity, charge conjugation and Hermitian conjugation as well as the chiral dimensions of chiral building blocks and baryon bilinears \(\bar B \Gamma B\) \protect\ct*{Oller2006}.}
\label{tab:blocksbar}
\setlength{\tabcolsep}{6pt}
\end{table}
The basis elements of the Dirac algebra forming the baryon bilinears transform as
\begin{align*} \label{eq:gammatransnum}
 &\gamma_0\Gamma\gamma_0 = (-1)^{p_\Gamma} \Gamma\,,\quad C^{-1}\Gamma C = (-1)^{c_\Gamma} \Gamma^\top\,,\\
 &\gamma_0\Gamma^\dagger\gamma_0 = (-1)^{h_\Gamma} \Gamma\,, \numberthis
\end{align*}
where the exponents \(p_\Gamma, c_\Gamma, h_\Gamma\in\{0,1\}\) can be found in \tab{subtab:clifford}.
As before, Lorentz indices of baryon bilinears transform with the matrix \({P^\mu}_\nu\) under parity.

Due to the identity
\begin{equation}
\left[D_\mu,D_\nu\right] A = \frac14 \left[\left[u_\mu,u_\nu\right],A\right] - \frac{\mathrm i}{2} \left[f^+_{\mu\nu},A\right]
\end{equation}
it is sufficient to use only totally symmetrized products of covariant derivatives, \(D^{\alpha\beta\gamma\dots}A\), for any building block \(A\) (or baryon field \(B\)).
Moreover, because of the relation
\begin{equation}
 D_\nu u_\mu - D_\mu u_\nu = f^-_{\mu\nu}\,,
\end{equation}
only the symmetrized covariant derivative acting on \(u_\mu\,\) need to be taken into account,
\begin{equation} \label{eq:defhmunu}
 h_{\mu\nu} = D_\mu u_\nu + D_\nu u_\mu\,.
\end{equation}

Finally, the chiral effective Lagrangian can be constructed by taking traces (and products of traces) of different polynomials in the building blocks, so that they are invariant under chiral symmetry, Lorentz transformations, \(C\) and \(P\).

\subsubsection{Leading-order meson Lagrangian}

As a first example, we show the leading-order purely mesonic Lagrangian.
From the general construction principles discussed above, one obtains for the leading-order effective Lagrangian
\begin{equation} \label{eq:fulllomesonlagr}
\mathscr{L}_\phi^{(2)} = \frac{f_0^2}4 \trace{u_\mu u^\mu + \chi_+} \,.
\end{equation}
Note that there is no contribution of order \(\mathcal O(q^0)\).
This is consistent with the vanishing interaction of the Goldstone bosons in the chiral limit at zero momenta.

Before we continue with the meson-baryon interaction Lagrangian, let us elaborate on the leading chiral Lagrangian in the purely mesonic sector without external fields, but with non-vanishing quark masses in the isospin limit: \(v^\mu(x)=a^\mu(x)=p(x)=0\) and \(s(x)=M=\diag\left(m,m,m_s\right)\).
Inserting the definitions of the building blocks, \eq{eq:fulllomesonlagr} becomes with these restrictions:%
\begin{equation} \label{eq:lomesonlagr}
\mathscr{L}_\phi^{(2)} = \frac{f_0^2}4 \trace{\partial_\mu U \partial^\mu U^\dagger} + \frac12 B_0 f_0^2 \trace{M U^\dagger+U M} \,.
\end{equation}
The physical decay constants \(f_\pi\neq f_K\neq f_\eta\) differ from the decay constant of the pseudoscalar Goldstone bosons in the chiral limit \(f_0\) in terms of order \((m,m_s)\):
\(f_\phi=f_0\left\{1+\mathcal{O}\left(m,m_s\right)\right\}\).
The constant \(B_0\) is related to the chiral quark condensate.
Already from this leading-order Lagrangian famous relations such as the (reformulated) Gell-Mann--Oakes--Renner relations
\begin{align*} \label{eq:GOL}
m^2_\pi &= 2 m B_0 + \mathcal{O}(m_q^2)\,,\\
m^2_K &= \left(m+m_s\right) B_0 + \mathcal{O}(m_q^2)\,,\\
m^2_\eta &= \frac{2}{3} \left(m+2m_s\right) B_0 + \mathcal{O}(m_q^2)\,, \numberthis
\end{align*}
or the Gell-Mann--Okubo mass formula, \(4 m_K^2 = 3 m_\eta^2 + m_\pi^2\), can be derived systematically.

\subsubsection{Leading-order meson-baryon interaction Lagrangian}

The leading-order meson-baryon interaction Lagrangian \(\mathscr{L}_\mathrm{B}^{(1)}\) is of order \(\mathcal{O}(q)\) and reads%
\footnote{Note that an overall plus sign in front of the constants \(D\) and \(F\) is chosen, consistent with the conventions in SU(2) \cheft \ct*{Epelbaum2009}.}
\begin{align*} \label{eq:barmeslagr}
 \mathscr{L}_\mathrm{B}^{(1)}=&\trace{\bar B \left(i \slashed{D} -M_\mathrm B\right) B} + \frac D 2 \trace{\bar B \gamma^\mu \gamma_5 \lbrace u_\mu,B\rbrace}\\
& + \frac F 2 \trace{\bar B \gamma^\mu \gamma_5 \left[u_\mu,B\right]}\,. \numberthis
\end{align*}
The constant \(M_\mathrm B\) is the mass of the baryon octet in  the chiral limit.
The two new constants \(D\) and \(F\) are called axial-vector coupling constants.
Their values can be obtained from semi-leptonic hyperon decays and are roughly \(D\approx0.8\) and \(F\approx0.5\) \ct*{Borasoy1999a}.
The sum of the two constants is related to the axial-vector coupling constant of nucleons, \(g_\mathrm A = D+F=1.27\), obtained from neutron beta decay.
At lowest order the pion-nucleon coupling constant \(g_{\pi N}\) is connected to the axial-vector coupling constant by the Goldberger-Treiman relation, \(g_{\pi N} f_\pi = g_\mathrm A M_N\).
The covariant derivative in \eq{eq:barmeslagr} includes the field \(\Gamma_\mu\), which leads to a vertex between two octet baryons and two mesons, whereas the terms containing \(u_\mu\) lead to a vertex between two octet baryons and one meson.
Different octet-baryon masses appear first in \(\mathscr{L}_\mathrm{B}^{(2)}\) due to explicit chiral symmetry breaking and renormalization and lead to corrections linear in the quark masses:
\begin{equation} \label{eq:barmasscorrection}
 M_i = M_\mathrm B + \mathcal{O}(m,m_s)\,.
\end{equation}

\subsection{Weinberg power counting scheme} \label{subsec:pwr}

As stated before, an effective field theory has an infinite number of terms in the effective Lagrangian and for a fixed process an infinite number of diagrams contribute.
Therefore, it is crucial to have a power counting scheme, to assign the importance of a term.
Then, to a certain order in the power counting, only a \emph{finite number} of terms contribute and the observables can be calculated to a given accuracy.

First, let us discuss the power counting scheme of \cheft in the pure meson sector, \ie only the pseudoscalar Goldstone bosons are explicit degrees of freedom.
The \emph{chiral dimension} \(\nu\) of a Feynman diagram represents the order in the low-momentum expansion, \((q/{\Lambda_\chi})^\nu\).
The symbol \(q\) is generic for a small external meson momentum or a small meson mass.
The scale of chiral symmetry breaking \(\Lambda_\chi\) is often estimated as \(4\pi f_\pi\approx 1\ \mathrm{GeV}\) or as the mass of the lowest-lying resonance, \(M_\rho \approx 770\ \mathrm{MeV}\).
A simple dimensional analysis leads to the following expression for the chiral dimension of a connected Feynman diagram \ct*{Weinberg1979}:
\begin{equation} \label{eq:mespwr}
\nu = 2 + 2L + \sum_i v_i \Delta_i\,, \qquad \Delta_i = d_i - 2 \,.
\end{equation}
The number of Goldstone boson loops is denoted by \(L\) and \(v_i\) is the number of vertices with vertex dimension \(\Delta_i\).
The symbol \(d_i\) stands for the number of derivatives or meson mass insertions at the vertex, \ie the vertex originates from a term of the Lagrangian of the order \(\mathcal O(q^{d_i})\).

With the introduction of baryons in the chiral effective Lagrangian, the power counting is more complicated.
The large baryon mass comes as an extra scale and destroys the one-to-one correspondence between the loop and the small momentum expansion.
Jenkins and Manohar used methods from heavy-quark effective field theory to solve this problem \ct*{Jenkins1991}.
Basically they considered baryons as heavy, static sources.
This leads to a description of the baryons in the extreme non-relativistic limit with an expansion in powers of the inverse baryon mass, called heavy-baryon chiral perturbation theory.
\begin{figure}
\centering
\(
\vcenter{\hbox{\includegraphics[scale=.4]{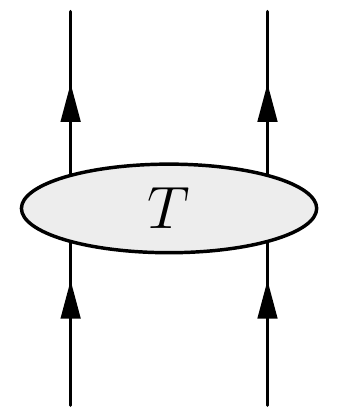}}}
\ =\
\vcenter{\hbox{\includegraphics[scale=.4]{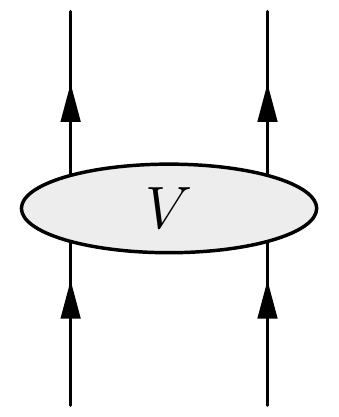}}}
\ +\
\vcenter{\hbox{\includegraphics[scale=.4]{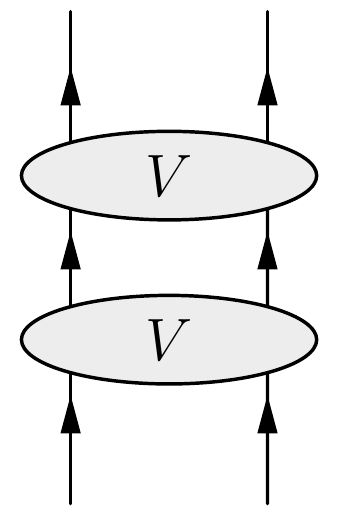}}}
\ +\
\vcenter{\hbox{\includegraphics[scale=.4]{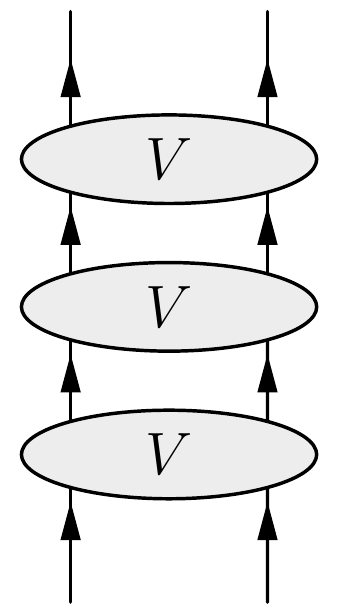}}}
\ +
\) \
{\large \(\cdots\)}
\caption{Graphical representation of the Lippmann-Schwinger equation.}
\label{fig:LSE}
\end{figure}
\begin{figure}
\centering
\vspace{\baselineskip}
$\vcenter{\hbox{
\begin{overpic}[scale=0.6]{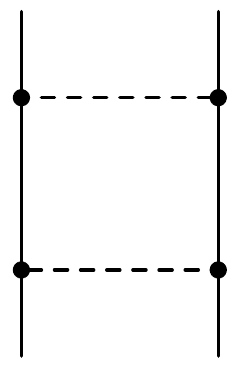}
\small
\put(-1,-9){$\Lambda$}
\put(52,-9){$N$}
\put(9,33){$\Lambda$}
\put(44,33){$N$}
\put(-1,102){$\Lambda$}
\put(52,102){$N$}
\put(28,15){$\eta$}
\put(28,80){$\eta$}
\linethickness{1.8pt}
\multiput(-8,50)(14,0){6}{\color{red}\line(1,0){10}}
\end{overpic}
}}$
\vspace{.3\baselineskip}
\caption{
Example of a planar box diagram.
It contains an reducible part equivalent to the iteration of two one-meson exchange diagrams, as generated by the Lippmann-Schwinger equation.
Additionally it contains a genuine irreducible contribution that is part of the effective potential.}
\label{fig:irrPb}
\end{figure}
Furthermore, in the two-baryon sector, additional features arise.
Reducible Feynman diagrams are enhanced due to the presence of small kinetic energy denominators resulting from purely baryonic intermediate states.
These graphs hint at the non-perturbative aspects in few-body problems, such as the existence of shallow bound states, and must be summed up to all orders.
As suggested by Weinberg \ct*{Weinberg1990,Weinberg1991}, the baryons can be treated non-relativistically and the power counting scheme can be applied to an effective potential \(V\), that contains only irreducible Feynman diagrams.
Terms with the inverse baryon mass \(M_\mathrm B^{-1}\) may be counted as
\begin{equation}
\frac{q}{M_\mathrm B} \propto \Big(\frac{q}{\Lambda_\chi}\Big)^2 \,.
\end{equation}

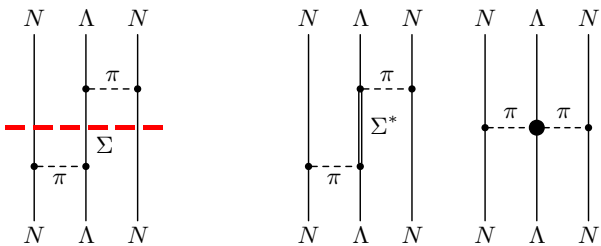
\begin{figure}
\centering
\vspace{\baselineskip}
$\vcenter{\hbox{
\begin{overpic}[scale=0.55]{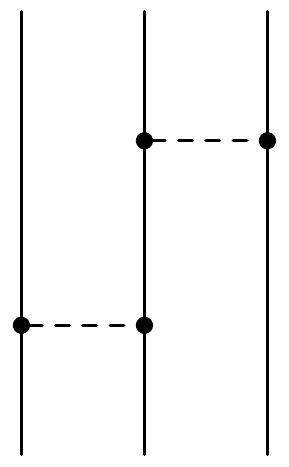}
\small
\put(-1,-9){$N$}
\put(27,-9){$\Lambda$}
\put(52,-9){$N$}
\put(-1,102){$N$}
\put(27,102){$\Lambda$}
\put(52,102){$N$}
\put(14,21){$\pi$}
\put(41,74){$\pi$}
\put(36,37){$\Sigma$}
\linethickness{1.8pt}
\multiput(-10,50)(14,0){6}{\color{red}\line(1,0){10}}
\end{overpic}
\qquad\qquad\qquad
\begin{overpic}[scale=0.55]{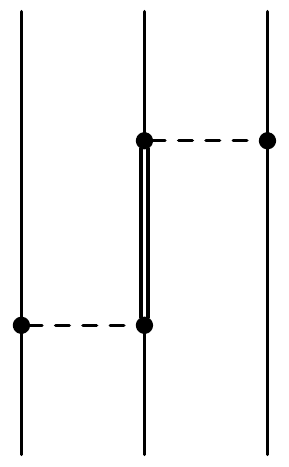}
\small
\put(-1,-9){$N$}
\put(27,-9){$\Lambda$}
\put(52,-9){$N$}
\put(-1,102){$N$}
\put(27,102){$\Lambda$}
\put(52,102){$N$}
\put(14,21){$\pi$}
\put(41,74){$\pi$}
\put(36,47){$\Sigma^*$}
\end{overpic}
\qquad
\begin{overpic}[scale=0.55]{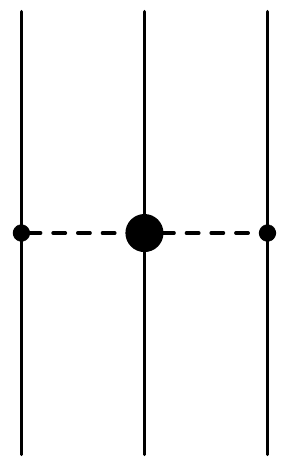}
\small
\put(-1,-9){$N$}
\put(27,-9){$\Lambda$}
\put(52,-9){$N$}
\put(-1,102){$N$}
\put(27,102){$\Lambda$}
\put(52,102){$N$}
\put(14,55){$\pi$}
\put(41,55){$\pi$}
\end{overpic}
}}$
\vspace{.3\baselineskip}
\caption{
Examples for reducible (left) and irreducible (right) three-baryon interactions for \(\Lambda NN\).
The thick dashed line cuts the reducible diagram in two two-body interaction parts.
}
\label{fig:irr3BF}
\end{figure}

The resulting effective potential is the input for quantum mechanical few-body calculations.
In case of the baryon-baryon interaction the effective potential is inserted into the Lippmann-Schwinger equation and solved for bound and scattering states.
This is graphically shown in \fig{fig:LSE} and \fig{fig:irrPb}.
The \(T\)-matrix is obtained from the infinite series of ladder diagrams with the effective potential \(V\).
In this way the omitted reducible diagrams are regained.
In the many-body sector, \eg Faddeev (or Yakubovsky) equations are typically solved within a coupled-channel approach.
In a similar way reducible diagrams such as on the left-hand side of \fig{fig:irr3BF}, are generated automatically and are not part of the effective potential.
One should distinguish such iterated two-body interactions, from irreducible three-baryon forces, as shown on the right-hand side of \fig{fig:irr3BF}.

\begin{figure*}
\centering
\newcommand{\hierscale}{.5}
\newcommand{\hierscaleBBB}{.4}
\setlength{\tabcolsep}{12pt}
\begin{tabular}{lcc}
\toprule \addlinespace[1ex]
& two-baryon force & three-baryon force \\ \addlinespace[1ex] \midrule \addlinespace[2ex]
LO & $\vcenter{\hbox{
\includegraphics[scale=\hierscale]{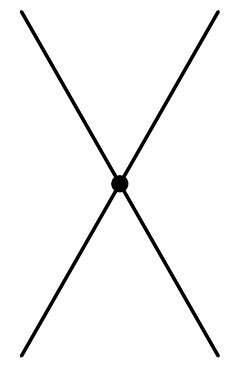}\quad
\includegraphics[scale=\hierscale]{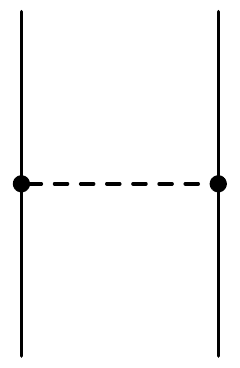}}}$ & \\ \addlinespace[2ex]
NLO &
$\vcenter{\hbox{
\includegraphics[scale=\hierscale]{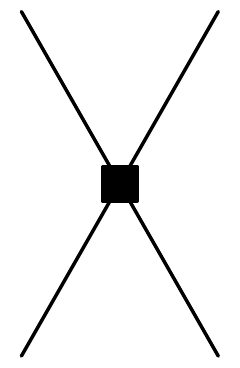}\quad
\includegraphics[scale=\hierscale]{files/Feynman/BF/FBBpb}\quad
\includegraphics[scale=\hierscale]{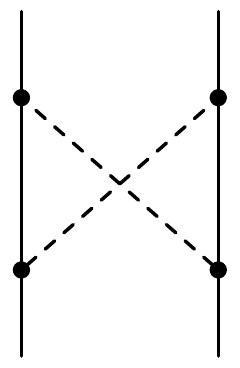}}}$ & \\ \addlinespace[2ex]
&
$\vcenter{\hbox{
\includegraphics[scale=\hierscale]{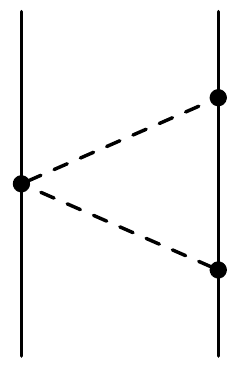}\quad
\includegraphics[scale=\hierscale]{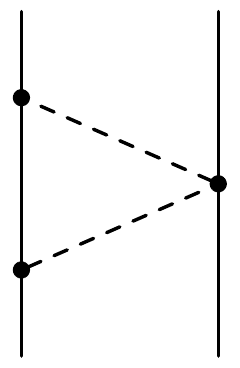}\quad
\includegraphics[scale=\hierscale]{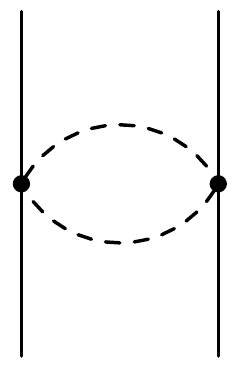}}}$ & \\ \addlinespace[2ex]
NNLO &
$\vcenter{\hbox{
\includegraphics[scale=\hierscale]{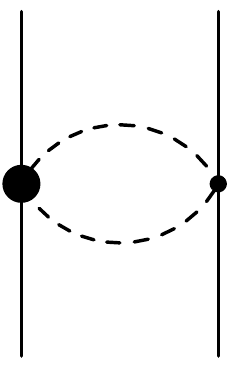}\quad
\includegraphics[scale=\hierscale]{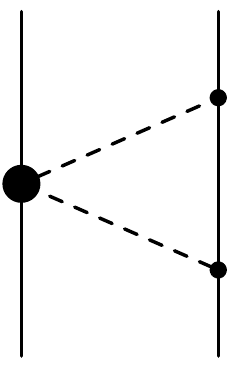}}}\quad \cdots$ &
$\vcenter{\hbox{
\includegraphics[scale=\hierscaleBBB]{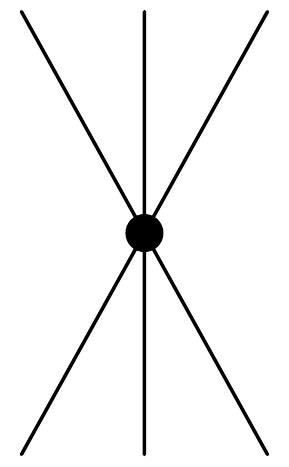}\quad
\includegraphics[scale=\hierscaleBBB]{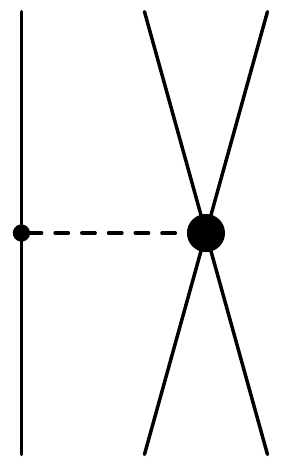}\quad
\includegraphics[scale=\hierscaleBBB]{files/Feynman/BF/FBBB2ME2}}}$ \\ \addlinespace[2ex]
\bottomrule
\end{tabular}
\caption{Hierarchy of baryonic forces.
Solid lines are baryons, dashed lines are pseudoscalar mesons.
Solid dots, filled circles and squares denote vertices with \(\Delta_i=0,1\text{ and }2\), respectively.}
\label{fig:hier}
\setlength{\tabcolsep}{6pt}
\end{figure*}

After these considerations, a consistent power counting scheme for the effective potential \(V\) is possible.
The soft scale \(q\) in the low-momentum expansion \((q/{\Lambda_\chi})^\nu\) denotes now small external meson four-momenta, small external baryon three-momenta or the small meson masses.
Naive dimensional analysis leads to the generalization of \eq{eq:mespwr}:
\begin{equation} \label{eq:pwrtmp}
\nu = 2 - B + 2L + \sum_i v_i \Delta_i\,, \quad \Delta_i = d_i + \frac12 b_i - 2 \,,
\end{equation}
where \(B\) is the number of external baryons and \(b_i\) is the number of internal baryon lines at the considered vertex.
However, \eq{eq:pwrtmp} has an unwanted dependence on the baryon number, due to the normalization of baryon states.
Such an effect can be avoided by assigning the chiral dimension to the transition operator instead of the matrix elements.
This leads to the addition of \(3B-6\) to the formula for the chiral dimension, which leaves the \(B=2\) case unaltered, and one obtains (see for example \cts{Bernard1995,Epelbaum2009,Machleidt2011,Scherer2012})
\begin{equation} \label{eq:barpwr}
\nu = -4 + 2B + 2L + \sum_i v_i \Delta_i\,, \quad \Delta_i = d_i + \frac12 b_i - 2 \,.
\end{equation}
Following this scheme one arrives at the hierarchy of baryonic forces shown in \fig{fig:hier}.
The leading-order (\(\nu=0\)) potential is given by one-meson-exchange diagrams and non-derivative four-baryon contact terms.
At next-to-leading order (\(\nu=2\)) higher order contact terms and two-meson-exchange diagrams with intermediate octet baryons contribute.
Finally, at next-to-next-to-leading order (\(\nu=3\)) the three-baryon forces start to contribute.
Diagrams that lead to mass and coupling constant renormalization are not shown.

\section{Baryon-baryon interaction potentials}  \label{sec:BB}

This section is devoted to the baryon-baryon interaction potentials up to next-to-leading order, constructed from the diagrams shown in \fig{fig:hier}.
Contributions arise from contact interaction, one- and two-Goldstone-boson exchange.
The constructed potentials serve not only as input for the description of baryon-baryon scattering, but are also basis for few- and many-body calculations.
We give also a brief introduction to common meson-exchange models and the difference to interaction potentials from \cheft.

\subsection{Baryon-baryon contact terms} \label{subsec:potct}

\begin{figure}
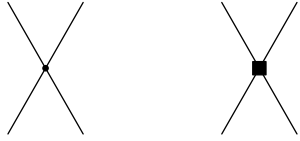

 \centering
 \includegraphics[scale=0.5]{files/Feynman/BF/FBBcont}\qquad\qquad\quad
 \includegraphics[scale=0.5]{files/Feynman/BF/FBBcontNLO}
 \caption{Leading-order and next-to-leading-order baryon-baryon contact vertices.} \label{fig:BBctLONLO}
\end{figure}

The chiral Lagrangian necessary for the contact vertices shown in \fig{fig:BBctLONLO} can be constructed straightforwardly according to the principles outlined in \sect{sec:ChEFT}.
For pure baryon-baryon scattering processes, no pseudoscalar mesons are involved in the contact vertices and almost all external fields can be dropped.
Covariant derivatives \(D_\mu\) reduce to ordinary derivatives \(\partial_\mu\).
The only surviving external field is \(\chi_+\), which is responsible for the inclusion of quark masses into the chiral Lagrangian:
\begin{align*} \label{eq:chi}
\frac{\chi_+}2 &= \chi = 2B_0 \begin{pmatrix} m_u & 0 & 0 \\ 0 & m_d & 0 \\ 0 & 0 & m_s \end{pmatrix} \\
&\approx \begin{pmatrix} m_\pi^2 & 0 & 0 \\ 0 & m_\pi^2 & 0 \\ 0 & 0 & 2m_K^2-m_\pi^2 \end{pmatrix} \,, \numberthis
\end{align*}
where in the last step the Gell-Mann--Oakes--Renner relations, \eq{eq:GOL}, have been used.
In flavor space the possible terms are of the schematic form
\begin{equation}
 \trace{\bar B B \bar B B}\,,\
 \trace{\bar B\bar B BB}\,,\
 \trace{\bar BB} \trace{\bar BB}\,,\
 \trace{\bar B\bar B}\trace{BB}\,,
\end{equation}
and terms where the field \(\chi\) is inserted such as
\begin{equation}
 \trace{\bar B\chi B \bar B B}\,,\
 \trace{\bar B B \chi \bar BB}\,,\
 \trace{\bar B \chi B} \trace{\bar BB}\,,\
 \dots
\end{equation}
where in both cases appropriate structures in Dirac space have to be inserted.
For the case of the non-relativistic power counting it would also be sufficient, to insert the corresponding structures in spin-momentum space.
The terms involving \(\chi\) lead to explicit SU(3) symmetry breaking at NLO linear in the quark masses.
A set of linearly independent Lagrangian terms up to \(\mathcal O(q^2)\) for pure baryon-baryon interaction in non-relativistic power counting can be found in \ct{Petschauer2013a}.

After a non-relativistic expansion up to \(\mathcal O (q^2)\) the four-baryon contact Lagrangian leads to potentials in spin and momentum space. 
A convenient operator basis is given by \ct*{Polinder2006}:%
\begin{align*} \label{eq:BBbasis}
 P_1 &= \mathbbm{1}\,, \\
 P_2 &= \vec\sigma_1\cdot\vec\sigma_2\,, \displaybreak[0]\\
 P_3 &= (\vec\sigma_1\cdot\vec q\,)(\vec\sigma_2\cdot\vec q\,) - \frac13(\vec\sigma_1\cdot\vec\sigma_2)\vec q^{\,2} \,, \displaybreak[0]\\
 P_4 &= \frac{\mathrm i}2(\vec\sigma_1+\vec\sigma_2)\cdot\vec n \,, \displaybreak[0]\\
 P_5 &= (\vec\sigma_1\cdot\vec n\,)(\vec\sigma_2\cdot\vec n\,)\,, \displaybreak[0]\\
 P_6 &= \frac{\mathrm i}2(\vec\sigma_1-\vec\sigma_2)\cdot\vec n \,, \displaybreak[0]\\
 P_7 &= (\vec\sigma_1\cdot\vec k\,)(\vec\sigma_2\cdot\vec q\,) + (\vec\sigma_1\cdot\vec q\,)(\vec\sigma_2\cdot\vec k\,)\,, \\
 P_8 &= (\vec\sigma_1\cdot\vec k\,)(\vec\sigma_2\cdot\vec q\,) - (\vec\sigma_1\cdot\vec q\,)(\vec\sigma_2\cdot\vec k\,)\,, \numberthis
\end{align*}
with \(\vec\sigma_{1,2}\) the Pauli spin matrices and with the vectors
\begin{equation}
 \vec k = \frac12(\vec p_f+\vec p_i)\,,\quad \vec q = \vec p_f - \vec p_i\,,\quad \vec n = \vec p_i\times\vec p_f\,.
\end{equation}
The momenta \(\vec p_f\) and \(\vec p_i\) are the initial and final state momenta in the center-of-mass frame.
%
In order to obtain the minimal set of Lagrangian terms in the non-relativistic power counting of \ct{Petschauer2013a}, the potentials have been decomposed into partial waves.
The formulas for the partial wave projection of a general interaction \(V=\sum_{j=1}^8 V_j P_j\) can be found in the appendix of \ct{Polinder2006}.
For each partial wave one produces a non-square matrix which connects the Lagrangian constants with the different baryon-baryon channels.
Lagrangian terms are considered as redundant if their omission does not lower the rank of this matrix.
For the determination of the potential not only direct contributions have to be considered, but also additional structures from exchanged final state baryons, where the negative spin-exchange operator \(-P^{(\sigma)}=-\frac12\left(\mathbbm1+\vec\sigma_1\cdot\vec\sigma_2\right)\) is applied.
In the end 6 momentum-independent terms at LO contribute, and are therefore only visible in \({}^1S_0\) and \({}^3S_1\) partial waves.
At NLO 22 terms contribute that contain only baryon fields and derivatives, and are therefore SU(3) symmetric.
The other 12 terms terms at NLO include the diagonal matrix \(\chi\) and produce explicit SU(3) symmetry breaking.

In \tab{tab:PWDBB} the non-vanishing transitions projected onto partial waves in the isospin basis are shown, \cf \cts{Polinder2006,Polinder2007,Haidenbauer2010a,Petschauer2013a}.
 The pertinent constants are redefined according to the relevant irreducible SU(3) representations.
This comes about in the following way.
Baryons form a flavor octet and the tensor product of two baryons decomposes into irreducible representations as follows:%
\begin{equation}
\mathbf8\otimes\mathbf8 = \mathbf{27}_s\oplus\mathbf{10}_a\oplus\mathbf{10}_a^\mathbf*\oplus\mathbf{8}_s\oplus\mathbf{8}_a\oplus\mathbf1_s \,,
\end{equation}
where the irreducible representations \(\mathbf{27}_s,\ \mathbf{8}_s,\ \mathbf 1_s\) are symmetric and \(\mathbf{10}_a,\ \mathbf{10}^*_a,\ \mathbf{8}_a\) are antisymmetric with respect to the exchange of both baryons.
Due to the generalized Pauli principle, the symmetric flavor representations \(\mathbf{27}_s,\ \mathbf{8}_s,\ \mathbf 1_s\) have to combine with the space-spin antisymmetric partial waves \({}^1S_0,\ {}^3P_0,\ {}^3P_1,\ {}^3P_2,\ \dots\) (\(L+S\) even).
The antisymmetric flavor representations \(\mathbf{10}_a,\ \mathbf{10}^*_a,\ \mathbf{8}_a\) combine with the space-spin symmetric partial waves \({}^3S_1,\ {}^1P_1,\ {}^3D_1\leftrightarrow{}^3S_1,\ \dots\) (\(L+S\) odd).
Transitions can only occur between equal irreducible representations.
Hence, transitions between space-spin antisymmetric partial waves up to \(\mathcal{O}(q^2)\) involve the 15 constants \(\tilde c_{{}^1S_0}^{27,8s,1}\), \(c_{{}^1S_0}^{27,8s,1}\), \(c_{{}^3P_0}^{27,8s,1}\), \(c_{{}^3P_1}^{27,8s,1}\) and \(c_{{}^3P_2}^{27,8s,1}\), whereas
transitions between space-spin symmetric partial waves involve the 12 constants \(\tilde c_{{}^3S_1}^{8a,10,10^*}\), \(c_{{}^3S_1}^{8a,10,10^*}\), \(c_{{}^1P_1}^{8a,10,10^*}\) and \(c_{{}^3D_1\text-{}^3S_1}^{8a,10,10^*}\).
The constants with a tilde denote leading-order constants, whereas the ones without tilde are at NLO\@.
The spin singlet-triplet transitions \({}^1P_1\leftrightarrow{}^3P_1\) is perfectly allowed by SU(3) symmetry since it is related to transitions between the irreducible representations \(\mathbf 8_a\) and \(\mathbf 8_s\).
Such a transition originated from the antisymmetric spin-orbit operator \(P_6\) and its Fierz-transformed counterpart \(P_8\) and the single corresponding low-energy constant is denoted by \(c^{8as}\).
In case of the \(NN\) interaction such transitions are forbidden by isospin symmetry.
The constants \(\tilde c_{{}^1S_0}^{27,8s,1}\) and \(\tilde c_{{}^3S_1}^{8a,10,10^*}\) fulfill the same SU(3) relations as the constants \(c_{{}^1S_0}^{27,8s,1}\) and \(c_{{}^3S_1}^{8a,10,10^*}\) in \tab{tab:PWDBB}.
SU(3) breaking terms linear in the quark masses appears only in the S-waves, \({}^1S_0,\ {}^3S_1\), and are proportional \(m_K^2-m_\pi^2\).
The corresponding 12 constants are \(c_\chi^{1,\dots,12}\).
The SU(3) symmetry relations in \tab{tab:PWDBB} can also be derived by group theoretical considerations \cts*{Polinder2006,Iwao1964,Dover1990,Dover1992}.
Clearly, for the SU(3)-breaking part this is not possible and these contributions have to be derived from the chiral Lagrangian.

In order to obtain the complete partial-wave projected potentials, some entries in \tab{tab:PWDBB} have to be multiplied with additional momentum factors.
The leading order constants \(\tilde c^{\,i}_j\) receive no further factor.
For the next-to-leading-order constants (without tilde and without \(\chi\)) the contributions to the partial waves \({}^1S_0,\ {}^3S_1\) have to be multiplied with a factor  \(p^2_i+p_f^2\).
The contribution to the partial waves \({}^1S_0,\ {}^3S_1\) from constants \(c^j_\chi\) has to be multiplied with \((m_K^2-m_\pi^2)\).
The partial waves \({}^3P_0,\ {}^3P_1,\ {}^3P_2,\ {}^1P_1,\ {}^1P_1\leftrightarrow{}^3P_1\) get multiplied with the factor \(p_ip_f\).
The entries for \({}^3S_1\rightarrow{}^3D_1\) and \({}^3D_1\rightarrow{}^3S_1\) have to be multiplied with \(p_i^2\) and \(p_f^2\), respectively.
For example, one obtains for the \(NN\) interaction in the \({}^1S_0\) partial wave:
\begin{align*} \label{pot:su3b}
 &\langle NN, {}^1S_0|\hat V|NN, {}^1S_0\rangle \\
 &\qquad= \tilde c^{27}_{{}^1S_0} + c^{27}_{{}^1S_0} (p^2_i+p_f^2) + \frac12 c^1_\chi(m_K^2-m_\pi^2)  \,, \numberthis
\end{align*}
or for the \(\Xi N\rightarrow\Sigma\Sigma\) interaction with total isospin \(I=0\) in the \({}^1P_1\rightarrow{}^3P_1\) partial wave:
\begin{equation}
 \langle \Sigma\Sigma, {}^3P_1|\hat V|\Xi N, {}^1P_1\rangle = 2\sqrt3 c^{8as} p_ip_f \,.
\end{equation}
When restricting to the \(NN\) channel the well-known two leading and seven next-to-leading order low-energy constants of \ct{Epelbaum2004} are recovered, which contribute to the partial waves \({}^1S_0\), \({}^3S_1\), \({}^1P_1\), \({}^3P_{0}\), \({}^3P_{1}\), \({}^3P_{2}\), \({}^3S_1\leftrightarrow {}^3D_1 \).

Note, that the SU(3) relations in \tab{tab:PWDBB} are general relations that have to be fulfilled by the baryon-baryon potential in the SU(3) limit, \ie \(m_\pi = m_K = m_\eta\).
This feature can be used as a check for the inclusion of the loop diagrams.
Another feature is, that the SU(3) relations contain only a few constants in each partial wave.
For example, in the \({}^1S_{0}\) partial wave only the constants \(\tilde c_{{}^1S_0}^{27}\), \(\tilde c_{{}^1S_0}^{8s}\), \(\tilde c_{{}^1S_0}^{1}\) are present.
If these constants are fixed in some of the baryons channels, predictions for other channels can be made.
This has, for instance, been used in \ct{Haidenbauer2015}, where the existence of \(\Sigma\Sigma\), \(\Sigma\Xi\) and \(\Xi\Xi\) bound states has been studied within SU(3) \cheft.

\begin{turnpage}
\begin{table*}
\centering
\resizebox*{.99\textheight}{!}{
\begin{tabular}{>{\(}c<{\)}>{\(}c<{\)}>{\(}c<{\)}>{\(}c<{\)}>{\(}c<{\)}>{\(}c<{\)}>{\(}c<{\)}>{\(}c<{\)}>{\(}c<{\)}}
\toprule
 S & I & \text{transition} & j\in\{{}^1S_0, {}^3P_0, {}^3P_1, {}^3P_2\} & j\in\{{}^3S_1, {}^1P_1, {}^3S_1\leftrightarrow{}^3D_1\} & {}^1P_1\rightarrow {}^3P_1 & {}^3P_1\rightarrow {}^1P_1 & {}^1S_0 \ \chi & {}^3S_1 \ \chi \\
\cmidrule(lr){1-3}\cmidrule(lr){4-9}
  0 & 0 & NN\rightarrow NN & 0 & c^{10^*}_{j} & 0 & 0 & 0 & \frac{c_\chi^7}{2} \\
   & 1 & NN\rightarrow NN & c^{27}_{j} & 0 & 0 & 0 & \frac{c_\chi^1}{2} & 0 \\
\cmidrule(lr){1-3}\cmidrule(lr){4-9}
 -1  & \frac12 & \Lambda N\rightarrow \Lambda N & \frac{1}{10} (9 c^{27}_{j}+c^{8s}_{j}) & \frac{1}{2}(c^{10^*}_{j}+c^{8a}_{j}) & -c^{8as} & -c^{8as} & c_\chi^2 & c_\chi^8 \\
  & \frac12 & \Lambda N\rightarrow \Sigma N & -\frac{3}{10} (c^{27}_{j}-c^{8s}_{j}) & \frac{1}{2}(c^{10^*}_{j}-c^{8a}_{j}) & -3 c^{8as} & c^{8as} & -c_\chi^3 & -c_\chi^9 \\
  & \frac12 & \Sigma N\rightarrow \Sigma N & \frac{1}{10} (c^{27}_{j}+9 c^{8s}_{j}) & \frac{1}{2}(c^{10^*}_{j}+c^{8a}_{j}) & 3 c^{8as} & 3 c^{8as} & c_\chi^4 & c_\chi^{10} \\
  & \frac32 & \Sigma N\rightarrow \Sigma N & c^{27}_{j} & c^{10}_{j} & 0 & 0 & \frac{c_\chi^1}{4} & -\frac{c_\chi^7}{4} \\
\cmidrule(lr){1-3}\cmidrule(lr){4-9}
  -2 & 0 & \Lambda \Lambda \rightarrow \Lambda \Lambda & \frac{1}{40} (5 c^{1}_{j}+27 c^{27}_{j}+8 c^{8s}_{j}) & 0 & 0 & 0 & \frac{c_\chi^5}{2} & 0 \\
  & 0 & \Lambda \Lambda \rightarrow \Xi N & \frac{1}{20} (5 c^{1}_{j}-9 c^{27}_{j}+4 c^{8s}_{j}) & 0 & 0 & 2 c^{8as} & \frac{3 c_\chi^1}{4}-3 c_\chi^2-c_\chi^3+\frac{3 c_\chi^5}{4} & 0 \\
  & 0 & \Lambda \Lambda \rightarrow \Sigma \Sigma & -\frac{\sqrt{3}}{40}  (5 c^{1}_{j}+3 c^{27}_{j}-8 c^{8s}_{j}) & 0 & 0 & 0 & 0 & 0 \\
  & 0 & \Xi N \rightarrow \Xi N & \frac{1}{10} (5 c^{1}_{j}+3 c^{27}_{j}+2 c^{8s}_{j}) & c^{8a}_{j} & 2 c^{8as} & 2 c^{8as} & \frac{2 c_\chi^1}{3}-3 c_\chi^2+\frac{c_\chi^4}{3}+\frac{9 c_\chi^5}{8} & c_\chi^{11} \\
  & 0 & \Xi N \rightarrow \Sigma \Sigma & \frac{\sqrt{3}}{20}  (-5 c^{1}_{j}+c^{27}_{j}+4 c^{8s}_{j}) & 0 & 2 \sqrt{3} c^{8as} & 0 & -\frac{c_\chi^1}{4 \sqrt{3}}+\sqrt{3} c_\chi^3+\frac{c_\chi^4}{\sqrt{3}} & 0 \\
  & 0 & \Sigma \Sigma \rightarrow \Sigma \Sigma & \frac{1}{40} (15 c^{1}_{j}+c^{27}_{j}+24 c^{8s}_{j}) & 0 & 0 & 0 & 0 & 0 \\
  & 1 & \Xi N \rightarrow \Xi N & \frac{1}{5} (2 c^{27}_{j}+3 c^{8s}_{j}) & \frac{1}{3} (c^{10}_{j}+c^{10^*}_{j}+c^{8a}_{j}) & -2 c^{8as} & -2 c^{8as} & c_\chi^6 & c_\chi^{12} \\
  & 1 & \Xi N \rightarrow \Sigma \Sigma  & 0 & \frac{1}{3 \sqrt{2}}(c^{10}_{j}+c^{10^*}_{j}-2 c^{8a}_{j}) & 0 & 2 \sqrt{2} c^{8as} & 0 & \sqrt{2} c_\chi^{10}-\frac{c_\chi^7}{2 \sqrt{2}}-\sqrt{2} c_\chi^9 \\
  & 1 & \Xi N \rightarrow \Sigma \Lambda  & \frac{\sqrt{6}}{5}  (c^{27}_{j}-c^{8s}_{j}) & \frac{1}{\sqrt{6}}(c^{10}_{j}-c^{10^*}_{j}) & 2 \sqrt{\frac{2}{3}} c^{8as} & 0 & -\frac{1}{3} \sqrt{\frac{2}{3}} c_\chi^1+\sqrt{\frac{3}{2}} c_\chi^2-\frac{c_\chi^4}{3 \sqrt{6}}-\sqrt{\frac{2}{3}} c_\chi^6 & \frac{c_\chi^{10}}{\sqrt{6}}+\sqrt{\frac{2}{3}} c_\chi^{12}+\frac{c_\chi^7}{2 \sqrt{6}}-\sqrt{\frac{3}{2}} c_\chi^8+\sqrt{\frac{2}{3}} c_\chi^9 \\
  & 1 & \Sigma \Lambda \rightarrow \Sigma \Lambda  & \frac{1}{5} (3 c^{27}_{j}+2 c^{8s}_{j}) & \frac{1}{2}(c^{10}_{j}+c^{10^*}_{j}) & 0 & 0 & -\frac{c_\chi^1}{9}+\frac{4 c_\chi^3}{3}+\frac{4 c_\chi^4}{9}+\frac{2 c_\chi^6}{3} & \frac{4 c_\chi^{10}}{3}+\frac{2 c_\chi^{12}}{3}-\frac{c_\chi^7}{3}-\frac{4 c_\chi^9}{3} \\
  & 1 & \Sigma \Lambda \rightarrow \Sigma \Sigma  & 0 & \frac{1}{2 \sqrt{3}}(c^{10}_{j}-c^{10^*}_{j}) & 0 & \frac{4}{\sqrt{3}} c^{8as} & 0 & 0 \\
  & 1 & \Sigma \Sigma \rightarrow \Sigma \Sigma  & 0 & \frac{1}{6} (c^{10}_{j}+c^{10^*}_{j}+4 c^{8a}_{j}) & 0 & 0 & 0 & 0 \\
  & 2 & \Sigma \Sigma \rightarrow \Sigma \Sigma  & c^{27}_{j} & 0 & 0 & 0 & 0 & 0 \\
\cmidrule(lr){1-3}\cmidrule(lr){4-9}
  -3 & \frac12 & \Xi \Lambda \rightarrow \Xi \Lambda & \frac{1}{10} (9 c^{27}_{j}+c^{8s}_{j}) & \frac{1}{2}(c^{10}_{j}+c^{8a}_{j}) & -c^{8as} & -c^{8as} & -\frac{55 c_\chi^1}{72}+2 c_\chi^2+\frac{7 c_\chi^3}{6}+\frac{c_\chi^4}{18}+\frac{3 c_\chi^5}{32}+\frac{c_\chi^6}{12} & \frac{11 c_\chi^{10}}{12}+\frac{3 c_\chi^{11}}{4}+\frac{25 c_\chi^{12}}{12}+\frac{5 c_\chi^7}{24}-\frac{7 c_\chi^8}{4}-\frac{c_\chi^9}{6} \\
  & \frac12 & \Xi \Lambda \rightarrow \Xi \Sigma & -\frac{3}{10} (c^{27}_{j}-c^{8s}_{j}) & \frac{1}{2}(c^{10}_{j}-c^{8a}_{j}) & -3 c^{8as} & c^{8as} & \frac{11 c_\chi^1}{24}-\frac{3 c_\chi^2}{2}-\frac{c_\chi^3}{2}-\frac{c_\chi^4}{3}+\frac{9 c_\chi^5}{32}+\frac{c_\chi^6}{4} & \frac{9 c_\chi^{10}}{4}-\frac{3 c_\chi^{11}}{4}+\frac{5 c_\chi^{12}}{4}-\frac{c_\chi^7}{8}-\frac{3 c_\chi^8}{4}-\frac{c_\chi^9}{2} \\
  & \frac12 & \Xi \Sigma \rightarrow \Xi \Sigma & \frac{1}{10} (c^{27}_{j}+9 c^{8s}_{j}) & \frac{1}{2}(c^{10}_{j}+c^{8a}_{j}) & 3 c^{8as} & 3 c^{8as} & \frac{11 c_\chi^1}{24}-3 c_\chi^2+\frac{5 c_\chi^3}{2}+\frac{c_\chi^4}{6}+\frac{27 c_\chi^5}{32}+\frac{3 c_\chi^6}{4} & \frac{5 c_\chi^{10}}{4}+\frac{3 c_\chi^{11}}{4}+\frac{3 c_\chi^{12}}{4}-\frac{c_\chi^7}{8}-\frac{3 c_\chi^8}{4}-\frac{3 c_\chi^9}{2} \\
  & \frac32 & \Xi \Sigma \rightarrow \Xi \Sigma & c^{27}_{j} & c^{10^*}_{j} & 0 & 0 & -\frac{2 c_\chi^1}{3}+\frac{3 c_\chi^2}{2}+c_\chi^3+\frac{c_\chi^4}{6} & \frac{3 c_\chi^{10}}{2}-c_\chi^7+\frac{3 c_\chi^8}{2}-3 c_\chi^9 \\
\cmidrule(lr){1-3}\cmidrule(lr){4-9}
  -4 & 0 & \Xi \Xi \rightarrow \Xi \Xi & 0 & c^{10}_{j} & 0 & 0 & 0 & 5 c_\chi^{10}+4 c_\chi^{12}-3 c_\chi^8-2 c_\chi^9 \\
  & 1 & \Xi \Xi \rightarrow \Xi \Xi & c^{27}_{j} & 0 & 0 & 0 & -\frac{4 c_\chi^1}{3}+3 c_\chi^2+2 c_\chi^3+\frac{c_\chi^4}{3} & 0 \\
\bottomrule
\end{tabular}
}
\caption{SU(3) relations of pure baryon-baryon contact terms for non-vanishing partial waves up to \(\mathcal O(q^2)\) in non-relativistic power counting for channels described by strangeness \(S\) and total isospin \(I\) \ct*{Petschauer2013a}.
} \label{tab:PWDBB}
\end{table*}
\end{turnpage}

\subsection{One- and two-meson-exchange contributions} \label{subsec:potme}

In the last section, we have addressed the short-range part of the baryon-baryon interaction via contact terms.
Let us now analyze the long- and mid-range part of the interaction, generated by one- and two-meson-exchange as determined in \ct{Haidenbauer2013a}.
The contributing diagrams up to NLO are shown in \fig{fig:hier}, which displays the hierarchy of baryonic forces.

The vertices, necessary for the construction of these diagrams stem from the leading-order meson-baryon interaction Lagrangian \(\mathscr{L}_\mathrm{B}^{(1)}\) in \eq{eq:barmeslagr}.
The vertex between two baryons and one meson emerges from the part
\begin{align*} \label{eq:3vert}
 & \frac D 2  \langle \bar B \gamma^\mu \gamma_5 \lbrace u_\mu,B\rbrace\rangle + \frac F 2 \langle\bar B \gamma^\mu \gamma_5 \left[u_\mu,B\right]\rangle \\
 & = - \frac{1}{2f_0} \sum_{i,j,k=1}^8 N_{B_iB_j\phi_k} (\bar B_i \gamma^\mu \gamma_5 B_j) (\partial_\mu \phi_k) + \mathcal O(\phi^3)\,, \numberthis
\end{align*}
where we have used \(u_\mu = -\frac1{f_0}\partial_\mu\phi+\mathcal O(\phi^3)\) and have rewritten the pertinent part of the Lagrangian in terms of the physical meson and baryon fields
\begin{align*}
 &\phi_i \in \left\{\pi^0,\pi^+,\pi^-,K^+,K^-,K^0,\bar K^0,\eta\right\}\,, \\
 &B_i \in \left\{n,p,\Sigma^0,\Sigma^+,\Sigma^-,\Lambda,\Xi^0,\Xi^-\right\}\,. \numberthis
\end{align*}
The factors \(N_{B_iB_j\phi_k}\) are linear combinations of the axial vector coupling constants \(D\) and \(F\) with certain SU(3) coefficients.
These factors vary for different combinations of the involved baryons and mesons and can be obtained easily by multiplying out the baryon and meson flavor matrices.
In a similar way, we obtain the (Weinberg-Tomozawa) vertex between two baryons and two mesons from the covariant derivative in  \(\mathscr{L}_\mathrm{B}^{(1)}\), leading to
\begin{align*} \label{eq:4vert}
 & \langle\bar B \mathrm i\gamma^\mu \left[\Gamma_\mu,B\right]\rangle \\
 &= \frac{\mathrm i}{8f_0^2} \sum_{i,j,k,l=1}^8 N_{B_i\phi_kB_j\phi_l} (\bar B_i \gamma^\mu B_j) (\phi_k\partial_\mu \phi_l) + \mathcal O(\phi^4)\,, \numberthis
\end{align*}
where \(\Gamma_\mu = \frac1{8f_0^2}[\phi,\partial_\mu\phi] + \mathcal O(\phi^4) \) was used.

The calculation of the baryon-baryon potentials is done in the center-of-mass frame and in the isospin limit \(m_u=m_d\).
To obtain the contribution of the Feynman diagrams to the non-relativistic potential, we perform an expansion in the inverse baryon mass \(1/M_\mathrm B\).
If loops are involved, the integrand is expanded before integrating over the loop momenta.
This produces results that are equivalent to the usual heavy-baryon formalism.
In the case of the two-meson-exchange diagrams at one-loop level, ultraviolet divergences are treated by dimensional regularization, which introduces a scale \(\lambda\).
In dimensional regularization divergences are isolated as terms proportional to
\begin{equation}
 R = \frac 2 {d-4} + \gamma_\mathrm E - 1 - \ln\left( 4\pi \right)\,,
\end{equation}
with \(d\neq4\) the space-time dimension and the Euler-Mascheroni constant \(\gamma_\mathrm E\apr0.5772\).
These terms can be absorbed by the contact terms.

\begin{figure*}
 \centering
 \begin{subfigure}[b]{.3\textwidth}
  \centering
  \(\vcenter{\hbox{\includegraphics[scale=.9]{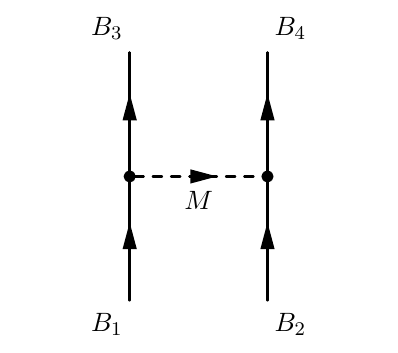}}}\) \\
  \(N=N_{B_3B_1\bar M}N_{B_4B_2M}\) \\ \mbox{}
  \caption{One-meson exchange}\label{fig:ome}
 \end{subfigure}
 \begin{subfigure}[b]{.3\textwidth}
  \centering
  \(\vcenter{\hbox{\includegraphics[scale=0.9]{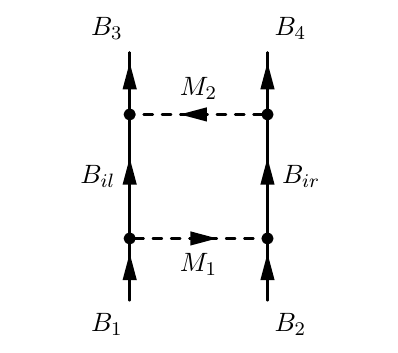}}}\) \\
  \(N=N_{B_{il}B_1\bar M_1}N_{B_3B_{il}M_2}\) \\
  \(\qquad\times N_{B_{ir}B_2M_1}N_{B_4B_{ir}\bar M_2}\)
  \caption{Planar box}\label{fig:pb}
 \end{subfigure}
 \begin{subfigure}[b]{.3\textwidth}
  \centering
  \(\vcenter{\hbox{\includegraphics[scale=0.9]{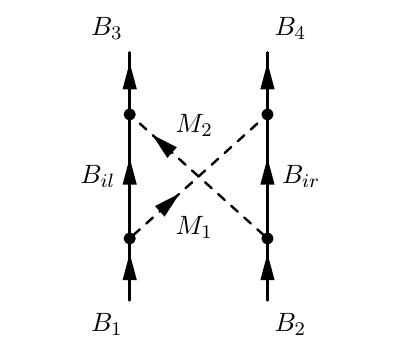}}}\) \\
  \(N=N_{B_{il}B_1\bar M_1}N_{B_3B_{il}M_2}\) \\
  \(\qquad\times N_{B_{ir}B_2\bar M_2}N_{B_4B_{ir}M_1}\)
  \caption{Crossed box}\label{fig:cb}
 \end{subfigure}
 \begin{subfigure}[b]{.3\textwidth}
  \centering
  \(\includegraphics[scale=0.9]{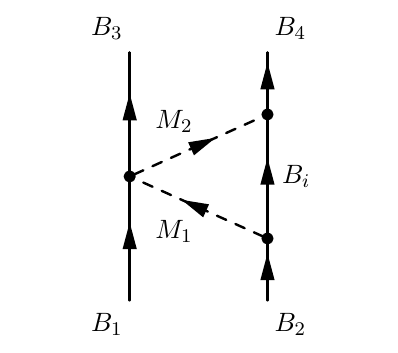}\) \\
  \( N=N_{B_3\bar M_2B_1M_1}N_{B_iB_2\bar M_1}N_{B_4B_iM_2} \)
  \caption{Left triangle}\label{fig:triL}
 \end{subfigure}
 \begin{subfigure}[b]{.3\textwidth}
  \centering
  \(\includegraphics[scale=0.9]{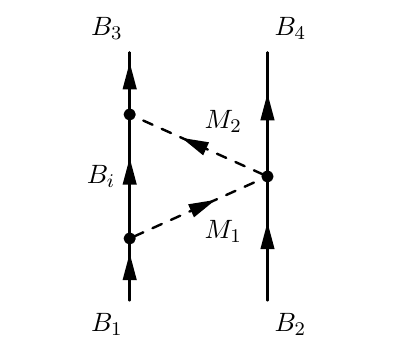}\) \\
  \( N=N_{B_iB_1\bar M_1}N_{B_3B_iM_2}N_{B_4\bar M_2B_2M_1} \)
  \caption{Right triangle}\label{fig:triR}
 \end{subfigure}
 \begin{subfigure}[b]{.3\textwidth}
  \centering
  \(\vcenter{\hbox{\includegraphics[scale=0.9]{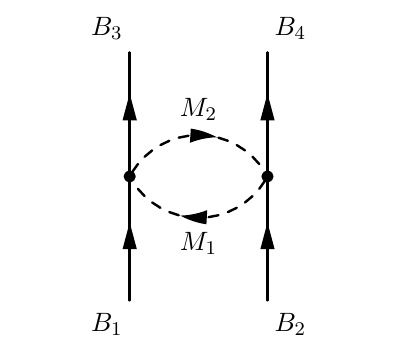}}}\) \\
  \(N=N_{B_3\bar M_2B_1M_1}N_{B_4\bar M_1B_2M_2}\)
  \caption{Football diagram}\label{fig:foot}
 \end{subfigure}
 \caption{One- and two-meson-exchange contributions and corresponding SU(3) factors.}\label{fig:mesN}
\end{figure*}

According to \eqs{eq:3vert} and \eq*{eq:4vert} the vertices have the same form for different combinations of baryons and mesons, just their prefactors change.
Therefore, the one- and two-pseudoscalar-meson exchange potentials can be given by a master formula, where the proper masses of the exchanged mesons have to be inserted, and which has to be multiplied with an appropriate SU(3) factor \(N\).
In the following we will present the analytic formulas for the one- and two-meson-exchange diagrams, introduced in \ct{Haidenbauer2013a}.
The pertinent SU(3) factors will be displayed next to the considered Feynman diagram, \cf \fig{fig:mesN}.
The results will be given in terms of a central potential (\(V_\mathrm C\)), a spin-spin potential $(\vec\sigma_1\cdot\vec\sigma_2\, V_S)$ and a tensor-type potential $(\vec\sigma_1\cdot\vec q \ \vec \sigma_2\cdot{\vec q}\ V_T)$.
The momentum transfer is \(q=\left|\vec p_f - \vec{p}_i\,\right|\), with \(\vec p_i\) and \(\vec p_f\) the initial and final state momenta in the center-of-mass frame.

Note that the presented results apply only to direct diagrams.
This is for example the case for the leading-order one-eta exchange in the \(\Lambda n\) interaction, \ie for
\(\Lambda(\vec p_i)n(-\vec p_i) \xrightarrow{\eta} \Lambda(\vec p_f)n(-\vec p_f)\).
An example of a crossed diagram is the one-kaon exchange in the process
\(\Lambda(\vec p_i)n(-\vec p_i) \xrightarrow{K} n(-\vec p_f)\Lambda(\vec p_f)\),
where the nucleon and the hyperon in the final state are interchanged and strangeness is exchanged.
In such cases, \(\vec p_f\) is replaced by \(-\vec p_f\) and the momentum transfer in the potentials is \(q=\left|\vec p_f + \vec{p}_i\,\right|\).
Due to the exchange of fermions in the final states a minus sign arises, and additionally the spin-exchange operator \(P^{(\sigma)} = \frac12(\mathbbm1+\vec\sigma_1\cdot\vec\sigma_2)\) has to be applied.
The remaining structure of the potentials stays the same (see also the discussion in \sect{sec:BBB}).

The leading-order contribution comes from the \emph{one-meson exchange} diagram in \fig{fig:ome}.
It contributes only to the tensor-type potential:
\begin{equation} \label{pot:ome} 
 V^\text{ome}_\mathrm{T}(q) = -\frac{N}{4f_0^2} \frac1{q^2+m^2-\mathrm i \epsilon} \,.
\end{equation}
The symbol \(\bar M\) in the SU(3) coefficient \(N\) denotes the charge-conjugated meson of meson \(M\) in particle basis (\eg \(\pi^+\leftrightarrow\pi^-\)).

At next-to-leading order the two-meson exchange diagrams start to contribute.
The \emph{planar box} in \fig{fig:pb} contains an irreducible part and a reducible part coming from the iteration of the one-meson exchange to second order.
Inserting the potential into the Lippmann-Schwinger equation generates the reducible part;
it is therefore not part of the potential, see also \ssect{subsec:pwr}.
The irreducible part is obtained from the residues at the poles of the meson propagators,
disregarding the (far distant) poles of the baryon propagators.
With the masses of the two exchanged mesons set to \(m_1\) and \(m_2\), the irreducible potentials can be written in closed analytical form,
\begin{align*}
\label{P1}
 &V^\text{planar box}_\mathrm{irr,\,C}(q) =\frac{N}{3072 \pi ^2f_0^4}\Bigg\{ \frac{5}{3}q^2 \\
 &+\frac{\left(m_1^2-m_2^2\right)^2}{q^2}+ 16 \left(m_1^2+m_2^2\right) \\
 &+ \left[23 q^2+45 \left(m_1^2+m_2^2\right)\right]\left(R+2\ln \frac{\sqrt{m_1 m_2}}{\lambda }\right)\\
 &+\frac{m_1^2-m_2^2}{q^4} \bigg[12 q^4+\left(m_1^2-m_2^2\right)^2 \\
 &\qquad\qquad\quad -9 q^2 \left(m_1^2+m_2^2\right)\bigg] \ln \frac{m_1}{m_2} \displaybreak[0]\\
 &+\frac{2}{w^2\left(q\right)} \bigg[23 q^4-\frac{\left(m_1^2-m_2^2\right)^4}{q^4} +56 \left(m_1^2+m_2^2\right) q^2 \\
 &\qquad\qquad\quad +8\frac{ m_1^2+m_2^2}{q^2}\left(m_1^2-m_2^2\right)^2 \\
 &\qquad\qquad\quad + 2 \left(21 m_1^4+22 m_1^2 m_2^2+21 m_2^4\right)\bigg] L\left(q\right) \Bigg\}\,, \numberthis
\end{align*}
\vspace{-\baselineskip}
\begin{align*}
\label{P2}
 &V^\text{planar box}_\mathrm{irr,\,T}\left(q\right) = - \frac1{q^2} V^\text{planar box}_\mathrm{irr,\,S}(q) \\
 &=-\frac{N}{128 \pi ^2 f_0^4} \Bigg[\ L\left(q\right)-\frac{1}{2}-\frac{m_1^2-m_2^2}{2 q^2}\ln \frac{m_1}{m_2} \\
 &\qquad\qquad\qquad\quad +\frac{R}{2}+\ln \frac{\sqrt{m_1 m_2}}{\lambda }\ \Bigg] \numberthis
\end{align*}
where we have defined the functions
\begin{align*}
 w\left(q\right) &= \frac1q\sqrt{\left(q^2+\left(m_1+m_2\right)^2\right)\left(q^{2}+\left(m_1-m_2\right)^2\right)}\,, \\
 L\left(q\right) &= \frac{w\left(q\right)}{2q} \ln \frac{\left[ qw\left(q\right) + q^2\right]^2 - \left(m_1^2-m_2^2\right)^2}{4m_1m_2q^2}\,. \numberthis
\end{align*}
The relation between the spin-spin and tensor-type potential follows from the identity
$(\vec\sigma_1\times {\vec q}\,) \cdot (\vec\sigma_2\times {\vec q}\,) =
{q}^2 \vec\sigma_1\cdot\vec\sigma_2 -
(\vec\sigma_1\cdot{\vec q}\,) \, (\vec\sigma_2\cdot{\vec q}\,)$.

One should note that all potentials shown above are finite also in the limit \(q\rightarrow0\).
Terms proportional to $1/q^2$ or $1/q^4$ are canceled by opposite terms in the functions $L(q)$ and $w(q)$ in the limit of small $q$.
For numerical calculations it is advantageous to perform an expansion of the potentials in a power series for small $q$ in order to implement directly this cancellation.
For equal meson masses the expressions for the potentials reduce to the results in \cts{Kaiser1997}.
This is the case for the $NN$ interaction of \cts{Epelbaum2004,Epelbaum1999,Epelbaum1999a,Entem2003} based on \cheft, where only contributions from two-pion exchange need to be taken into account.

In actual applications of these potentials such as in \ct{Haidenbauer2013a}, only the non-polynomial part of \eqs{P1} and \eq*{P2} is taken into account, \ie the pieces proportional to $L(q)$ and to $1/q^2$ and $1/q^4$.
The polynomial part is equivalent to the LO and NLO contact terms and, therefore, does not need to be considered.
The contributions proportional to the divergence \(R\) are likewise omitted.
Their effect is absorbed by the contact terms or a renormalization of the coupling constants, see, \eg the corresponding discussion in Appendix A of \ct{Epelbaum1999} for the $NN$ case.

These statements above apply also to the other contributions to the potential described below.

The \emph{crossed box} diagrams in \fig{fig:cb} contribute to the central, spin-spin, and tensor-type potentials.
The similar structure with some differences in the kinematics of the planar and crossed box diagram leads to relations between them.
Obviously, the crossed box has no iterated part.
The potentials of the crossed box are equal to the potentials of the irreducible part of the planar box, up to a sign in the central potential:
\begin{align*}
 V^\text{crossed box}_\mathrm{C}(q) &= - V^\text{planar box}_\mathrm{C,\,irr}(q)\,,\\
 V^\text{crossed box}_\mathrm{T}(q) &= - \frac1{q^2} V^\text{crossed box}_\mathrm{S}(q) = V^\text{planar box}_\mathrm{T,\,irr}(q)\,. \numberthis
\end{align*}

The two \emph{triangle} diagrams, \figs{fig:triL} and \fig*{fig:triR}, constitute potentials, that are of equal form with different SU(3) factors \(N\).
The corresponding central potential reads
\begin{align*}
 &V^\text{triangle}_\mathrm C (q) = -\frac{N}{3072 \pi ^2 f_0^4} \Bigg\{-2 \left(m_1^2+m_2^2\right)\\
 &+\frac{\left(m_1^2-m_2^2\right)^2}{q^2}-\frac{13}{3} q^2 \\
 & +\left[8 \left(m_1^2+m_2^2\right)-\frac{2 \left(m_1^2-m_2^2\right)^2}{q^2}+10 q^2\right] L\left(q\right)\\
 & +\frac{m_1^2-m_2^2}{q^4} \left[\left(m_1^2-m_2^2\right)^2-3 \left(m_1^2+m_2^2\right) q^2\right] \ln \frac{m_1}{m_2} \\
 & + \left[9 \left(m_1^2+m_2^2\right)+5 q^2\right] \left(R+2\ln \frac{\sqrt{m_1 m_2}}{\lambda }\right) \Bigg\} \,. \numberthis
\end{align*}

The \emph{football} diagrams in Fig.~\ref{fig:foot} also contributes only to the central potential.
One finds
\begin{align*}
 &V^\text{football}_\mathrm C (q) = \frac{N}{3072 \pi ^2 f_0^4}\Bigg\{-2 \left(m_1^2+m_2^2\right)\\
 &-\frac{\left(m_1^2-m_2^2\right)^2}{2 q^2}-\frac{5}{6} q^2  + w^2\left(q\right) L\left(q\right) \\
 & + \frac12\left[3 \left(m_1^2+m_2^2\right)+q^2\right] \left(R + 2 \ln \frac{\sqrt{m_1 m_2}}{\lambda }\right) \\
 & - \frac{m_1^2-m_2^2}{2 q^4} \left[\left(m_1^2-m_2^2\right)^2+3 \left(m_1^2+m_2^2\right) q^2\right] \ln \frac{m_1}{m_2}\ \Bigg\}\,. \numberthis
\end{align*}

\subsection{Meson-exchange models} \label{subsec:mesonex}

Earlier investigations of the baryon-baryon interactions has been done within phenomenological meson-exchange
potentials such as the J\"ulich \cite{Holzenkamp1989,Reuber1994a,Haidenbauer2005},
Nijmegen \cite{Rijken1998,Rijken2010,Nagels2019}, 
or Ehime \cite{Tominaga1998,Tominaga2001} potentials.
As we use two of them for comparison, we give a brief introduction to these type of models.

Conventional meson-exchange models of the \(YN\) interaction are 
usually also based on the assumption of SU(3) flavor symmetry for the 
occurring coupling constants, and in some cases even on the SU(6) symmetry 
of the quark model \cite{Holzenkamp1989,Reuber1994a}. 
In the derivation of the meson-exchange contributions one follows 
essentially the same procedure as outlined in Sect.~\ref{subsec:potme}  
for the case of pseudoscalar mesons. 
Besides the lowest pseudoscalar-meson multiplet also the 
exchanges of vector mesons ($\rho$, $\omega$, $K^*$), of scalar mesons 
($\sigma$ ($f_0(500)$), ...), or even of axial-vector mesons ($a_1(1270)$, ...) 
\cite{Rijken2010,Nagels2019} are included. The spin-space structure of the 
corresponding Lagrangians that enter into Eq. (\ref{eq:barmeslagr}) and 
subsequently into Eq. (\ref{eq:3vert}) differ and, accordingly, the final 
expressions for the corresponding contributions to the \(Y N\) interaction 
potentials differ too. 
Details can be found in Refs. ~\cite{Holzenkamp1989,Rijken1998,Rijken2010}.
We want to emphasize that even for pseudoscalar mesons the final result for 
the interaction potentials differs, in general, from the expression given in 
Eq. (\ref{pot:ome}). Contrary to the chiral EFT
approach, recoil and relativistic corrections are often kept in meson-exchange
models because no power counting rules are applied. Moreover, in case of the
J\"ulich potential pseudoscalar coupling is assumed for the meson-baryon
interaction Lagrangian for the pseudoscalar mesons instead of the 
pseudovector coupling (\ref{eq:barmeslagr}) dictated by chiral symmetry.
Note that in some \(YN\) potentials of the J\"ulich group \cite{Holzenkamp1989,Reuber1994a}
contributions from two-meson exchanges are included. The ESC08 and ESC16 potentials 
\cite{Rijken2010,Nagels2019} include likewise contributions from two-meson exchange, 
in particular, so-called meson-pair diagrams analog to the ones shown in 
Figs. \ref{fig:triL}, \ref{fig:triR}, and \ref{fig:foot}. 

The major conceptual difference between the various meson-exchange
models consists in the treatment of the scalar-meson sector. This simply 
reflects the fact that, unlike for pseudoscalar and vector mesons, so far there is
no general agreement about what are the actual members of the lowest lying
scalar-meson SU(3) multiplet.
Therefore, besides the question of the masses of the exchange
particles it also remains unclear whether and how the relations for the
coupling constants should be specified. As a consequence, different
prescriptions for describing the scalar sector, whose contributions play
a crucial role in any baryon-baryon interaction at intermediate ranges, were
adopted by the various authors who published meson-exchange models of the
\(Y N\) interaction.
For example, the Nijmegen group views this interaction as being
generated by genuine scalar-meson exchange. In their models NSC97
\cite{Rijken1998} and ESC08 (ESC16) \cite{Rijken2010,Nagels2019}
a scalar SU(3) nonet is exchanged - namely, two 
isospin-$0$ mesons (an $\epsilon$(760) and the $f_0(980)$) 
an isospin-$1$ meson ($a_0(980)$) and an isospin-1/2 strange meson
$\kappa$ with a mass of 1000 MeV. 
In the initial \(Y N\) models of the J\"ulich group \cite{Holzenkamp1989,Reuber1994a}
a $\sigma$ (with a mass of $\approx 550$ MeV) is included which is viewed as arising 
from correlated $\pi\pi$ exchange. In practice, however, the coupling strength of this 
fictitious $\sigma$ to the baryons is treated as a free parameter and fitted to the data.
In the latest meson-exchange \(Y N\) potential presented by the J\"ulich
group \cite{Haidenbauer2005} a microscopic model of correlated $\pi\pi$ and $K\bar K$ 
exchange \cite{Reuber1995} is utilized to fix the contributions in the 
scalar-isoscalar ($\sigma$) and vector-isovector ($\rho$) channels.

Let us mention for completeness that meson-exchange models are typically
equipped with phenomenological form factors in order to cut off the potential
for large momenta (short distances). 
For example, in case of the \(Y N\) models of the J\"ulich group the
interaction is supplemented with form factors for each meson-baryon-baryon 
vertex, cf.~\cite{Holzenkamp1989,Reuber1994a} for details. 
Those form factors are meant to take into account the extended
hadron structure and are parametrized in the conventional monopole or dipole
form. In case of the Nijmegen potentials a Gaussian form factor is used.
In addition there is some additional sophisticated short-range phenomenology
that controls the interaction at short distances \cite{Rijken2010,Nagels2019}. 

\section{Three-baryon interaction potentials}  \label{sec:BBB}

Three-nucleon forces are an essential ingredient for a proper description of nuclei and nuclear matter with low-momentum two-body interactions.
Similarly, three-baryon forces, especially the \(\Lambda NN\) interaction, are expected to play an important role in nuclear systems with strangeness.
Their introduction in calculations of light hypernuclei seems to be required.
Furthermore, the introduction of 3BF is traded as a possible solution to the hyperon puzzle (see \sect{sec:intro}).
However, so far only phenomenological 3BF have been employed.
In this section we present the leading irreducible three-baryon interactions from SU(3) chiral effective field theory as derived in \ct{Petschauer2016}.
We show the minimal effective Lagrangian required for the pertinent vertices.
Furthermore the estimation of the corresponding LECs through decuplet saturation and an effective density-dependent two-baryon potential will be covered \ct*{Petschauer2017a}.

\begin{figure}
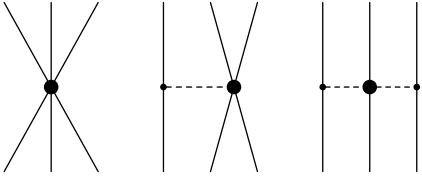

\centering
\includegraphics[scale=0.5]{files/Feynman/BF/FBBBcont}\qquad
\includegraphics[scale=0.5]{files/Feynman/BF/FBBB1ME}\qquad
\includegraphics[scale=0.5]{files/Feynman/BF/FBBB2ME2}
\caption{
Leading three-baryon interactions: contact term, one-meson exchange and two-meson exchange.
Filled circles and solid dots denote vertices with \(\Delta_i=1\) and \(\Delta_i=0\), respectively.
\label{fig:3BF}
}
\end{figure}

According to the power counting in \eq{eq:barpwr} the 3BF arise formally at NNLO in the chiral expansion, as can be seen from the hierarchy of baryonic forces in \fig{fig:hier}.
Three types of diagrams contribute: three-baryon contact terms, one-meson and two-meson exchange diagrams, \cf \fig{fig:3BF}.
Note that a two-meson exchange diagram, such as in \fig{fig:3BF}, with a (leading order) Weinberg-Tomozawa vertex in the middle, would formally be a NLO contribution.
However, as in the nucleonic sector, this contribution is kinematically suppressed due to the fact that the involved meson energies are differences of baryon kinetic energies.
Anyway, parts of these NNLO contributions get promoted to NLO by the introduction of intermediate decuplet baryons, so that it becomes appropriate to use these three-body interactions together with the NLO two-body interaction of \sect{sec:BB}.
As already stated, the irreducible contributions to the chiral potential are presented.
In contrast to typical phenomenological calculations, diagrams as on the left side of \fig{fig:irr3BF} do not lead to a genuine three-body potential, but are an iteration of the two-baryon potential.
Such diagrams will be incorporated automatically when solving, \eg the Faddeev (or Yakubovsky) equations within a coupled-channel approach.
The three-body potentials derived from SU(3) \cheft are expected to shed light on the effect of 3BFs in hypernuclear systems.
Especially in calculations about light hypernuclei these potentials can be implemented within 
reliable few-body techniques \ct*{Nogga2002,Nogga2014a,Wirth2014,Wirth2016}.

\subsection{Contact interaction} \label{subsec:BBBct}

In the following we consider the leading three-baryon contact interaction.
Following the discussion in \ssect{subsec:constchlag} the corresponding Lagrangian can be constructed.
The inclusion of external fields is not necessary, as we are interested in the purely baryonic contact term.
One ends up with the following possible structures in flavor space \ct*{Petschauer2016}
\begin{align*}
 &\trace{\bar B\bar B\bar B BBB}\,,
 \trace{\bar B\bar B B \bar BBB}\,,
 \trace{\bar B\bar B B B \bar BB}\,, \\&
 \trace{\bar B B\bar B B \bar BB}\,,
 \trace{\bar B \bar BBB}\trace{\bar B B}\,,
 \trace{\bar B B \bar BB}\trace{\bar B B}\,, \\&
 \trace{\bar B \bar B \bar BB}\trace{B B}\,,
 \trace{\bar B\bar B\bar B}\trace{BBB}\,,
 \trace{\bar B\bar BB}\trace{B\bar BB}\,, \\&
 \trace{\bar BB}\trace{\bar BB}\trace{\bar BB}\,,
 \trace{\bar B\bar B}\trace{\bar BB}\trace{BB}\,, \numberthis
\end{align*}
with possible Dirac structures
\begin{align*}
 &\mathbbm1\otimes\mathbbm1\otimes\mathbbm1\,,\
 \mathbbm1 \otimes \gamma_5\gamma^\mu \otimes \gamma_5\gamma_\mu\,,\
 \gamma_5\gamma^\mu \otimes \mathbbm1 \otimes \gamma_5\gamma_\mu\,,\\
 &\gamma_5\gamma^\mu \otimes \gamma_5\gamma_\mu \otimes \mathbbm1\,,\
 \gamma_5\gamma_\mu \otimes \mathrm i\;\sigma^{\mu\nu} \otimes \gamma_5\gamma_\nu\,, \numberthis
\end{align*}
leading to the following operators in the three-body spin space
\begin{equation} \label{eq:BBBbasis}
\mathbbm1\,,\
\vec\sigma_1\cdot\vec\sigma_2\,,\
\vec\sigma_1\cdot\vec\sigma_3\,,\
\vec\sigma_2\cdot\vec\sigma_3\,,\
\mathrm i\;\vec\sigma_1\cdot(\vec\sigma_2\times\vec\sigma_3) \,.
\end{equation}
All combinations of these possibilities leads to a (largely overcomplete) set of terms for the leading covariant Lagrangian.
Note that in \ct{Petschauer2016} the starting point is a covariant Lagrangian, but the minimal non-relativistic Lagrangian is the goal.
Hence, only Dirac structures leading to independent (non-relativistic) spin operators are relevant.

Let us consider the process \(B_1 B_2 B_3 \rightarrow B_4 B_5 B_6\), where the \(B_i\) are baryons in the particle basis, \(B_i\in\{n,p,\Lambda,\Sigma^+,\Sigma^0,\Sigma^-,\Xi^0,\Xi^-\}\).
The contact potential \(V\) has to be derived within a threefold spin space for this process.
The operators in spin-space 1 is defined to act between the two-component Pauli spinors of \(B_1\) and \(B_4\).
In the same way, spin-space 2 belongs to \(B_2\) and \(B_5\), and spin-space 3 to \(B_3\) and \(B_6\).
For a fixed spin configuration the potential can be calculated from
\begin{equation} \label{eq:spinpot}
{\chi_{B_4}^{(1)}}^\dagger {\chi_{B_5}^{(2)}}^\dagger {\chi_{B_6}^{(3)}}^\dagger \, V \, \chi_{B_1}^{(1)}\chi_{B_2}^{(2)}\chi_{B_3}^{(3)} \,,
\end{equation} 
where the superscript of a spinor denotes the spin space and the subscript denotes the baryon to which the spinor belongs.
The potential is obtained as
\(
V = -\langle B_4 B_5 B_6\vert \ \mathscr L \ \vert B_1 B_2 B_3\rangle
\),
where the contact Lagrangian \(\mathscr L\) has to be inserted, and the 36 Wick contractions need to be performed.
The number 36 corresponds to the \(3!\times3!\) possibilities to arrange the three initial and three final baryons into Dirac bilinears.
One obtains six direct terms, where the baryon bilinears combine the baryon pairs 1--4, 2--5 and 3--6, as shown in \eq{eq:spinpot}.
For the other 30 Wick contractions, the resulting potential is not fitting to the form of \eq{eq:spinpot}, because the wrong baryon pairs are connected in a separate spin space.
Hence, an appropriate exchange of the spin wave functions in the final state  has to be performed.
This is achieved by multiplying the potential with the well-known spin-exchange operators \(\Ps_{ij}=\frac12(\mathbbm1+\vec\sigma_i\cdot\vec\sigma_j)\).
Furthermore additional minus signs arise from the interchange of anticommuting baryon fields.
The full potential is then obtained by adding up all 36 contributions to the potential.
One obtains a potential that fulfills automatically the generalized Pauli principle and that is fully antisymmetrized.

In order to obtain a minimal set of Lagrangian terms of the final potential matrix have been eliminated until the rank of the final potential matrix
(consisting of multiple Lagrangian terms and the spin structures in \eq{eq:BBBbasis}) matches the number of terms in the Lagrangian.
The minimal non-relativistic six-baryon contact Lagrangian is \ct*{Petschauer2016}
\begin{align*} \label{eq:minct}
 \mathscr L =
 -\,&\lc_1 \trace{\bar B_a\bar B_b\bar B_c B_a B_b B_c} \\
 +\,&\lc_2 \trace{\bar B_a\bar B_b B_a\bar B_c B_b B_c} \displaybreak[0]\\
 -\,&\lc_3 \trace{\bar B_a\bar B_b B_a B_b\bar B_c B_c} \displaybreak[0]\\
 +\,&\lc_4 \trace{\bar B_a B_a\bar B_b B_b\bar B_c B_c} \displaybreak[0]\\
 -\,&\lc_5 \trace{\bar B_a\bar B_b B_a B_b}\; \trace{\bar B_c B_c} \displaybreak[0]\\
 -\,&\lc_6 \Big(\trace{\bar B_a\bar B_b\bar B_c B_a(\sigma^i B)_b(\sigma^i B)_c}\\
 &\qquad + \trace{\bar B_c\bar B_b\bar B_a(\sigma^i B)_c(\sigma^i B)_b B_a}\Big)\displaybreak[0]\\
 +\,&\lc_7 \Big(\trace{\bar B_a\bar B_b B_a\bar B_c(\sigma^i B)_b(\sigma^i B)_c}\\
&\qquad + \trace{\bar B_c\bar B_b(\sigma^i B)_c\bar B_a(\sigma^i B)_b B_a}\Big) \displaybreak[0]\\
 -\,&\lc_8 \Big(\trace{\bar B_a\bar B_b B_a(\sigma^i B)_b\bar B_c(\sigma^i B)_c}\\
&\qquad + \trace{\bar B_b\bar B_a(\sigma^i B)_b B_a\bar B_c(\sigma^i B)_c}\Big) \displaybreak[0]\\
 +\,&\lc_9 \trace{\bar B_a B_a\bar B_b(\sigma^i B)_b\bar B_c(\sigma^i B)_c} \displaybreak[0]\\
 -\,&\lc_{10} \Big(\trace{\bar B_a\bar B_b B_a(\sigma^i B)_b}\; \trace{\bar B_c(\sigma^i B)_c}\\
&\qquad + \trace{\bar B_b\bar B_a(\sigma^i B)_b B_a}\; \trace{\bar B_c(\sigma^i B)_c}\Big) \displaybreak[0]\\
 -\,&\lc_{11} \trace{\bar B_a\bar B_b\bar B_c(\sigma^i B)_a B_b(\sigma^i B)_c} \displaybreak[0]\\
 +\,&\lc_{12} \trace{\bar B_a\bar B_b(\sigma^i B)_a\bar B_c B_b(\sigma^i B)_c} \displaybreak[0]\\
 -\,&\lc_{13} \trace{\bar B_a\bar B_b(\sigma^i B)_a(\sigma^i B)_b\bar B_c B_c} \displaybreak[0]\\
 -\,&\lc_{14} \trace{\bar B_a\bar B_b(\sigma^i B)_a(\sigma^i B)_b}\; \trace{\bar B_c B_c} \displaybreak[0]\\
 -\,&\mathrm i\, \epsilon^{ijk}\lc_{15}\trace{\bar B_a\bar B_b\bar B_c(\sigma^i B)_a(\sigma^j B)_b(\sigma^k B)_c} \displaybreak[0]\\
 +\,&\mathrm i\, \epsilon^{ijk}\lc_{16}\trace{\bar B_a\bar B_b(\sigma^i B)_a\bar B_c(\sigma^j B)_b(\sigma^k B)_c} \displaybreak[0]\\
 -\,&\mathrm i\, \epsilon^{ijk}\lc_{17}\trace{\bar B_a\bar B_b(\sigma^i B)_a(\sigma^j B)_b\bar B_c(\sigma^k B)_c} \\
 +\,&\mathrm i\, \epsilon^{ijk}\lc_{18}\trace{\bar B_a(\sigma^i B)_a\bar B_b(\sigma^j B)_b\bar B_c(\sigma^k B)_c} \,, \numberthis
\end{align*}
with vector indices \(i,j,k\) and two-component spinor indices \(a,b,c\).
In total 18 low-energy constants \(\lc_1\dots \lc_{18}\) are present.
The low-energy constant \(E\) of the six-nucleon contact term (\cf \ct{Epelbaum2002}) can be expressed through these LECs by \(E=2(\lc_4-\lc_9)\).

\begin{table*}
\centering
\vspace{.3\baselineskip}

\begin{tabular}{>{$}c<{$}>{$}c<{$}>{$}c<{$}>{$}c<{$}}
\toprule
\text{states} & (S,I) & {}^2S_{1/2} & {}^4S_{3/2} \\
\cmidrule(lr){1-2} \cmidrule(lr){3-4}
NNN & (0,\frac12) & \ir{\overline{35}}\\
\cmidrule(lr){1-2} \cmidrule(lr){3-4}
\Lambda NN,\Sigma NN & (-1,0) & \ir{\overline{10}},\ir{\overline{35}} & \ir{\overline{10}}_a \\
\Lambda NN,\Sigma NN & (-1,1) & \ir{27},\ir{\overline{35}} & \ir{27}_a \\
\Sigma NN & (-1,2) & \ir{35}\\
\cmidrule(lr){1-2} \cmidrule(lr){3-4}
\Lambda\Lambda N,\Sigma\Lambda N,\Sigma\Sigma N,\Xi NN & (-2,\frac12) & \ir{8},\ir{\overline{10}},\ir{27},\ir{\overline{35}} & \ir{8}_a,\ir{\overline{10}}_a,\ir{27}_a\\
\Sigma\Lambda N,\Sigma\Sigma N,\Xi NN & (-2,\frac32) & \ir{10},\ir{27},\ir{35},\ir{\overline{35}} & \ir{10}_a,\ir{27}_a\\
\Sigma\Sigma N & (-2,\frac52) & \ir{35} \\
\cmidrule(lr){1-2} \cmidrule(lr){3-4}
\Lambda\Lambda\Lambda,\Sigma\Sigma\Lambda,\Sigma\Sigma\Sigma,\Xi\Lambda N,\Xi\Sigma N & (-3,0) & \ir{8},\ir{27} & \ir{1}_a,\ir{8}_a,\ir{27}_a\\
\Sigma\Lambda\Lambda,\Sigma\Sigma\Lambda,\Sigma\Sigma\Sigma,\Xi\Lambda N,\Xi\Sigma N & (-3,1) & \ir{8},\ir{10},\ir{\overline{10}},\ir{27},\ir{35},\ir{\overline{35}} & \ir{8}_a,\ir{10}_a,\ir{\overline{10}}_a,\ir{27}_a\\
\Sigma\Sigma\Lambda,\Sigma\Sigma\Sigma,\Xi\Sigma N & (-3,2) & \ir{27},\ir{35},\ir{\overline{35}} & \ir{27}_a \\
\cmidrule(lr){1-2} \cmidrule(lr){3-4}
\Xi\Lambda\Lambda,\Xi\Sigma\Lambda,\Xi\Sigma\Sigma,\Xi\Xi N & (-4,\frac12) & \ir{8},\ir{10},\ir{27},\ir{35} & \ir{8}_a,\ir{10}_a,\ir{27}_a\\
\Xi\Sigma\Lambda,\Xi\Sigma\Sigma,\Xi\Xi N & (-4,\frac32) & \ir{\overline{10}},\ir{27},\ir{35},\ir{\overline{35}} & \ir{\overline{10}}_a,\ir{27}_a \\
\Xi\Sigma\Sigma & (-4,\frac52) & \ir{\overline{35}} \\
\cmidrule(lr){1-2} \cmidrule(lr){3-4}
\Xi\Xi\Lambda,\Xi\Xi\Sigma & (-5,0) & \ir{10},\ir{35} & \ir{10}_a\\
\Xi\Xi\Lambda,\Xi\Xi\Sigma & (-5,1) & \ir{27},\ir{35} & \ir{27}_a \\
\Xi\Xi\Sigma & (-5,2) & \ir{\overline{35}}\\
\cmidrule(lr){1-2} \cmidrule(lr){3-4}
\Xi\Xi\Xi & (-6,\frac12) & \ir{35}\\
\bottomrule
\end{tabular}
\caption{Irreducible representations for three-baryon states with strangeness \(S\) and isospin \(I\) in partial waves \(\vert {}^{2S+1}L_J \rangle\),
with the total spin \(S=\frac12,\frac32\), the angular momentum \(L=0\) and the total angular momentum \(J=\frac12,\frac32\) \ct*{Petschauer2016}.
\label{tab:isoBBB}
}
\end{table*}

As in the two-body sector, group theoretical considerations can deliver valueable constrains on the resulting potentials.
In flavor space the three octet baryons form the 512-dimensional tensor product \(\ir8 \otimes \ir8 \otimes \ir8\), which decomposes into the following irreducible SU(3) representations
\begin{align*} \label{eq:irrBBB}
& \ir8 \otimes \ir8 \otimes \ir8 ={} \\
& \quad\ir{64} \oplus (\ir{35} \oplus \ir{\overline{35}})_2 \oplus \ir{27}_6
\oplus (\ir{10} \oplus \ir{\overline{10}})_4 \oplus \ir{8}_8 \oplus \ir{1}_2\,, \numberthis
\end{align*}
where the multiplicity of an irreducible representations is denoted by subscripts.
In spin space one obtain for the product of three doublets
\begin{equation}
\ir2 \otimes \ir2 \otimes \ir2 = \ir{2}_2 \oplus \ir{4}\,.
\end{equation}
Transitions are only allowed between irreducible representations of the same type.
Analogous to \ct{Dover1990} for the two-baryon sector, the contributions of different irreducible representations to three-baryon multiplets in \tab{tab:isoBBB} can be established.
At leading order only transitions between \(S\)-waves are possible, since the potentials are momentum-independent.
Due to the Pauli principle the totally symmetric spin-quartet \(\mathbf4\) must combine with the totally antisymmetric part of \(\ir8 \otimes \ir8 \otimes \ir8\) in flavor space,%
\begin{equation} \label{eq:alt38}
 \text{Alt}_3(\mathbf8) = \ir{56}_a = \ir{27}_a+\ir{10}_a+\ir{\overline{10}}_a+\ir{8}_a+\ir{1}_a \,.
\end{equation}
It follows, that these totally antisymmetric irreducible representations are present only in states with total spin 3/2.
The totally symmetric part of \(\ir8 \otimes \ir8 \otimes \ir8\) leads to
\begin{equation}
\text{Sym}_3(\mathbf8) = \ir{120}_s = \ir{64}_s+\ir{27}_s+\ir{10}_s+\ir{\overline{10}}_s+\ir{8}_s+\ir{1}_s \,.
\end{equation}
However, the totally symmetric flavor part has no totally antisymmetric counterpart in spin space, hence these representations do not contribute to the potential.
In \tab{tab:isoBBB} these restrictions obtained by the generalized Pauli principle have already be incorporated.
The potentials of \ct{Petschauer2016} (decomposed in isospin basis and partial waves) fulfill the restrictions of \tab{tab:isoBBB}.
For example the combination of LECs related to the representation \(\ir{\overline{35}}\) is present in the \(NNN\) interaction as well as in  the \(\Xi\Xi\Sigma\ (-5,2)\) interaction.

\subsection{One-meson exchange component} \label{subsec:BBBome}

The meson-baryon couplings in the one-meson exchange diagram of \fig{fig:3BF} emerges from the leading-order chiral Lagrangian \(\mathscr{L}_\mathrm{B}^{(1)}\), see \eq{eq:3vert}.
The other vertex involves four baryon fields and one pseudo\-scalar-meson field.
In \ct{Petschauer2016} an overcomplete set of terms for the corresponding Lagrangian has been constructed.
In order to obtain the complete minimal Lagrangian from the overcomplete set of terms, the matrix elements of the process \(B_1B_2\to B_3B_4\phi_1\) has been considered in \ct{Petschauer2016}.
The corresponding spin operators in the potential are
\begin{equation}
\vec\sigma_1\cdot\vec q\,,\quad
\vec\sigma_2\cdot\vec q\,,\quad
\mathrm i\,(\vec\sigma_1\times\vec\sigma_2)\cdot\vec q \,,
\end{equation}
where \(\vec q\) denotes the momentum of the emitted meson.
Redundant term are removed until the rank of the potential matrix formed by all transitions and spin operators matches the number of terms in the Lagrangian.
One ends up with the minimal non-relativistic chiral Lagrangian
\begin{align*} \label{eq:LBBMBBmin}
 \mathscr L ={}
 &\ld_1/f_0 \trace{\bar B_a \extfield B_a\bar B_b(\sigma^i B)_b}\displaybreak[0]\\
 &+\ld_2/f_0 \Big( \trace{\bar B_a  B_a \extfield\bar B_b(\sigma^i B)_b} \\
&\qquad\qquad + \trace{\bar B_a  B_a\bar B_b(\sigma^i B)_b \extfield}\Big)\displaybreak[0]\\
 &+\ld_3/f_0 \trace{\bar B_b \extfield(\sigma^i B)_b\bar B_a B_a}\displaybreak[0]\\
 &-\ld_4/f_0 \Big( \trace{\bar B_a\extfield\bar B_b B_a(\sigma^i B)_b} \\
&\qquad\qquad + \trace{\bar B_b \bar B_a (\sigma^i B)_b\extfield B_a}\Big)\displaybreak[0]\\
 &-\ld_5/f_0 \Big( \trace{\bar B_a\bar B_b\extfield B_a(\sigma^i B)_b} \\
&\qquad\qquad + \trace{\bar B_b \bar B_a \extfield (\sigma^i B)_b B_a}\Big)\displaybreak[0]\\
 &-\ld_6/f_0 \Big( \trace{\bar B_b\extfield\bar B_a(\sigma^i B)_b B_a} \\
&\qquad\qquad + \trace{\bar B_a \bar B_b B_a\extfield (\sigma^i B)_b}\Big)\displaybreak[0]\\
 &-\ld_7/f_0 \Big( \trace{\bar B_a\bar B_b B_a(\sigma^i B)_b\extfield} \\
&\qquad\qquad + \trace{\bar B_b \bar B_a (\sigma^i B)_b B_a\extfield}\Big)\displaybreak[0]\\
 &+\ld_8/f_0 \trace{\bar B_a\extfield B_a}\trace{\bar B_b(\sigma^i B)_b}\displaybreak[0]\\
 &+\ld_9/f_0 \trace{\bar B_a B_a\extfield}\trace{\bar B_b(\sigma^i B)_b}\displaybreak[0]\\
 &+\ld_{10}/f_0 \trace{\bar B_b\extfield(\sigma^i B)_b}\trace{\bar B_a B_a}\displaybreak[0]\\
 &+\mathrm i\,\epsilon^{ijk}\ld_{11}/f_0 \trace{\bar B_a (\sigma^i B)_a (\nabla^k \phi)\bar B_b(\sigma^j B)_b}\displaybreak[0]\\
 &-\mathrm i\,\epsilon^{ijk}\ld_{12}/f_0 \Big( \trace{\bar B_a(\nabla^k \phi)\bar B_b(\sigma^i B)_a(\sigma^j B)_b} \\
&\qquad\qquad - \trace{\bar B_b \bar B_a (\sigma^j B)_b(\nabla^k \phi) (\sigma^i B)_a}\Big)\displaybreak[0]\\
 &-\mathrm i\,\epsilon^{ijk}\ld_{13}/f_0 \trace{\bar B_a\bar B_b(\nabla^k \phi)(\sigma^i B)_a(\sigma^j B)_b}\displaybreak[0]\\
 &-\mathrm i\,\epsilon^{ijk}\ld_{14}/f_0 \trace{\bar B_a\bar B_b(\sigma^i B)_a(\sigma^j B)_b(\nabla^k \phi)}\,, \numberthis
\end{align*}
with two-component spinor indices  \(a\) and \(b\) and 3-vector indices \(i\), \(j\) and \(k\).
For all possible strangeness sectors \(S=-4\ldots0\) one obtains in total 14 low-energy constants \(\ld_1\dots \ld_{14}\) .
The low-energy constant of the corresponding vertex in the nucleonic sector \(D\) is related to the LECs above by \(D=4 (\ld_1 - \ld_3 + \ld_8 - \ld_{10})\).%
\footnote{This LEC \(D\) has not to be confused with the axial-vector coupling constant \(D\) in \eq{eq:3vert}.}

\begin{figure*}
\centering
\hfill
\begin{subfigure}[t]{.45\textwidth}
\centering
\vspace{.3\baselineskip}
\begin{overpic}[scale=.6]{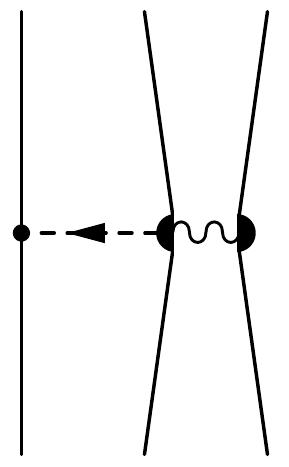}
\put(2,103){$l$}\put(27,103){$m$}\put(55,103){$n$}
\put(2,-10){$i$}\put(27,-10){$j$}\put(55,-10){$k$}
\put(0,-25){$A$}\put(25,-25){$B$}\put(53,-25){$C$}
\put(16,59){$\phi$}
\end{overpic}
\vspace{1.7\baselineskip}
\caption{
Generic one-meson exchange diagram
\label{fig:ome-gen}
}
\end{subfigure}
\hfill
\begin{subfigure}[t]{.45\textwidth}
\centering
\vspace{.3\baselineskip}
\begin{overpic}[scale=.6]{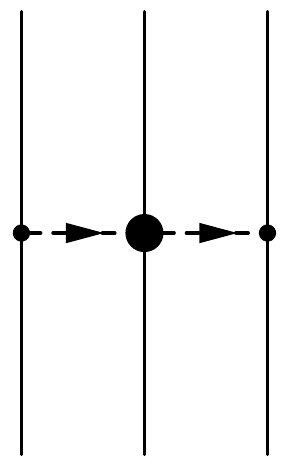}
\put(2,103){$l$}\put(27,103){$m$}\put(55,103){$n$}
\put(2,-10){$i$}\put(27,-10){$j$}\put(55,-10){$k$}
\put(0,-25){$A$}\put(25,-25){$B$}\put(53,-25){$C$}
\put(13,59){$\phi_1$} \put(39,59){$\phi_2$}
\end{overpic}
\vspace{1.7\baselineskip}
\caption{
Generic two-meson exchange diagram
\label{fig:tme-gen}
}
\end{subfigure}
\hfill\mbox{}
\caption{Generic meson-exchange diagrams.
The wiggly line symbolized the four-baryon contact vertex, to illustrate the baryon bilinears.}
\end{figure*}
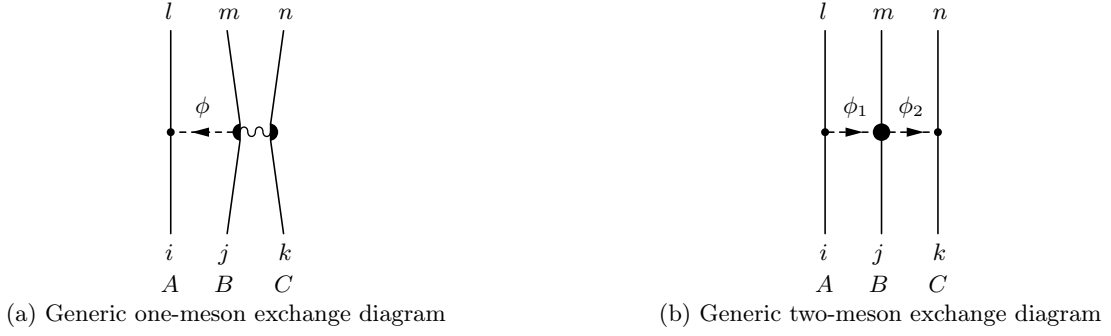

To obtain the 3BF one-meson-exchange diagram, the generic one-meson-exchange diagram in \fig{fig:ome-gen} can be investigated.
It involves the baryons \(i,j,k\) in the initial state, the baryons \(l,m,n\) in the final state and an exchanged meson \(\phi\).
The contact vertex on the right is pictorially separated into two parts to indicate that baryon j--m and k--n are in the same bilinear.
The spin spaces corresponding to the baryon bilinears are denoted by \(A,B,C\).

On obtains a generic potential of the form
\begin{equation}
V = \frac{1}{2f_0^2} \frac{\vec\sigma_A\cdot\vec q_{li}}{\vec q_{li}^{\,2}+m_{\phi}^2} \Big(
N_1 \vec\sigma_C\cdot\vec q_{li}
+N_2 \mathrm i\,(\vec\sigma_B\times\vec\sigma_C)\cdot\vec q_{li}
\Big)\,,
\end{equation}
with the momentum transfer \(\vec q_{li} = \vec p_l-\vec p_i\) carried by the exchanged meson.
The constants \(N_1\) and \(N_2\) are linear combinations of low-energy constants.

The complete one-meson exchange three-baryon potential for the process \(B_1B_2B_3\to B_4B_5B_6\) is finally obtained by summing up the 36 permutations of initial-state and final-state baryons for a fixed meson and by summing over all mesons \(\phi \in \left\{\pi^0,\pi^+,\pi^-,K^+,K^-,K^0,\bar K^0,\eta\right\}\).
Additional minus signs arise from interchanging fermions and some diagrams need to be multiplied by spin exchange operators in order to be consistent with the form set up in \eq{eq:spinpot}.
As defined before, the baryons \(B_1\), \(B_2\) and \(B_3\) belong to the spin-spaces 1, 2 and 3, respectively.

\subsection{Two-meson exchange component} \label{subsec:2MEtme}

The two-meson exchange diagram of \fig{fig:3BF} includes the vertex arising from the Lagrangian in \eq{eq:3vert}.
Furthermore the the well-known \(\mathcal O(q^2)\) meson-baryon Lagrangian \ct*{Krause1990} is necessary.
For the two-meson exchange diagram of \fig{fig:3BF} we need in addition to the Lagrangian in \eq{eq:3vert} the well-known \(\mathcal O(q^2)\) meson-baryon Lagrangian \ct*{Krause1990}.
The relevant terms are \ct*{Oller2006}
\begin{align*} \label{eq:MBMBLagr}
\mathscr L ={}
& b_D\langle\bar B\{\chi_+,B\}\rangle
+ b_F\langle\bar B[\chi_+,B]\rangle
+ b_0\langle\bar BB\rangle\,\langle\chi_+\rangle \\
&+ b_1\langle\bar B[u^\mu,[u_\mu,B]]\rangle
+ b_2\langle\bar B\{u^\mu,\{u_\mu,B\}\}\rangle \\
&+ b_3\langle\bar B\{u^\mu,[u_\mu,B]\}\rangle
+ b_4\langle\bar BB\rangle\,\langle u^\mu u_\mu\rangle \\
&+\mathrm i d_1\langle\bar B\{[u^\mu,u^\nu],\sigma_{\mu\nu}B\}\rangle\\
&+\mathrm i d_2\langle\bar B[[u^\mu,u^\nu],\sigma_{\mu\nu}B]\rangle \\
&+\mathrm i d_3\langle\bar Bu^\mu\rangle\langle u^\nu\sigma_{\mu\nu}B\rangle \,, \numberthis
\end{align*}
with \(u_\mu = -\frac1{f_0}\partial_\mu\phi+\mathcal O(\phi^3)\) and \(\chi_+ = 2\chi-\frac1{4f_0^2}\{\phi,\{\phi,\chi\}\}\linebreak[0]+\mathcal O(\phi^4)\), where
\begin{equation}
 \chi = \begin{pmatrix} m_\pi^2 & 0 & 0 \\ 0 & m_\pi^2 & 0 \\ 0 & 0 & 2m_K^2-m_\pi^2 \end{pmatrix} \,.
\end{equation}
The terms proportional to \(b_D,b_F,b_0\) break explicitly SU(3) flavor symmetry, because of different meson masses \(m_K\neq m_\pi\).
The LECs of \eq{eq:MBMBLagr} are related to the conventional LECs of the nucleonic sector by \cts*{Frink2004,Mai2009}
\begin{align*} \label{eq:LECc134}
c_1 &= \frac{1}{2} (2 b_0+b_D+b_F)\,,\\
c_3 &= b_1+b_2+b_3+2 b_4\,,\\
c_4 &= 4 (d_1+d_2)\,. \numberthis
\end{align*}

To obtain the potential of the two-meson exchange diagram of \fig{fig:3BF}, the generic diagram of \fig{fig:tme-gen} can be considered.
It includes the baryons \(i,j,k\) in the initial state, the baryons \(l,m,n\) in the final state, and two exchanged mesons \(\phi_1\) and \(\phi_2\).
The spin spaces corresponding to the baryon bilinears are denoted by \(A,B,C\) and they are aligned with the three initial baryons.
The momentum transfers carried by the virtual mesons are \(\vec q_{li} = \vec p_l-\vec p_i\) and \(\vec q_{nk} = \vec p_n-\vec p_k\).
One obtains the generic transition amplitude
\begin{align*}
V ={} & -\frac{1}{4f_0^4} \frac{\vec\sigma_A\cdot\vec q_{li}\ \vec\sigma_C\cdot\vec q_{nk}}{(\vec q_{li}^{\,2}+m_{\phi_1}^2)(\vec q_{nk}^{\,2}+m_{\phi_2}^2)} \\
&\qquad\times\Big(N'_1 + N'_2\,\vec q_{li}\cdot\vec q_{nk} +N'_3\,\mathrm i\,(\vec q_{li}\times\vec q_{nk})\cdot\vec\sigma_B\Big) \,, \numberthis
\end{align*}
with \(N^\prime_i\) linear combinations of the low-energy constants of the three involved vertices.
The complete three-body potential for a transition \(B_1B_2B_3\rightarrow B_4B_5B_6\) can be calculated by summing up the contributions of all 18 distinguable Feynman diagrams and by summing over all possible exchanged mesons.
If the baryon lines are not in the configuration 1--4, 2--5 and 3--6 additional (negative) spin-exchange operators have to be included.

\subsection{\texorpdfstring{$\Lambda NN$}{Lambda-N-N} three-baryon potentials} \label{subsec:BBBpotex}

In order to give a concrete example the explicit expression for the \(\Lambda NN\) three-body potentials in spin-, isospin- and momentum-space are presented
for the contact interaction and one- and two-pion exchange contributions \ct*{Petschauer2016}.
The potentials are calculated in the particle basis and afterwards rewritten into isospin operators.

The \(\Lambda NN\) contact interaction is described by the following potential
\begin{align*}
V^{\Lambda NN}_\mathrm{ct} ={}
& \phantom{{}+{}} \lc'_1\ (\mathbbm1 - \vec\sigma_2\cdot\vec\sigma_3 ) ( 3 + \vec\tau_2\cdot\vec\tau_3 ) \\
& + \lc'_2\ \vec\sigma_1\cdot(\vec\sigma_2+\vec\sigma_3)\,(\mathbbm1 - \vec\tau_2\cdot\vec\tau_3) \\
& + \lc'_3\ (3 + \vec\sigma_2\cdot\vec\sigma_3 ) ( \mathbbm1 - \vec\tau_2\cdot\vec\tau_3 ) \,, \numberthis
\end{align*}
where the primed constants are linear combinations of \(\lc_1\dots \lc_{18}\) of \eq{eq:minct}.
The symbols \(\vec\sigma\) and \(\vec\tau\) denote the usual Pauli matrices in spin and isospin space.
The constant \(\lc'_1\) appears only in the transition with total isospin \(I=1\).
The constants \(\lc'_2\) and \(\lc'_3\) contribute for total isospin \(I=0\).

For the \(\Lambda NN\) one-pion exchange three-body potentials, various diagrams are absent due to the vanishing \(\Lambda\Lambda\pi\)-vertex, which is forbidden by isospin symmetry.
One obtains the following potential
\begin{align*} \label{eq:LNNope}
V^{\Lambda NN}_\mathrm{OPE} =&{} -\frac{g_A}{2f_0^2} \\
\times\bigg(&\frac{\vec\sigma_2\cdot\vec q_{52}}{\vec q_{52}^{\,2}+m_\pi^2} \vec\tau_2\cdot\vec\tau_3
\Big[  (\ld'_1\vec\sigma_1+\ld'_2\vec\sigma_3)\cdot\vec q_{52} \Big] \\
&+\frac{\vec\sigma_3\cdot\vec q_{63}}{\vec q_{63}^{\,2}+m_\pi^2} \vec\tau_2\cdot\vec\tau_3
\Big[  (\ld'_1\vec\sigma_1+\ld'_2\vec\sigma_2)\cdot\vec q_{63} \Big] \\
&+\Ps_{23}\Pt_{23} \Ps_{13}\frac{\vec\sigma_2\cdot\vec q_{62}}{\vec q_{62}^{\,2}+m_\pi^2} \vec\tau_2\cdot\vec\tau_3 \\
&\qquad\times\Big[  -\frac{\ld'_1+\ld'_2}2 (\vec\sigma_1+\vec\sigma_3)\cdot\vec q_{62} \\
&\qquad\qquad+ \frac{\ld'_1-\ld'_2}2\,\mathrm i\,(\vec\sigma_3\times\vec\sigma_1)\cdot\vec q_{62}  \Big] \\
&+\Ps_{23}\Pt_{23} \Ps_{12}\frac{\vec\sigma_3\cdot\vec q_{53}}{\vec q_{53}^{\,2}+m_\pi^2} \vec\tau_2\cdot\vec\tau_3 \\
&\qquad\times\Big[  -\frac{\ld'_1+\ld'_2}2 (\vec\sigma_1+\vec\sigma_2)\cdot\vec q_{53} \\
&\qquad\qquad- \frac{\ld'_1-\ld'_2}2\,\mathrm i\,(\vec\sigma_1\times\vec\sigma_2)\cdot\vec q_{53}  \Big]
\bigg)\,, \numberthis
\end{align*}
with only two constants \(\ld'_1\) and \(\ld'_2\), which are linear combinations of the constants \(\ld_1\dots\ld_{14}\).
Exchange operators in spin space \(\Ps_{ij}=\frac12(\mathbbm1+\vec\sigma_i\cdot\vec\sigma_j)\) and in isospin space \(\Pt_{ij}=\frac12(\mathbbm1+\vec\tau_i\cdot\vec\tau_j)\) have been introduced.

The \(\Lambda NN\) three-body interaction generated by two-pion exchange is given by
\begin{align*}
&V^{\Lambda NN}_\mathrm{TPE} ={}
\frac{g_A^2}{3f_0^4}
\frac{\vec\sigma_3\cdot\vec q_{63}\ \vec\sigma_2\cdot\vec q_{52}}{(\vec q_{63}^{\,2}+m_{\pi}^2)(\vec q_{52}^{\,2}+m_{\pi}^2)} \vec\tau_2\cdot\vec\tau_3 \\
&\qquad \ \times \Big( -(3 b_0 + b_D) m_\pi^2      +      (2 b_2 + 3 b_4)      \,\vec q_{63}\cdot\vec q_{52}\Big) \\
&\quad- \Ps_{23}\Pt	_{23} \frac{g_A^2}{3f_0^4}
\frac{\vec\sigma_3\cdot\vec q_{53}\ \vec\sigma_2\cdot\vec q_{62}}{(\vec q_{53}^{\,2}+m_{\pi}^2)(\vec q_{62}^{\,2}+m_{\pi}^2)} \vec\tau_2\cdot\vec\tau_3 \\
&\qquad \ \times \Big( -(3 b_0 + b_D) m_\pi^2      +      (2 b_2 + 3 b_4)      \,\vec q_{53}\cdot\vec q_{62}\Big) \,. \numberthis
\end{align*}
Due to the vanishing of the \(\Lambda \Lambda \pi\) vertex, only those two diagrams contribute, where the (final and initial) \(\Lambda\) hyperon are attached to the central baryon line.

\subsection{Three-baryon force through decuplet saturation}  \label{subsec:BBBDec}

Low-energy two- and three-body interactions derived from SU(2) \cheft are used consistently in combination with each other in nuclear few- and many-body calculations.
The a priori unknown low-energy constants are fitted, for example, to \(NN\) scattering data and \(3N\) observables such as 3-body binding energies \ct*{Epelbaum2002}.
Some of these LECs are, however, large compared to their order of magnitude as expected from the hierarchy of nuclear forces in \fig{fig:hier}.
This feature has its physical origin in strong couplings of the \(\pi N\)-system to the low-lying \(\Delta(1232)\)-resonance.
It is therefore natural to include the \(\Delta(1232)\)-isobar as an explicit degree of freedom in the chiral Lagrangian (\cf \cts{Bernard1997,Kaiser1998,Krebs2007}).
The small mass difference between nucleons and deltas (\(293\ \mathrm{MeV}\)) introduces a small scale, which can be included consistently in the chiral power counting scheme and the hierarchy of nuclear forces.
The dominant parts of the three-nucleon interaction mediated by two-pion exchange at NNLO are then promoted to NLO through the delta contributions.
The appearance of the inverse mass splitting explains the large numerical values of the corresponding LECs \ct*{Epelbaum2009,Epelbaum2008a}.

\begin{figure*}
\centering
\newcommand{\dechierscale}{.5}
\setlength{\tabcolsep}{8pt}
\begin{tabular}{lcc}
\toprule \addlinespace[1ex]
& \multicolumn{2}{c}{three-baryon force} \\ \addlinespace[1ex]
& decuplet-less EFT & decuplet-contribution \\ \addlinespace[1ex] \midrule \addlinespace[2ex]
LO & $\vcenter{\hbox{\rule{0cm}{2cm}}}$ & \\ \addlinespace[2ex]
NLO & &
$\vcenter{\hbox{
\includegraphics[scale=\dechierscale]{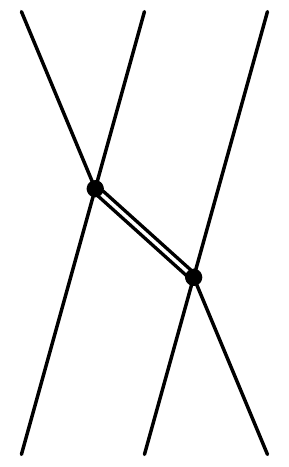}\ \
\includegraphics[scale=\dechierscale]{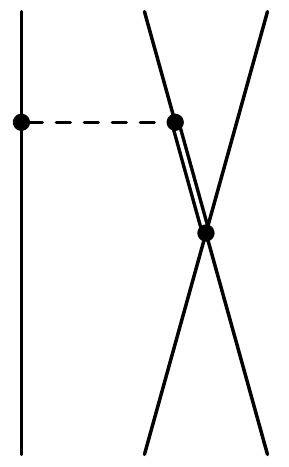}\ \
\includegraphics[scale=\dechierscale]{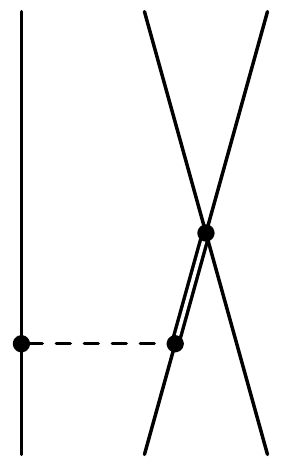}\ \
\includegraphics[scale=\dechierscale]{files/Feynman/BF/FBBBdec2ME}}}$ \\ \addlinespace[2ex]
NNLO &
$\vcenter{\hbox{
\includegraphics[scale=\dechierscale]{files/Feynman/BF/FBBBcont}\ \
\includegraphics[scale=\dechierscale]{files/Feynman/BF/FBBB1ME}\ \
\includegraphics[scale=\dechierscale]{files/Feynman/BF/FBBB2ME2}}}$ &
$\cdots$ \\ \addlinespace[2ex]
\bottomrule
\end{tabular}
\caption{Hierarchy of three-baryon forces with explicit introduction of the baryon decuplet (represented by double lines).} \label{fig:hierdec}
\setlength{\tabcolsep}{6pt}
\end{figure*}

In SU(3) \cheft the situation is similar.
In systems with strangeness \(S=-1\) like \(\Lambda NN\), resonances such as the spin-3/2 \(\Sigma^*\)(1385)-resonance play a similar role as the \(\Delta\) in the \(NNN\) system, as depicted in \fig{fig:irr3BF} on the right side.
The small decuplet-octet mass splitting (in the chiral limit), \(\Delta\defeq M_{10}-M_{8}\), is counted together with external momenta and meson masses as \(\mathcal O(q)\) and thus parts of the NNLO three-baryon interaction are promoted to NLO by the explicit inclusion of the baryon decuplet, as illustrated in \fig{fig:hierdec}. 
It is therefore likewise compelling to treat the three-baryon interaction together with the NLO hyperon-nucleon interaction of \sect{sec:BB}.
Note that in the nucleonic sector, only the two-pion exchange diagram with an intermediate \(\Delta\)-isobar is allowed.
Other diagrams are forbidden due to the Pauli principle, as we will show later.
For three flavors more particles are involved and, in general, also the other diagrams (contact and one-meson exchange) with intermediate decuplet baryons in \fig{fig:hierdec} appear.

The large number of unknown LECs presented in the previous subsections is related to the multitude of three-baryon multiplets, with strangeness ranging from \(0\) to \(-6\).
For selected processes only a small subset of these constants contributes as has been exemplified for the \(\Lambda NN\) three-body interaction.
In this section we present the estimation of these LECs by resonance saturation as done in \ct{Petschauer2017a}.

The leading-order non-relativistic interaction Lagrangian between octet and decuplet baryons (see, \eg \ct{Sasaki2006}) is
\begin{align*} \label{eq:LMBD}
\mathscr{L} = \frac C{f_0} \sum_{a,b,c,d,e=1}^3 \epsilon_{abc} \Bigg(& \bar T_{ade} \vec S^{\,\dagger} \cdot \left(\vec\nabla \phi_{db}\right)B_{ec} \\
&+ \bar B_{ce} \vec S \cdot \left(\vec\nabla\phi_{bd}\right)T_{ade} \Bigg)\,, \numberthis
\end{align*}
where the decuplet baryons are represented by the totally symmetric three-index tensor \(T\), \cf \eq{eq:Tfields}.
At this order only a single LEC \(C\) appears.
Typically the (large-\(N_c\)) value \(C=\frac34g_A\approx 1\) is used, as it leads to a decay width \(\Gamma(\Delta\to\pi N)=110.6\ \mathrm{MeV}\) that is in good agreement with the empirical value of \(\Gamma(\Delta\to\pi N)=(115\pm 5)\ \mathrm{MeV}\) \ct*{Kaiser1998}.
The spin \(\frac12\) to \(\frac32\) transition operators \(\vec S\) connect the two-component spinors of octet baryons with the four-component spinors of decuplet baryons (see \eg \ct*{Weise1988}).
In their explicit form they are given as \(2\times4\) transition matrices
\begin{align*}
S_1 &= \begin{pmatrix}
-\frac{1}{\sqrt{2}} & 0 & \frac{1}{\sqrt{6}} & 0 \\
0 & -\frac{1}{\sqrt{6}} & 0 & \frac{1}{\sqrt{2}}
\end{pmatrix} ,\\
S_2 &= \begin{pmatrix}
-\frac{i}{\sqrt{2}} & 0 & -\frac{i}{\sqrt{6}} & 0 \\
0 & -\frac{i}{\sqrt{6}} & 0 & -\frac{i}{\sqrt{2}}
\end{pmatrix} ,\\
S_3 &= \begin{pmatrix}
0 & \sqrt{\frac{2}{3}} & 0 & 0 \\
0 & 0 & \sqrt{\frac{2}{3}} & 0
\end{pmatrix}. \numberthis
\end{align*}
These operators fulfill the relation \( S_i {S_j}^\dagger = \frac13 ( 2\delta_{ij}-\mathrm i\epsilon_{ijk} \sigma_k )\).

A non-relativistic \(B^*BBB\) Lagrangian with a minimal set of terms is given by \ct*{Petschauer2017a}:
\begin{align*} \label{eq:LDBBBmin}
\mathscr{L} = &\quad\,
\lC_1\sum_{\substack{a,b,c,\\d,e,f=1}}^3 \epsilon_{abc}
\big[
\left(\bar T_{ade}\vec S^\dagger B_{db}\right)\cdot\left(\bar B_{fc}\vec\sigma B_{ef}\right) \\
&\qquad\qquad\qquad\qquad +\left(\bar B_{bd}\vec S\, T_{ade}\right)\cdot\left(\bar B_{fe}\vec\sigma B_{cf}\right)
\big]\\
&+\lC_2
\sum_{\substack{a,b,c,\\d,e,f=1}}^3 \epsilon_{abc}
\big[
\left(\bar T_{ade}\vec S^\dagger B_{fb}\right)\cdot\left(\bar B_{dc}\vec\sigma B_{ef}\right) \\
&\qquad\qquad\qquad\qquad +\left(\bar B_{bf}\vec S\, T_{ade}\right)\cdot\left(\bar B_{fe}\vec\sigma B_{cd}\right)
\big] \,, \numberthis
\end{align*}
with the LECs \(\lC_1\) and \(\lC_2\).
Again one can employ group theory to justify the number of two constants for a transition \(BB\to B^*B\).
In flavor space the two initial octet baryons form the tensor product \(\mathbf8\otimes\mathbf8\), and in spin space they form the product \(\mathbf{2} \otimes \mathbf{2}\).
These tensor products can be decomposed into irreducible representations:
\begin{align*} \label{eq:BBgroup}
\mathbf8\otimes\mathbf8 &= \underbrace{{\mathbf{27}}\oplus{\mathbf{8}_s}\oplus{\mathbf1}}_\text{symmetric}\oplus\underbrace{\mathbf{10}\oplus\mathbf{10^*}\oplus\mathbf{8}_a}_\text{antisymmetric}\,,\\
\mathbf{2} \otimes \mathbf{2} &= \mathbf{1}_a \oplus \mathbf{3}_s \,. \numberthis
\end{align*}
In the final state, having a decuplet and an octet baryon, the situation is similar:
\begin{align*} \label{eq:DBgroup}
\mathbf{10}\otimes\mathbf8 &= \mathbf{35}\oplus\mathbf{27}\oplus\mathbf{10}\oplus\mathbf{8}\,,\\
\mathbf{4} \otimes \mathbf{2} &= \mathbf{3} \oplus \mathbf{5} \,.\numberthis
\end{align*}
As seen in the previous sections, at leading order only \(S\)-wave transitions occur, as no momenta are involved.
Transitions are only allowed between the same types of irreducible (flavor and spin) representations.
Therefore, in spin space the representation \(\mathbf{3}\) has to be chosen.
Because of the Pauli principle in the initial state, the symmetric \(\mathbf{3}\) in spin space combines with the antisymmetric representations \(\mathbf{10},\mathbf{10^*},\mathbf{8}_a\) in flavor space.
But only \(\mathbf{10}\) and \(\mathbf{8}_a\) have a counterpart in the final state flavor space.
This number of two allowed transitions matches the number of two LECs in the minimal Lagrangian.
Another interesting observation can be made from \eqs{eq:BBgroup} and \eq*{eq:DBgroup}.
For \(NN\) states only the representations \(\mathbf{27}\) and \(\mathbf{10^*}\) can contribute, as can be seen, \eg in \tab{tab:PWDBB}.
But these representations combine either with the wrong spin, or have no counterpart in the final state.
Therefore, \(NN\to \Delta N\) transitions in \(S\)-waves are not allowed because of the Pauli principle.

Having the above two interaction types at hand, one can estimate the low-energy constants of the leading three-baryon interaction by decuplet saturation using the diagrams shown in \fig{fig:hierdec}.
At this order, where no loops are involved, one just needs to evaluate the diagrams with an intermediate decuplet baryon and the diagrams without decuplet baryons and compare them with each other.

\begin{figure*}
\centering
\hfill
\begin{subfigure}[b]{.25\textwidth}
\centering
\(
\vcenter{\hbox{\includegraphics[scale=.45]{files/Feynman/BF/FBBBcont}}}
\ =\
\vcenter{\hbox{\includegraphics[scale=.45]{files/Feynman/BF/FBBBdeccont}}}
\)
\caption{Saturation of the six-baryon contact interaction}
\label{fig:3BFctDec}
\end{subfigure}
\hfill
\begin{subfigure}[b]{.3\textwidth}
\centering
\(
\vcenter{\hbox{\includegraphics[scale=.55]{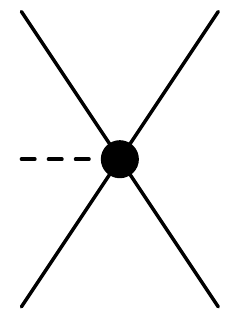}}} =
\vcenter{\hbox{\includegraphics[scale=.55]{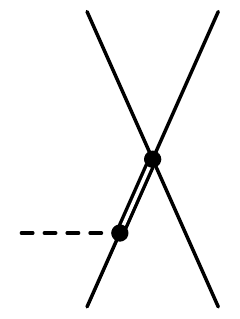}}} +
\vcenter{\hbox{\includegraphics[scale=.55]{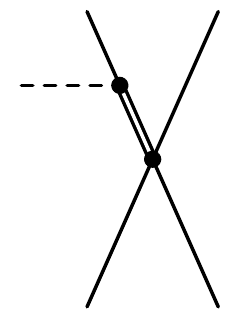}}}
\)
\vspace{.3\baselineskip}
\caption{Saturation of the \(BB\to BB\phi\) vertex}
\label{fig:3BF1MEDec}
\end{subfigure}
\hfill
\begin{subfigure}[b]{.4\textwidth}
\centering
\(
\vcenter{\hbox{\includegraphics[scale=.5]{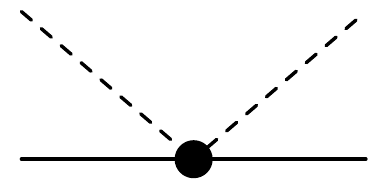}}}
\ = \
\vcenter{\hbox{\includegraphics[scale=.5]{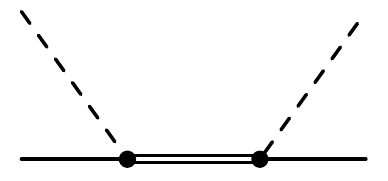}}}
\ + \
\vcenter{\hbox{\includegraphics[scale=.5]{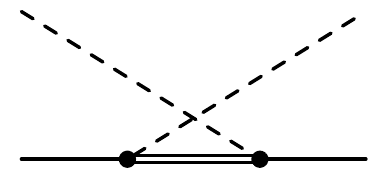}}}
\)
\caption{Saturation of the NLO baryon-meson vertex\\\mbox{}}
\label{fig:3BF2MEDec}
\end{subfigure}
\hfill\mbox{}
\caption{Saturation via decuplet resonances.}
\end{figure*}
In order to estimate the LECs of the six-baryon contact Lagrangian of \eq{eq:minct}, one can consider the process \(B_1B_2B_3\to B_4B_5B_6\) as depicted in \fig{fig:3BFctDec}.
The left side of \fig{fig:3BFctDec} has already been introduced in the previous subsection and can be obtained by performing all 36 Wick contractions.
For the diagrams on right side of \fig{fig:3BFctDec} the procedure is similar.
After summing over all intermediate decuplet baryons \(B^*\), the full three-body potential of all possible combinations of baryons on the left side of \fig{fig:3BFctDec} can be compared with the ones on the right side.
In the end the 18 LECs of the six-baryon contact Lagrangian \(\lc_1,\dots,\lc_{18}\) of \eq{eq:minct} can be expressed as linear combinations of the combinations \(\lC_1^2\), \(\lC_2^2\) and \(\lC_1\lC_2\)
and are proportional to the inverse average decuplet-octet baryon mass splitting \(1/\Delta\) \ct*{Petschauer2017a}.

Since we are at the leading order only tree-level diagrams are involved and we can estimate the LECs of the one-meson-exchange part of the three-baryon forces already on the level of the vertices, as depicted in \fig{fig:3BF1MEDec}.
We consider the transition matrix elements of the process \(B_1B_2\to B_3B_4\phi\) and start with the left side of \fig{fig:3BF1MEDec}.
After doing all possible Wick contractions, summing over all intermediate decuplet baryons, and comparing the left side of \fig{fig:3BF1MEDec} with the right hand side for all combinations of baryons and mesons, the LECs can be estimated.
The LECs of the minimal non-relativistic chiral Lagrangian for the four-baryon vertex including one meson of \eq{eq:LBBMBBmin} \(\ld_1,\dots,\ld_{14}\) are then proportional to \(C/\Delta\) and to linear combinations of \(\lC_1\) and \(\lC_2\) \ct*{Petschauer2017a}.

The last class of diagrams is the three-body interaction with two-meson exchange.
As done for the one-meson exchange, the unknown LECs can be saturated directly on the level of the vertex and one can consider the process \(B_1\phi_1\to B_2\phi_2\) as shown in \fig{fig:3BF2MEDec}.
A direct comparison of the transition matrix elements for all combinations of baryons and mesons after summing over all intermediate decuplet baryons \(B^*\) leads to the following contributions to the LECs of the meson-baryon Lagrangian in Eq.~\eqref{eq:MBMBLagr}:
\begin{align*}
b_D  &= 0 \,,\
b_F  = 0 \,,\
b_0  = 0 \,,\\
b_1  &= \frac{7 C^2}{36 \Delta } \,,\
b_2  = \frac{C^2}{4 \Delta } \,,\
b_3  = -\frac{C^2}{3 \Delta } \,,\
b_4  = -\frac{C^2}{2 \Delta } \,,\\
d_1  &= \frac{C^2}{12 \Delta } \,,\
d_2  = \frac{C^2}{36 \Delta } \,,\
d_3  = -\frac{C^2}{6 \Delta } \,, \numberthis
\end{align*}
These findings are consistent with the \(\Delta\)(1232) contribution to the LECs \(c_1,c_3,c_4\) (see \eq{eq:LECc134}) in the nucleonic sector \ct*{Bernard1997,Epelbaum2008a}:
\begin{equation}
c_1=0\,,\qquad c_3=-2c_4=-\frac{g_A^2}{2 \Delta } \,.
\end{equation}

Employing the LECs obtained via decuplet saturation, the constants of the \(\Lambda NN\) interaction (contact interaction, one-pion and two-pion exchange) of \ssect{subsec:BBBpotex} can be evaluated:
\begin{align*}
&\lc^\prime_1  = \lc^\prime_3 = {} \frac{(\lC_1+3 \lC_2)^2}{72 \Delta } \,, \quad
\lc^\prime_2  ={} 0 \,, \\
&\ld'_1  ={} 0 \,, \quad
\ld'_2  ={} \frac{2 C (\lC_1+3 \lC_2)}{9 \Delta } \,, \\
&3b_0+b_D ={} 0 \,, \quad
2b_2+3b_4 ={} -\frac{C^2}{\Delta } \,. \numberthis
\end{align*}
Obviously, the only unknown constant here is the combination \(\lC'=\lC_1+3 \lC_2\).
It is also interesting to see, that the (positive) sign of the constants \(\lc^\prime_i\) for the contact interaction is already fixed, independently of the values of the two LECs \(\lC_1\) and \(\lC_2\).

\subsection{Effective in-medium two-baryon interaction}

\begin{figure}
\begin{minipage}{.09\textwidth} \centering
\includegraphics[scale=.5]{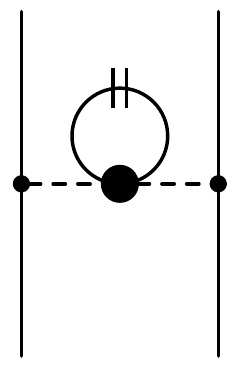} \\ (1)
\end{minipage} \quad
\begin{minipage}{.11\textwidth} \centering
\includegraphics[scale=.5]{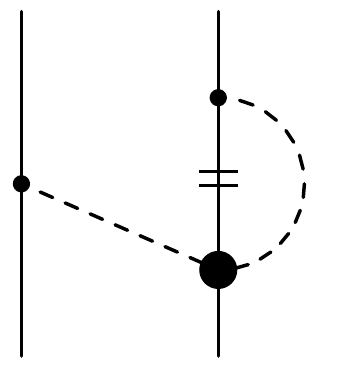} \\ (2a)
\end{minipage} \
\begin{minipage}{.11\textwidth} \centering
\includegraphics[scale=.5]{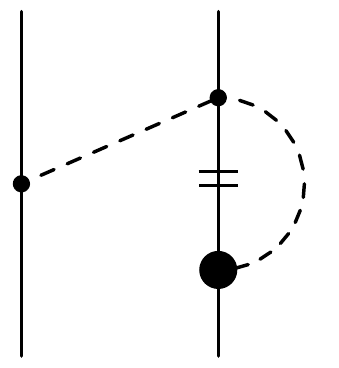} \\ (2b)
\end{minipage} \
\begin{minipage}{.09\textwidth} \centering
\includegraphics[scale=.5]{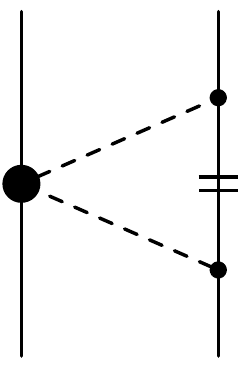} \\ (3)
\end{minipage} \\[.5\baselineskip]
\begin{minipage}{.09\textwidth} \centering
\includegraphics[scale=.5]{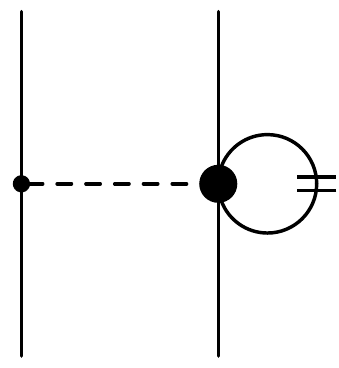} \\ (4)
\end{minipage} \
\begin{minipage}{.09\textwidth} \centering
\includegraphics[scale=.5]{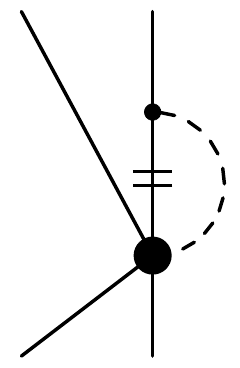} \\ (5a)
\end{minipage} \
\begin{minipage}{.09\textwidth} \centering
\includegraphics[scale=.5]{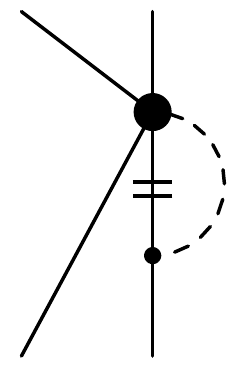} \\ (5b)
\end{minipage} \
\begin{minipage}{.09\textwidth} \centering
\includegraphics[scale=.5]{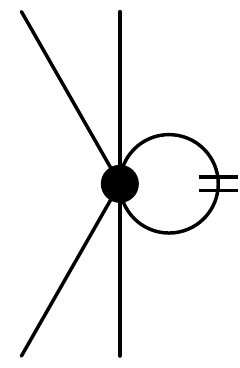} \\ (6)
\end{minipage}
\caption{
Effective two-baryon interaction from genuine three-baryon forces. Contributions arise from
two-pion exchange (1), (2a), (2b), (3), one-pion exchange (4), (5a), (5b) and the contact interaction (6).
\label{fig:med}
}
\end{figure}
In this subsection we summarize how the effect of three-body force in the presence of a (hyper)nuclear medium can be incorporated in an effective baryon-baryon potential.
In \ct{Holt:2009ty} the density-dependent corrections to the \(NN\) interaction have been calculated from the leading chiral three-nucleon forces.
This work has been extended to the strangeness sector in \ct{Petschauer2017a}.
In order to obtain an effective baryon-baryon interaction from the irreducible 3BFs in \fig{fig:3BF},
two baryon lines have been closed, which represents diagrammatically the sum over occupied states within the Fermi sea.
Such a ``medium insertion'' is symbolized by short double lines on a baryon propagator.
All types of diagrams arising this way are shown in \fig{fig:med}.

In \ct{Petschauer2017a} the calculation is restricted to the contact term and to the contributions from one- and two-pion exchange processes, as they are expected to be dominant.
When computing the diagrams of \fig{fig:3BF} the medium insertion corresponds to a factor \(-2\pi\delta(k_0)\theta(k_f-|\vec k|)\).
Furthermore, an additional minus sign comes from a closed fermion loop.
The effective two-body interaction can also be calculated from the expressions for the three-baryon potentials of \ct{Petschauer2016} via the relation
\begin{equation} \label{eq:red}
V_{12} = \sum_B\tr_{\sigma_3}\int_{|\vec k|\leq k_f^B}\frac{\mathrm d^3k}{(2\pi)^3} V_{123} \,,
\end{equation}
where \(\tr_{\sigma_3}\) denotes the spin trace over the third particle and where a summation over all baryon species \(B\) in the Fermi sea (with Fermi momentum \(k_f^B\)) is done.

As an example of such an effective interaction, we display the the effective \(\Lambda N\) interaction in nuclear matter (with \(\rho_p\neq\rho_n\)) derived in \ct{Petschauer2017a}.
It is determined from two-pion-exchange, one-pion-exchange and contact \(\Lambda NN\) three-body forces.
Only the expressions for the \(\Lambda n\) potential are shown as the \(\Lambda p\) potential can be obtained just by interchanging the Fermi momenta \(k_f^p\) with \(k_f^n\) (or the densities \(\rho_p\) with \(\rho_n\)) in the expressions for \(\Lambda n\).
In the following formulas the sum over the contributions from the protons and neutrons in the Fermi sea is already employed.
Furthermore, the values of the LECs are already estimated via decuplet saturation, see \ssect{subsec:BBBDec}.
The topologies (1), (2a) and (2b) vanish here because of the non-existence of an isospin-symmetric \(\Lambda\Lambda\pi\) vertex.
One obtains the density-dependent \(\Lambda n\) potential in a nuclear medium
\begin{align*} \label{eq:medDec2pe}
&V^{\mathrm{med,\pi\pi}}_{\Lambda n} = \frac{C^2g_A^2}{12\pi^2f_0^4\Delta}\Bigg\{ \\
&\quad \frac14\Big[ \frac83 ({k_f^n}^3 + 2 {k_f^p}^3) - 4(q^2+2m^2)\tilde\Gamma_0(p) - 2q^2 \tilde\Gamma_1(p) \\
&\quad \qquad+ (q^2+2m^2)^2\tilde G_0(p,q) \Big]  \\
&\quad + \frac{\mathrm i}2 (\vec q\times\vec p\,)\cdot\vec\sigma_2
\Big(2\tilde\Gamma_0(p)+2\tilde\Gamma_1(p) \\
&\quad \qquad-(q^2+2m^2)(\tilde G_0(p,q)+2\tilde G_1(p,q))\Big) \Bigg\} \,, \numberthis
\end{align*}
\begin{align*} \label{eq:medDec1pe}
V^{\mathrm{med,\pi}}_{\Lambda n} = \frac{g_A C \lC'}{54\pi^2f_0^2\Delta}
\big(2({k_f^n}^3+2{k_f^p}^3)-3m^2\tilde\Gamma_0(p)\big) \,, \numberthis
\end{align*}
\begin{align} \label{eq:medDecCont}
V^{\mathrm{med,ct}}_{\Lambda n} &= \frac{\lC'^2}{18\Delta}(\rho_n +2\rho_p) \,.
\end{align}
The different topologies related to two-pion exchange ((1), (2a), (2b), (3)) and one-pion exchange ((4), (5a), (5b)) have already been combined in $V{^\mathrm{med,\pi\pi}}$ and $V^{\mathrm{med,\pi}}$, respectively.
The density and momentum dependent loop functions \(\tilde\Gamma_i(p)\) and \(\tilde G_i(p,q)\) can be found in \ct{Petschauer2017a}.
The only spin-dependent term is the one proportional to \(\vec\sigma_2 = \frac12(\vec\sigma_1+\vec\sigma_2)-\frac12(\vec\sigma_1-\vec\sigma_2)\) and therefore one recognizes a symmetric and an antisymmetric spin-orbit potential of equal but opposite strength.

\section{Applications} \label{sec:application}
   \subsection{Hyperon-nucleon and hyperon-hyperon scattering} \label{subsec:YN}

With the hyperon-nucleon potentials outlined in \sect{sec:BB} hyperon-nucleon scattering processes can be investigated.
The very successful approach to the nucleon-nucleon interaction of \cts{Epelbaum2004,Epelbaum1999,Epelbaum1999a} within SU(2) \cheft, has been extended to the leading-order baryon-baryon interaction in \cts{Polinder2006,Polinder2007,Haidenbauer2010a} by the Bonn-J\"ulich group.
In \cts{Haidenbauer2013a,Haidenbauer2015,Haidenbauer2015c} this approach has been extended to next-to-leading order in SU(3) chiral effective field theory.
As mentioned in \ssect{subsec:pwr} the chiral power counting is applied to the potential, where only two-particle irreducible diagrams contribute.
These potentials are then inserted into a regularized Lippmann-Schwinger equation to obtain the reaction amplitude (or \(T\)-matrix).
In contrast to the \(NN\) interaction, the Lippmann-Schwinger equation for the \(YN\) interaction involves not only coupled partial waves, but also coupled two-baryon channels.
The coupled-channel Lippmann-Schwinger equation in the particle basis reads after partial-wave decomposition (see also \fig{fig:LSE})
\begin{align*} \label{eq:LSE}
T^{{\rho''\rho'},J}_{{\nu''\nu'}}(k'',k';\sqrt{s}) ={}& V^{{\rho''\rho'},J}_{{\nu''\nu'}}(k'',k') \\
&+ \sum_{{\rho},{\nu}}\int_0^\infty\! \frac{\mathrm dk\,k^2}{(2\pi)^3}\,
V^{{\rho''\rho}\, ,J}_{{\nu''\nu}}(k'',k)  \\
&\quad\times\frac{2\mu_{\nu}}{k_{\nu}^2-k^2+\mathrm i\epsilon}T^{{\rho\rho'},J}_{{\nu\nu'}}(k,k';\sqrt{s})\,, \numberthis
\end{align*}
where \(J\) denotes the conserved total angular momentum.
The coupled two-particle channels (\(\Lambda p, \Sigma^+n,\Sigma^0p,\)\dots) are labeled by \(\nu\), and the partial waves (\({}^1S_0, {}^3P_0,\dots\)) by \(\rho\).
Furthermore, \(\mu_{\nu}\) is the reduced baryon mass in channel \(\nu\).
In \ct{Haidenbauer2013a} a non-relativistic scattering equation has been chosen to ensure that the potential can also be applied consistently to Faddeev and Faddeev-Yakubovsky calculations in the few-body sector, and to (hyper-) nuclear matter calculations within the conventional Brueckner-Hartree-Fock formalism (see \ssect{subsec:YMatter}).
Nevertheless, the relativistic relation between the on-shell momentum \(k_\nu\) and the center-of-mass energy has been used,
\(\sqrt{s}=\sqrt{M^2_{B_{1,\nu}}+k_{\nu}^2}+\sqrt{M^2_{B_{2,\nu}}+k_{\nu}^2}\), in order to get the two-particle thresholds at their correct positions.
The physical baryon masses have been used in the Lippmann-Schwinger equation, which introduces some additional SU(3) symmetry breaking.
Relativistic kinematics has also been used to relate the laboratory momentum \(p_\mathrm{lab}\) of the hyperon to the center-of-mass energy \(\sqrt s\).
The Coulomb interaction has been implemented by the use of the Vincent-Phatak method \cts*{Epelbaum2004,Vincent1974}.
Similar to the nucleonic sector at NLO \ct*{Epelbaum2004}, a regulator function of the form \(f_R(\Lambda) = \exp[-(k'^4+k^4)/\Lambda^4]\) is employed to cut off the high-energy components of the potential.
For higher orders in the chiral power counting, higher powers than 4 in the exponent of \(f_R\) have to be used. 
This ensures that the regulator introduces only contributions, that are beyond the given order.
The cutoff \(\Lambda\) is varied in the range \((500\dots700)\ \mathrm{MeV}\),
\ie comparable to what was used for the \(NN\) interaction in \ct{Epelbaum2004}.
The resulting bands represent the cutoff dependence, after readjusting the contact parameters, and thus could be viewed 
as a lower bound on the theoretical uncertainty.
Recently, improved schemes to estimate the theoretical uncertainty were proposed and applied to the \(NN\) interaction 
\cts*{Furnstahl2014,Epelbaum2015,Furnstahl2015,Binder2016}.
Some illustrative results for \(YN\) scattering based on the method by Epelbaum et al.~\cite{Epelbaum2015,Binder2016}
have been included in Ref.~\cite{Haidenbauer2019b}. 
However, such schemes require higher orders than NLO in the chiral power counting if one wants to address 
questions like the convergence of the expansion.

\begin{figure*}
\centering
\includegraphics[width=0.35\textwidth]{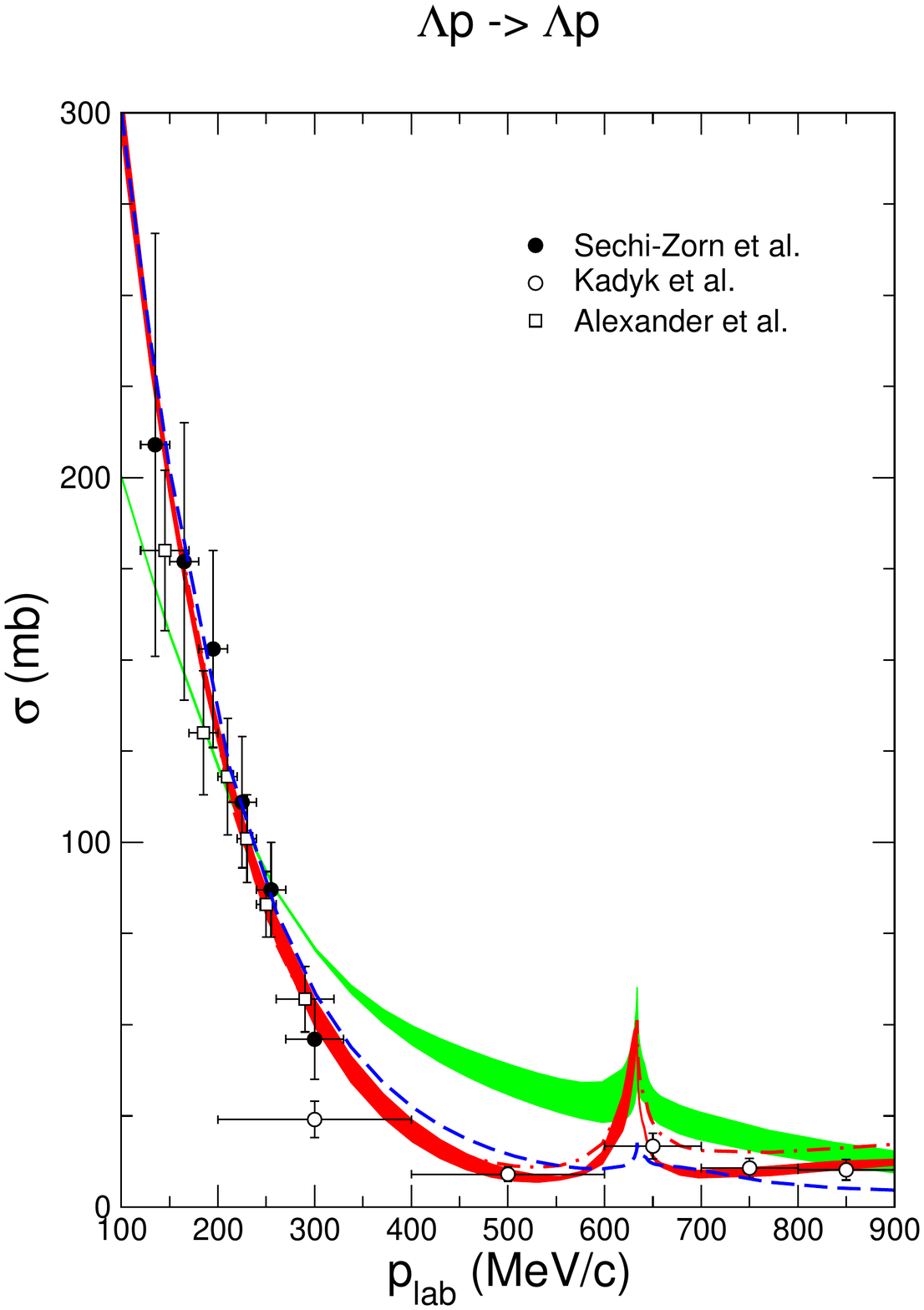} \hspace{-1cm}
\includegraphics[width=0.35\textwidth]{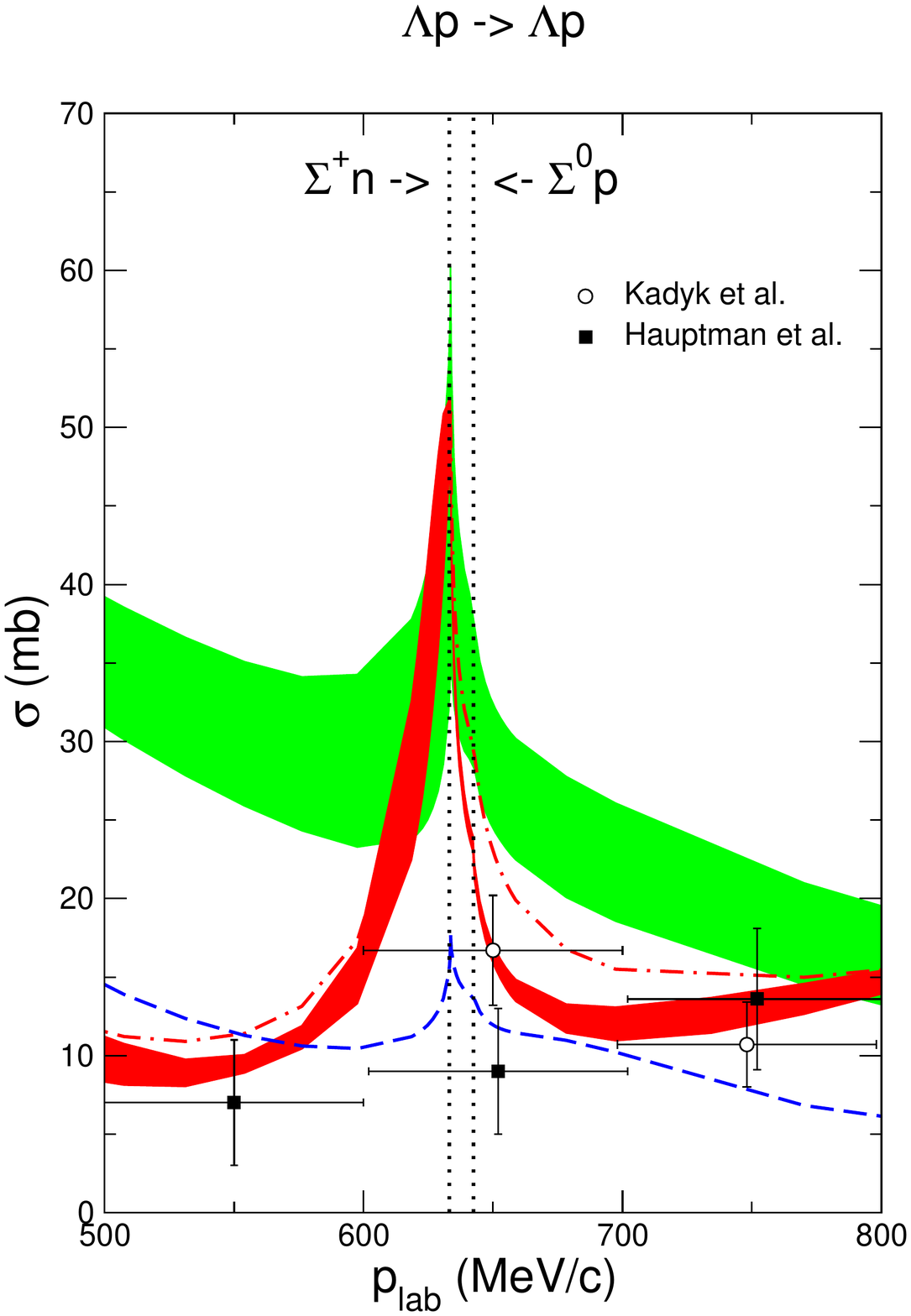} \hspace{-1cm}
\includegraphics[width=0.35\textwidth]{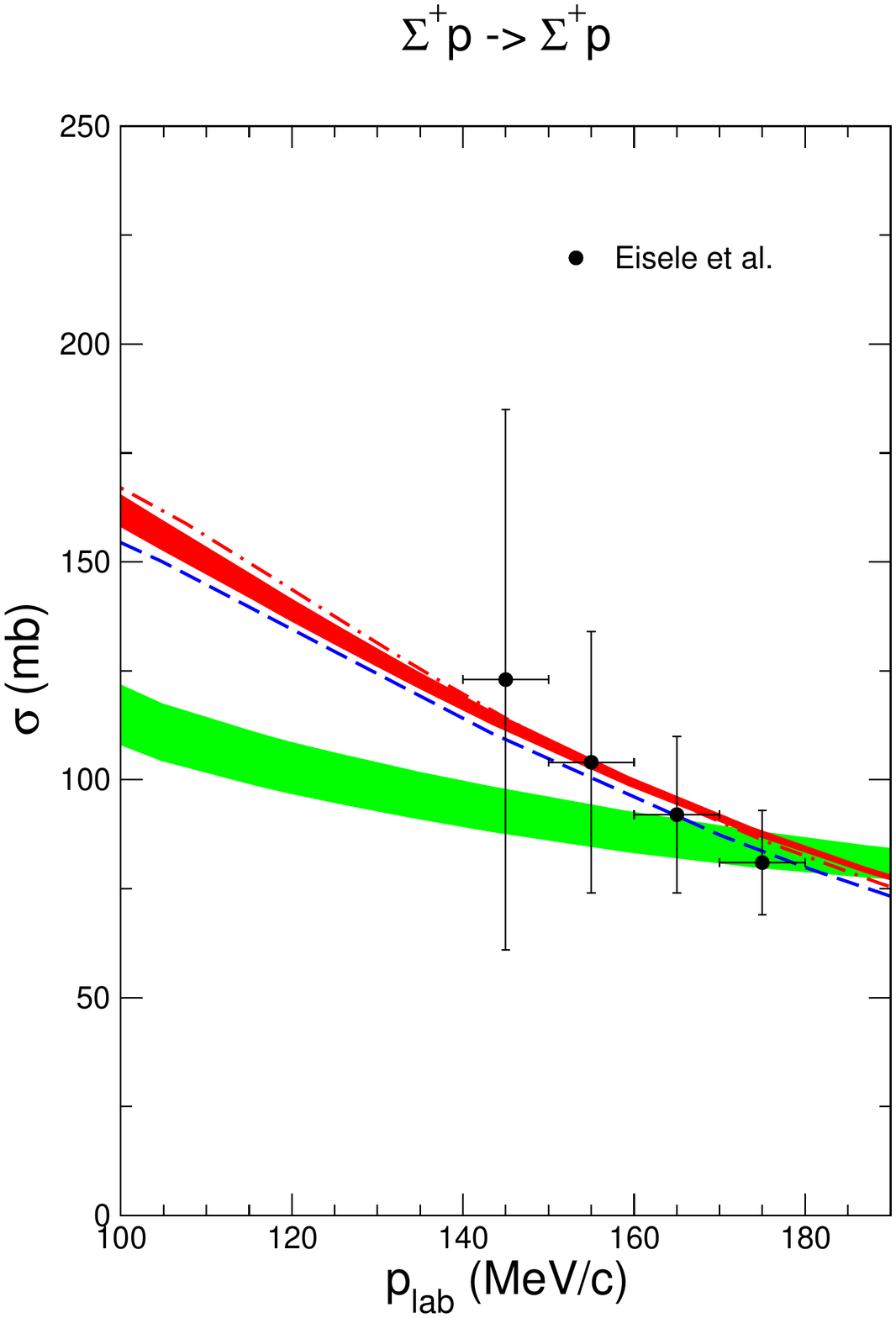}
\vspace{-1.1\baselineskip}
\caption{
Total cross section $\sigma$ as a function of $p_{\mathrm{lab}}$.
The red (dark) band shows the chiral EFT results to NLO for variations of the cutoff in the 
range $\Lambda = (500\ldots650)$~MeV \ct*{Haidenbauer2013a},
while the green (light) band are results to LO for $\Lambda =(550\ldots700)$~MeV \ct*{Polinder2006}.
The dashed curves are the result of the J{\"u}lich '04 meson-exchange potential \ct*{Haidenbauer2005},
the dash-dotted curves of the Nijmegen NSC97f potential \cite{Rijken1998}. 
}
\label{fig:tc}
\end{figure*}

The partial-wave contributions of the meson-exchange diagrams are obtained by employing the partial-wave decomposition formulas of \ct{Polinder2006}.
For further remarks on the employed approximations and the fitting strategy we refer the reader to \ct{Haidenbauer2013a}.
As can be seen in \tab{tab:PWDBB}, one gets for the \(YN\) contact terms five independent LO constants, acting in the \(S\)-waves, eight additional constants at NLO in the \(S\)-waves, and nine NLO constant acting in the \(P\)-waves.
The contact terms represent the unresolved short-distance dynamics, and the corresponding low-energy constants are fitted to the ``standard'' set of 36 \(YN\) empirical data points \cts*{SechiZorn1968,Alexander1968,Engelmann1966,Eisele1971,Hepp1968,Stephen1970}.
The hypertriton (\({}^3_\Lambda \mathrm{H}\)) binding energy has been chosen as a further input.
It determines the relative strength of the spin-singlet and spin-triplet \(S\)-wave contributions of the \(\Lambda p\) interaction.
Due to the sparse and inaccurate experimental data, the obtained fit of the low-energy constants is not unique.
For instance, the \(YN\) data can be described equally well with a repulsive or an attractive interaction in the \({}^3S_1\) partial wave of the \(\Sigma N\) interaction with isospin \(I=3/2\).
However, recent calculations from lattice QCD \ct*{Beane2012,Nemura2018} suggest a repulsive \({}^3S_1\) phase shift in the \(\Sigma N\ I=3/2\) channel, hence the repulsive solution has been adopted.
Furthermore, this is consistent with empirical information from \(\Sigma^-\)-production on nuclei, which point to a repulsive \(\Sigma\)-nucleus potential (see also \ssect{subsec:YMatter}).

In the following we present some of the results of \ct{Haidenbauer2013a}.
For comparison, results of the J\"ulich '04 \ct*{Haidenbauer2005} and the Nijmegen \cite{Rijken1998} meson-exchange models are also shown in the figures.
In \fig{fig:tc} the total cross sections as functions of $p_{\mathrm{lab}}$ for various \(YN\) interactions are presented.
The experimental data is well reproduced at NLO\@.
Especially the results in the \(\Lambda p\) channel are in line with the data points (also at higher energies) and the energy dependence in the \(\Sigma^+p\) channel is significantly improved at NLO\@.
It is also interesting to note that the NLO results are now closer to the phenomenological J\"ulich '04 model than at LO\@.
One expects the theoretical uncertainties to become smaller, when going to higher order in the chiral power counting.
This is reflected in the fact, that the bands at NLO are considerably smaller than at LO\@.
These bands represent only the cutoff dependence and therefore constitute a lower bound on the theoretical error.

\begin{table*}
\setlength\tabcolsep{5pt}
\centering
\centerline{
\begin{tabular}{crrrrrrrrr}
\toprule
& \multicolumn{6}{c}{NLO} & \ LO \ & \ J\"ul '04 \ & \ NSC97f \ \\
\midrule
${\Lambda}$ [MeV] & 450 & 500 & 550 & 600 & 650 & 700 & 600 & & \\
\midrule
$a^{\Lambda p}_s$ & $-2.90$ & $-2.91$ & $-2.91$ & $-2.91$ & $-2.90$ & $-2.90$ & $-1.91$ & $-2.56$ & $-2.60$ \\
$r^{\Lambda p}_s$ & $ 2.64$ & $ 2.86$ & $ 2.84$ & $ 2.78$ & $ 2.65$ & $ 2.56$ & $ 1.40$ & $ 2.74$ & $ 3.05$ \\
$a^{\Lambda p}_t$ & $-1.70$ & $-1.61$ & $-1.52$ & $-1.54$ & $-1.51$ & $-1.48$ & $-1.23$ & $-1.67$ & $-1.72$ \\
$r^{\Lambda p}_t$ & $ 3.44$ & $ 3.05$ & $ 2.83$ & $ 2.72$ & $ 2.64$ & $ 2.62$ & $ 2.13$ & $ 2.93$ & $ 3.32$ \\
\midrule
$a^{\Sigma^+ p}_s$ & $-3.58$ & $-3.59$ & $-3.60$ & $-3.56$ & $-3.46$ & $-3.49$ & $-2.32$ & $-3.60$ & $-4.35$ \\
$r^{\Sigma^+ p}_s$ & $ 3.49$ & $ 3.59$ & $ 3.56$ & $ 3.54$ & $ 3.53$ & $ 3.45$ & $ 3.60$ & $ 3.24$ & $ 3.16$ \\
$a^{\Sigma^+ p}_t$ & $ 0.48$ & $ 0.49$ & $ 0.49$ & $ 0.49$ & $ 0.48$ & $ 0.49$ & $ 0.65$ & $ 0.31$ & $-0.25$ \\
$r^{\Sigma^+ p}_t$ & $-4.98$ & $-5.18$ & $-5.03$ & $-5.08$ & $-5.41$ & $-5.18$ & $-2.78$ & $-12.2$ & $-28.9$ \\
\midrule
$(^3_\Lambda \textrm H)$ $E_B$ & $-2.39$ & $-2.33$ & $-2.30$ & $-2.30$ & $-2.30$ & $-2.32$ &$-2.34$ & $-2.27$ & $-2.30$ \\
\bottomrule
\end{tabular}
}
\caption{
The $YN$ singlet ($s$) and triplet ($t$) scattering length \(a\) and effective range \(r\) (in fm) 
and the hypertriton binding energy $E_B$ (in MeV) \ct*{Haidenbauer2013a}.
The binding energies for the hypertriton are calculated using
the Idaho-N3LO $NN$ potential \ct*{Entem2003}.
The experimental value for the $^3_\Lambda \textrm H$ binding energy is -2.354(50) MeV.
\label{tab:sl}
}
\setlength\tabcolsep{6pt}
\end{table*}

In \tab{tab:sl} the scattering lengths and effective range parameters for the \(\Lambda p\) and \(\Sigma^+p\) interactions in the \({}^1S_0\) and \({}^3S_1\) partial waves are given.
Result for LO \ct*{Polinder2006} and NLO \cheft \ct*{Haidenbauer2013a}, for the J\"ulich '04 model \ct*{Haidenbauer2005} and for the Nijmegen NSC97f potential \ct*{Rijken1998} are shown.
The NLO \(\Lambda p\) scattering lengths are larger than for the LO calculation, and closer to the values obtained by the meson-exchange models.
The triplet \(\Sigma^+p\) scattering length is positive in the LO as well as the NLO calculation, which indicates a repulsive interaction in this channel.
Also given in \tab{tab:sl} is the hypertriton binding energy, calculated with the corresponding chiral potentials.
As stated before, the hypertriton binding energy was part of the fitting procedure and values close to the experimental value could be achieved.
The predictions for the \({}^3_\Lambda H\) binding energy are based on the Faddeev equations in momentum space, as described in \cts{Nogga2014a,Nogga2012}.
Note that genuine (irreducible) three-baryon interactions were not included in this calculation.
However, in the employed coupled-channel formalism, effects like the important \(\Lambda\)-\(\Sigma\) conversion process are naturally included.
It is important to distinguish such iterated two-body interactions, from irreducible three-baryon forces, as exemplified in \fig{fig:irr3BF}.

\begin{figure*}
    \centering
    \includegraphics[width=0.35\textwidth]{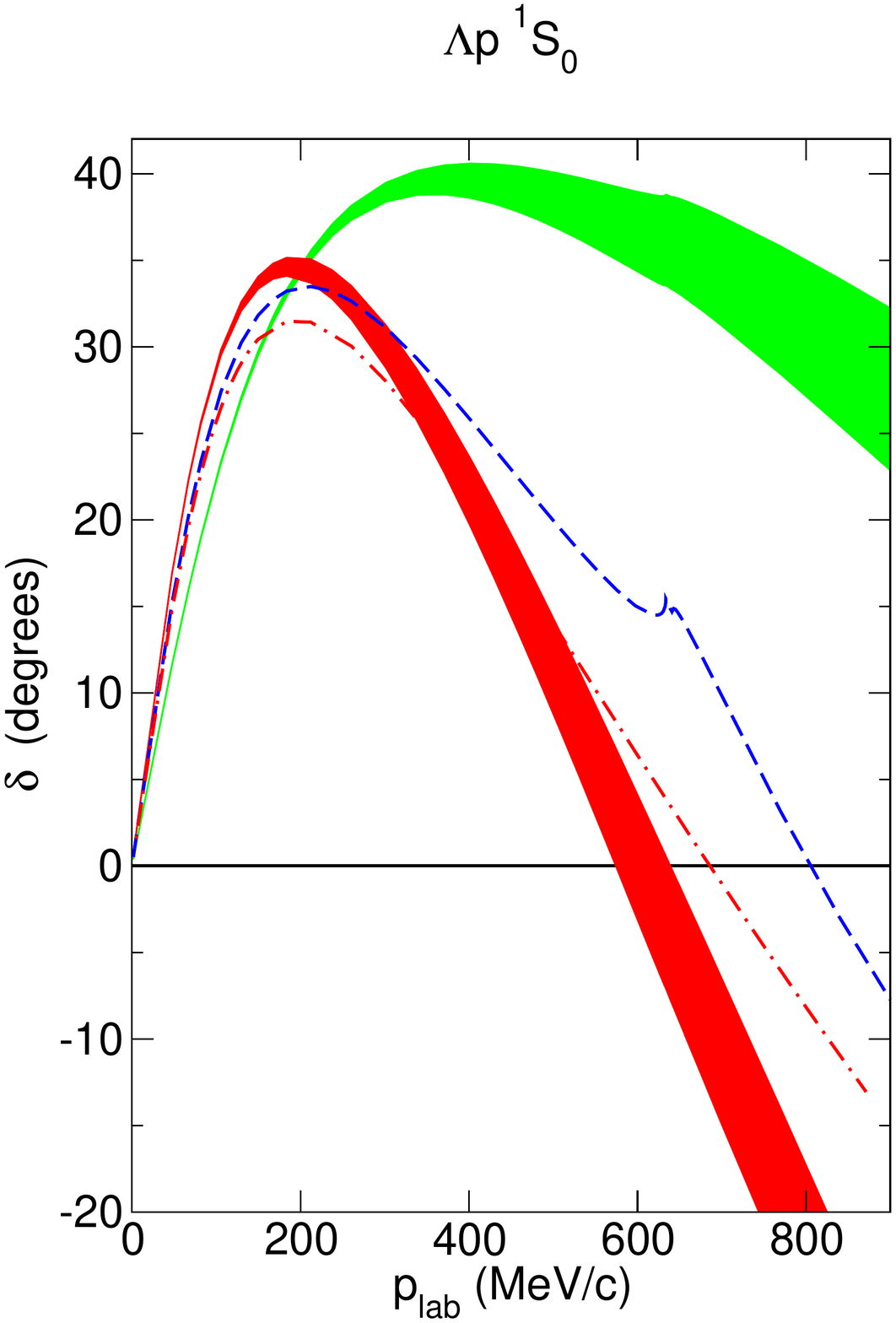} \hspace{-1cm}
    \includegraphics[width=0.35\textwidth]{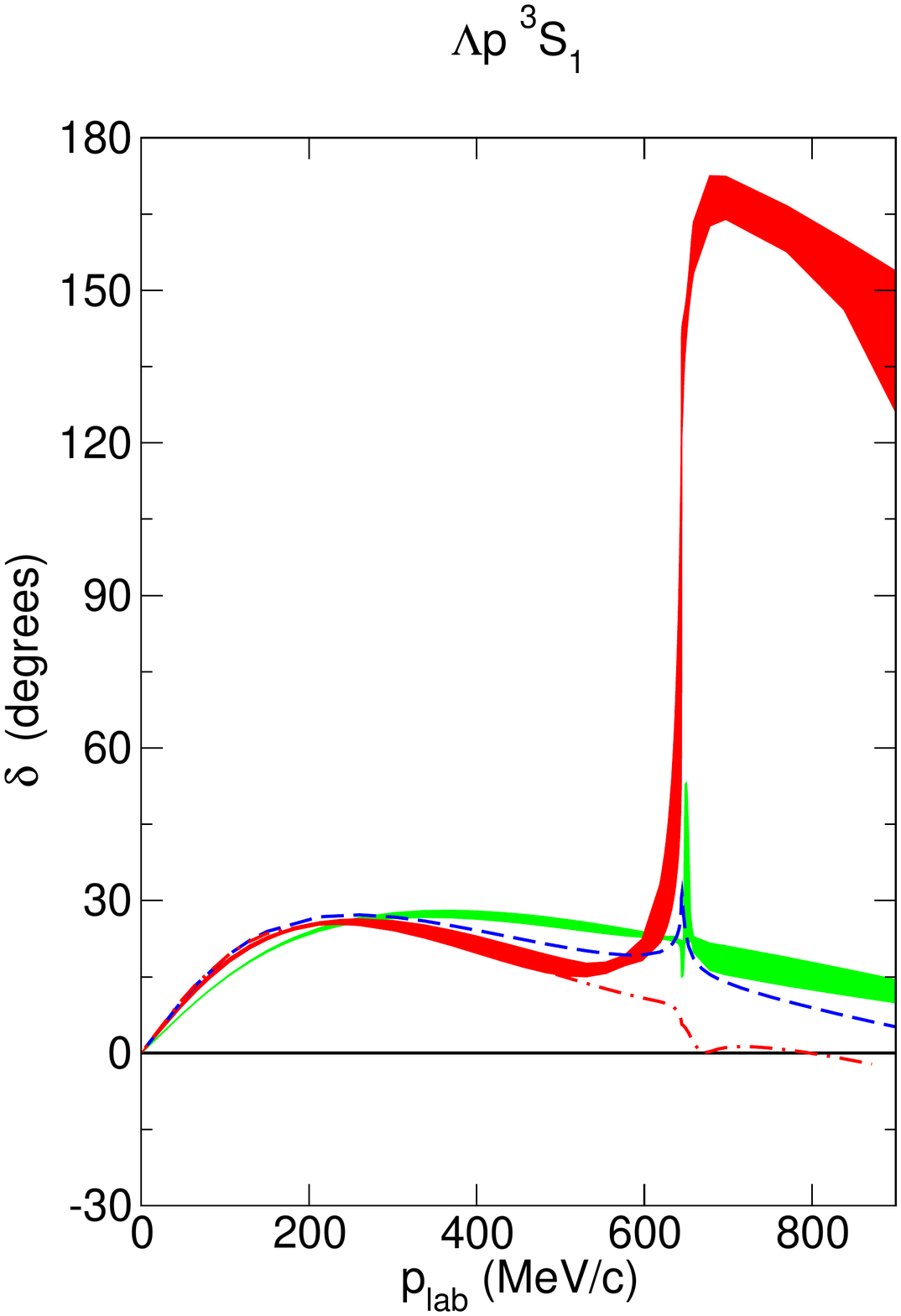} \hspace{-1cm}
    \includegraphics[width=0.35\textwidth]{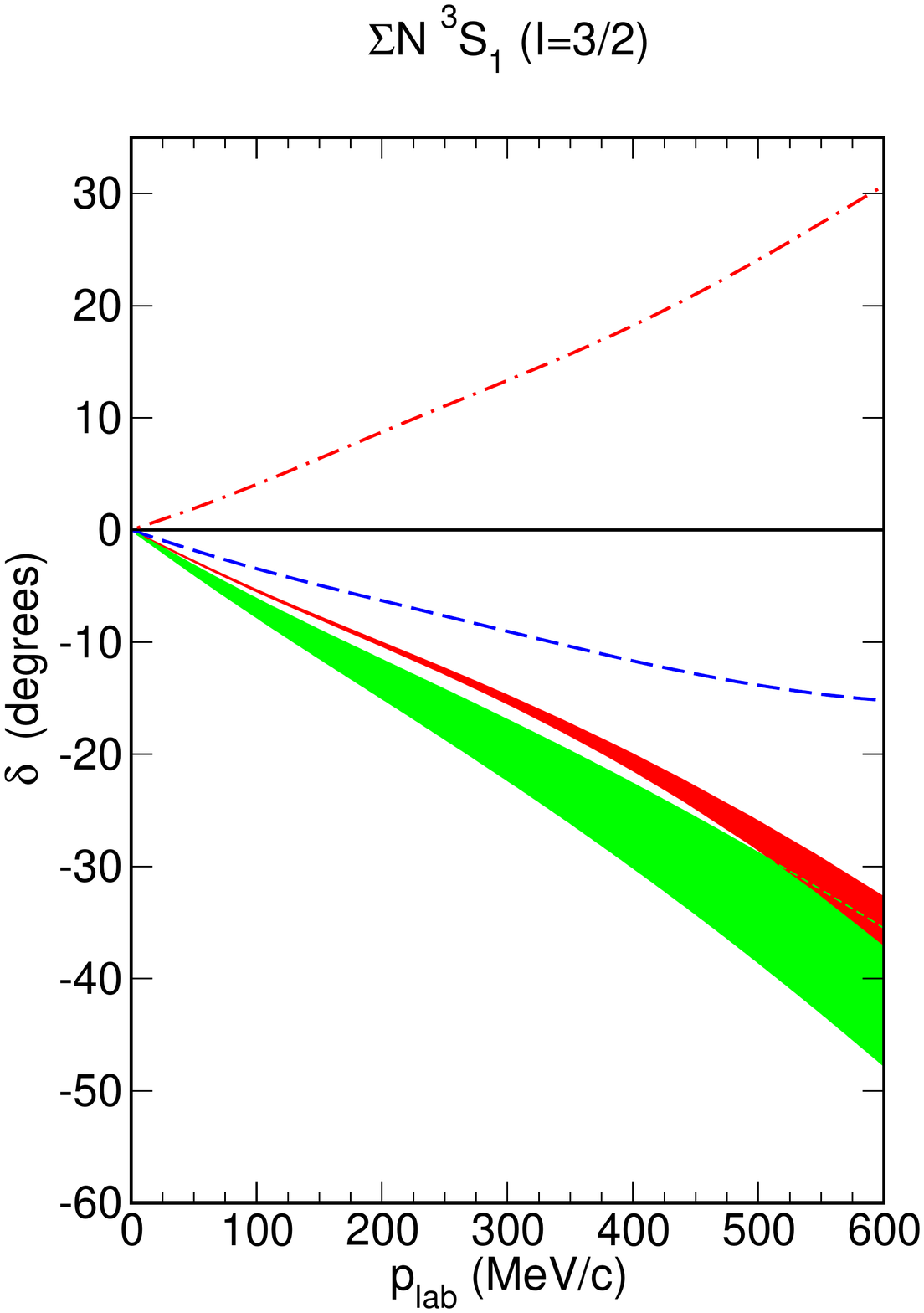}
    \vspace{-1.1\baselineskip}
    \caption{
    Various \(S\)- and \(P\)-wave phase shifts $\delta$ as a function of $p_{\mathrm{lab}}$ for the $\Lambda p$ and $\Sigma^+ p$ interaction \ct*{Haidenbauer2013a}.
    Same description of curves as in \fig{fig:tc}.
    }
    \label{fig:ph}
\end{figure*}

Predictions for \(S\)- and \(P\)-wave phase shifts $\delta$ as a function of $p_{\mathrm{lab}}$ for $\Lambda p$ and $\Sigma^+ p$ scattering are shown in \fig{fig:ph}.
The \({}^1S_0\) \(\Lambda p\) phase shift from the NLO \cheft calculation is closer to the phenomenological J\"ulich '04 model than the LO result.
It points to moderate attraction at low momenta and strong repulsion at higher momenta.
At NLO the phase shift has a stronger downward bending at higher momenta compared to LO or the J\"ulich '04 model.
As stated before, more repulsion at higher energies is a welcome feature in view of neutron star matter with \(\Lambda\)-hyperons as additional baryonic degree of freedom.
The \({}^3S_1\) \(\Lambda p\) phase shift, part of the \(S\)-matrix for the coupled \({}^3S_1\)-\({}^3D_1\) system, changes qualitatively from LO to NLO\@.
The \({}^3S_1\) phase shift of the NLO interaction passes through \(90^\circ\) slightly below the \(\Sigma N\) threshold, which indicates the presence of an unstable bound state in the \(\Sigma N\) system.
For the LO interaction and the J\"ulich '04 model no passing through \(90^\circ\) occurs and a cusp is predicted, that is caused by an inelastic virtual state in the \(\Sigma N\) system.
These effects are also reflected by a strong increase of the \(\Lambda p\) cross section close to the \(\Sigma N\) threshold, see \fig{fig:tc}.
The \({}^3S_1\) \(\Sigma N\) phase shift for the NLO interaction is moderately repulsive and comparable to the LO phase shift.

Recently an alternative NLO $\chi$EFT potential for $YN$ scattering has been presented \cite{Haidenbauer2019b}.
In that work a different strategy for fixing the low-energy constants that determine the strength of the contact 
interactions is adopted.
The objective of that exploration was to reduce the number of LECs that need to be
fixed in a fit to the $\Lambda N$ and $\Sigma N$ data by inferring some of them from the $NN$ sector via the
underlying SU(3) symmetry, cf. Sect.~\ref{subsec:potct}.
Indeed, correlations between the LO and NLO LECs of the $S$-waves, \ie between the $\tilde c$'s and $c$'s,
had been observed already in the initial $YN$ study \cite{Haidenbauer2013a} and
indicated that a unique determination of them by considering the existing $\Lambda N$ and $\Sigma N$ data 
alone is not possible.
It may be not unexpected in view of those correlations,
that the variant considered in \cite{Haidenbauer2019b} yields practically equivalent results for $\Lambda N$
and $\Sigma N$ scattering observables.
However, it differs considerable in the strength of the $\Lambda N \to \Sigma N$ transition potential and that becomes manifest in applications to few- and many-body systems \cite{Haidenbauer2017,Haidenbauer2019b}.
 
\begin{figure*}
\centering
\includegraphics[height=55mm]{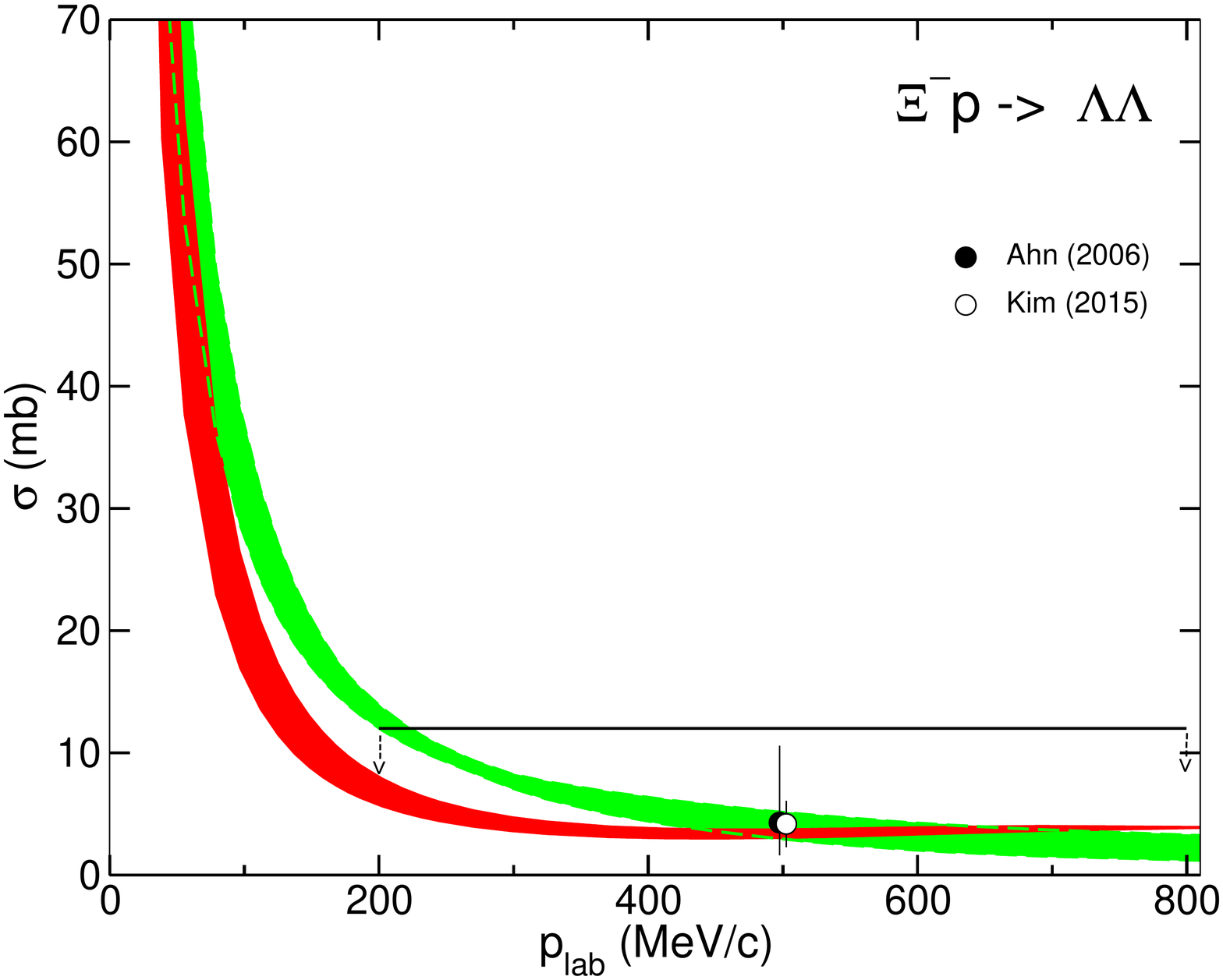}\includegraphics[height=55mm]{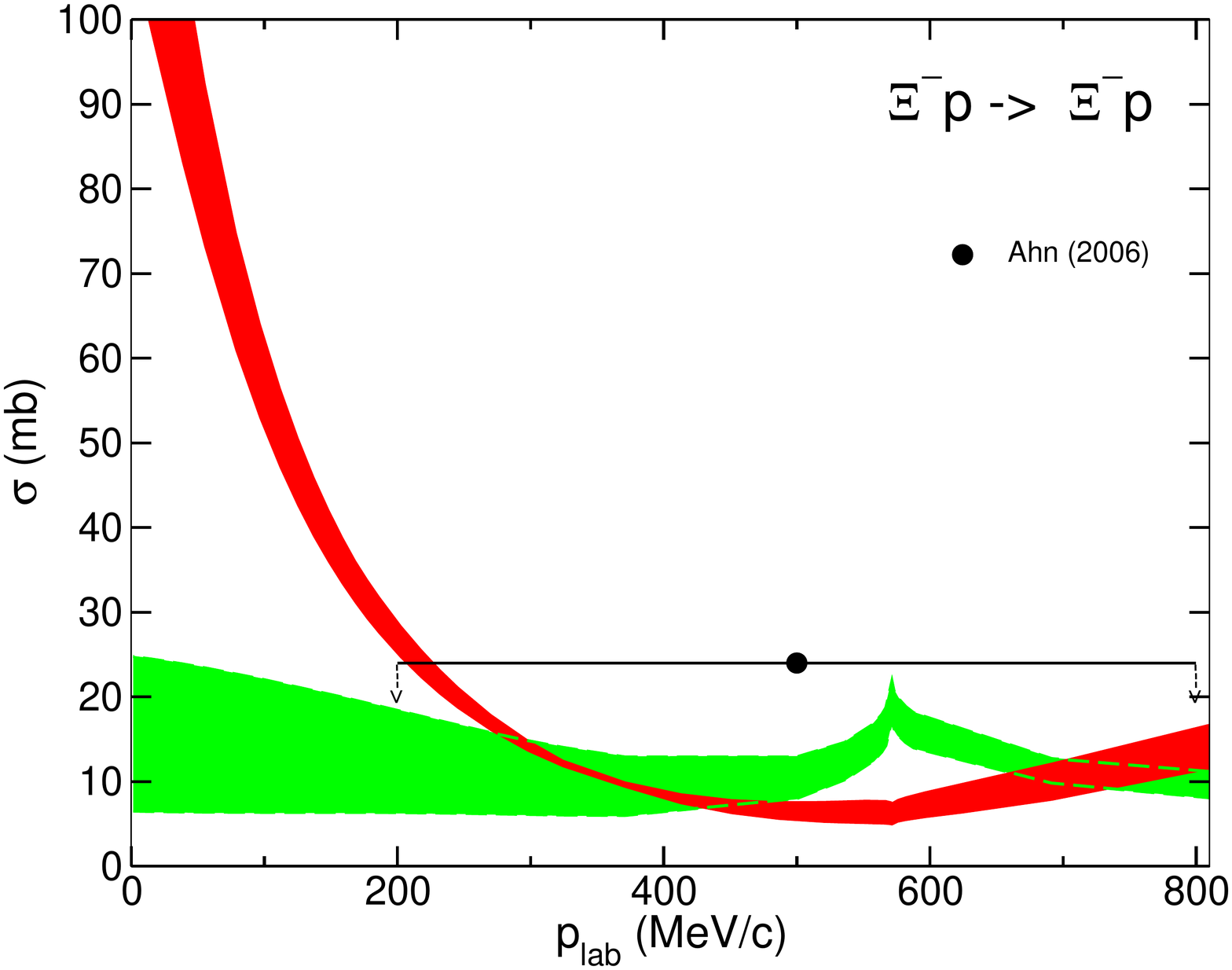}

\includegraphics[height=55mm]{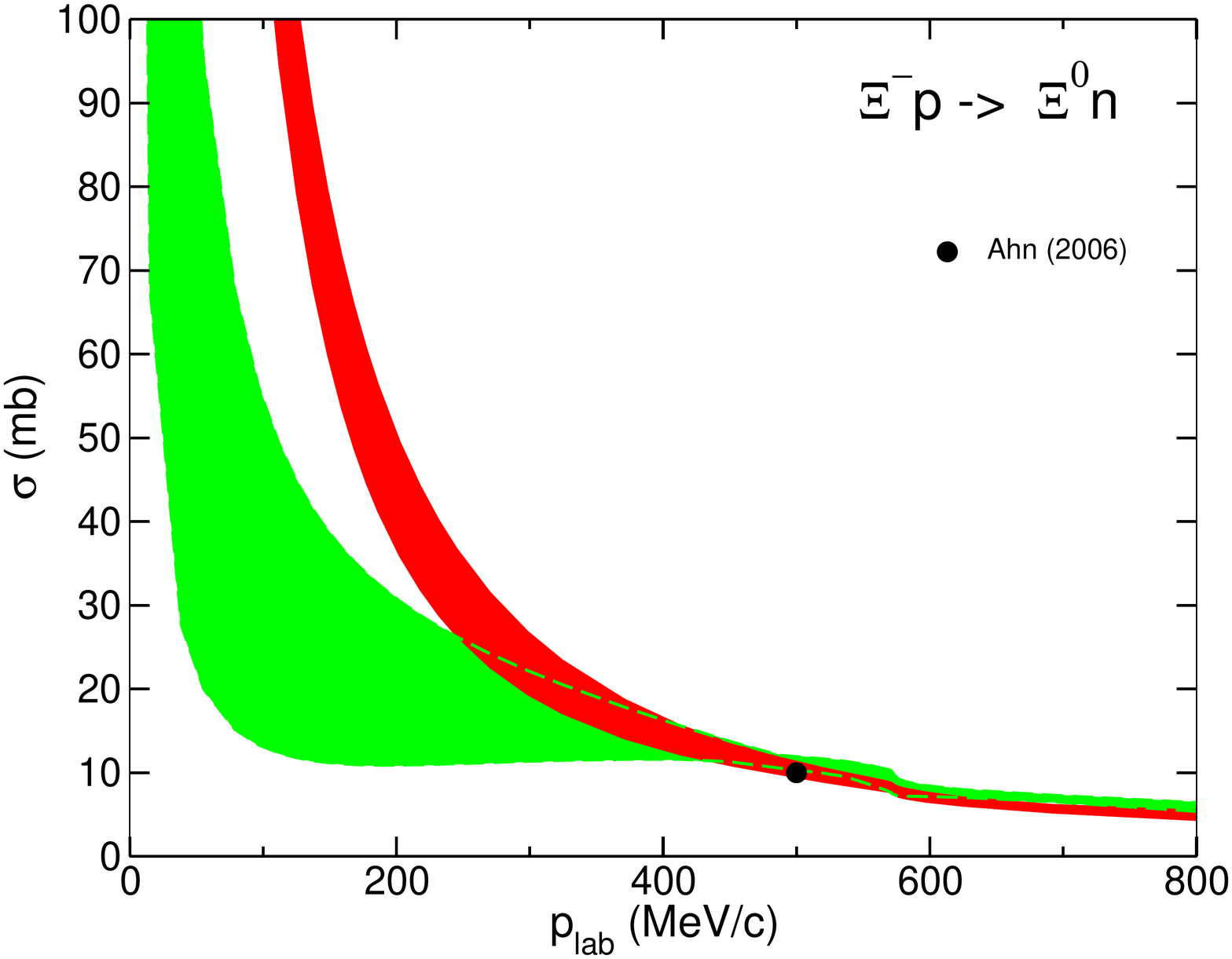}\includegraphics[height=55mm]{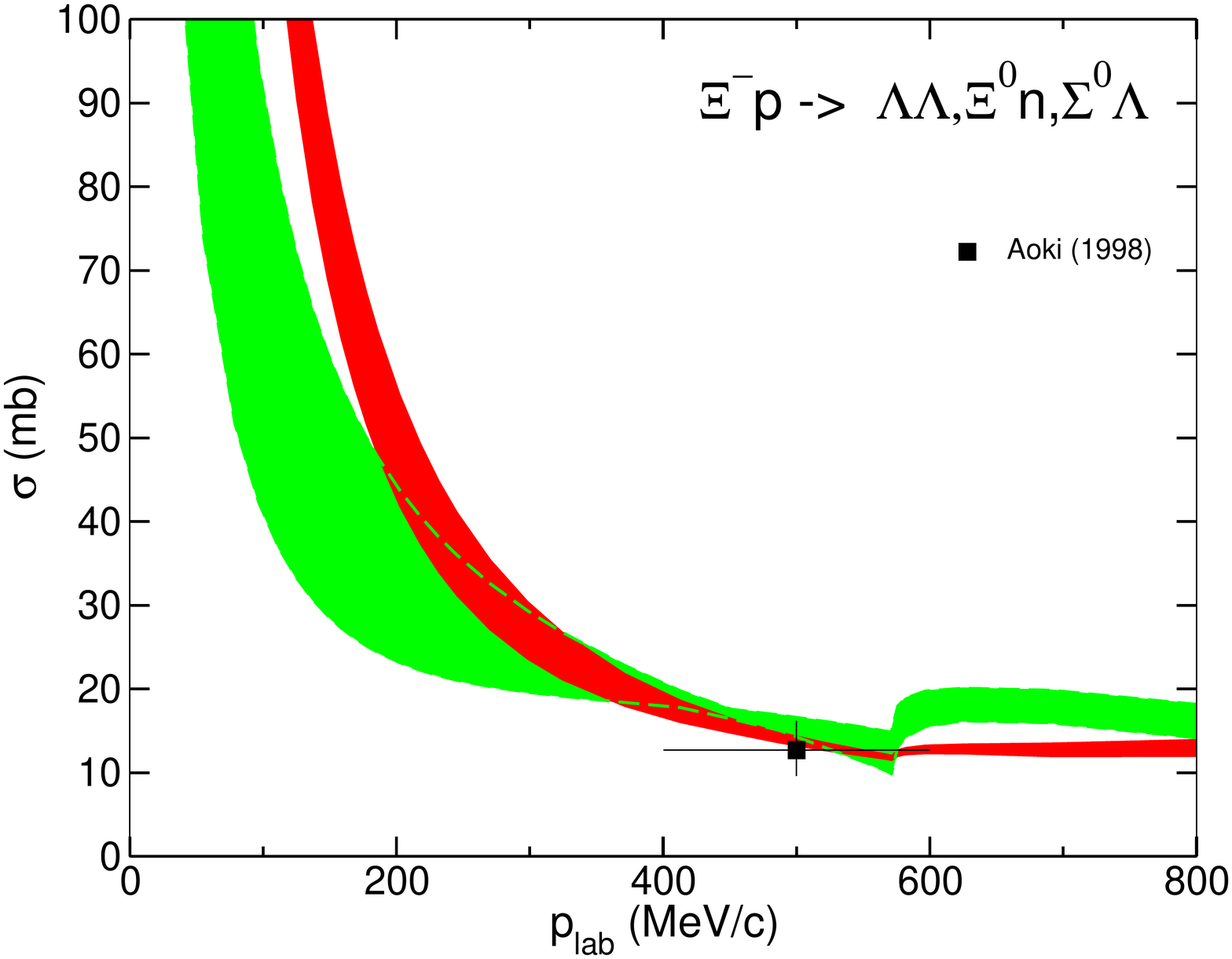}
\caption{
$\Xi^- p$ induced cross sections.
The bands represent results at NLO (red/black) \cite{Haidenbauer2015c} 
and LO (green/grey) \cite{Polinder2007} .
Experiments are from Ahn et al.~\cite{Ahn2005a}, Kim et al.~\cite{Kim2015},
and Aoki et al. \cite{Aoki1998}.
Upper limits are indicated by arrows.
}
\label{fig:t2c}
\end{figure*}

There is very little empirical information about baryon-baryon systems
with $S=-2$, \ie about the interaction in the $\Lambda\Lambda$, $\Sigma\Sigma$,
$\Lambda\Sigma$, and $\Xi N$ channels. Actually, all one can find in the 
literature \cite{Haidenbauer2015c} are a few values and upper bounds for the 
$\Xi^- p$ elastic and inelastic cross sections \cite{Aoki1998,Ahn2005a}.
In addition there are constraints on the strength of the $\Lambda\Lambda$ interaction from 
the separation energy of the ${}_{\Lambda\Lambda}^{\;\;\;6}{\rm He}$ hypernucleus \cite{Takahashi2001}.
Furthermore estimates for the $\Lambda\Lambda$ $^1S_0$ scattering length exist 
from analyses of the $\Lambda\Lambda$ invariant mass measured in the reaction 
$^{12}C(K^-,K^+\Lambda\Lambda X)$ \cite{Gasparyan2011}
and of $\Lambda\Lambda$ correlations measured in relativistic heavy-ion collisions \cite{Ohnishi2016}.

Despite the rather poor experimental situation, 
it turned out that SU(3)-symmetry breaking contact terms that
arise at NLO, see Sect.~\ref{subsec:potct}, need to be taken into account
when going from strangeness $S=-1$ to $S=-2$ in order to achieve agreement
with the available measurements and upper bounds for the $\Lambda \Lambda$ 
and $\Xi N$ cross sections \cite{Haidenbauer2015c}. 
This concerns, in particular, the LEC $c^1_\chi$ that appears in the $^1S_0$ 
partial wave, cf. Eq.~(\ref{pot:su3b}). Actually, its value can be fixed by 
considering the $pp$ and $\Sigma^+ p$ systems, as shown in Ref.~\cite{Haidenbauer2015}, 
and then employed in the $\Lambda\Lambda$ system.

Selected results for the strangeness $S=-2$ sector are presented in Fig.~\ref{fig:t2c}.
Further results and a detailed description of the interactions can be found
in Refs.~\cite{Polinder2007,Haidenbauer2015,Haidenbauer2015c,Haidenbauer2019}.
Interestingly, the results based on the LO interaction from Ref.~\cite{Polinder2007}
(green/grey bands) are consistent with all empirical constraints. The cross sections at LO
are basically genuine predictions that follow from SU(3) symmetry utilizing LECs fixed 
from a fit to the $\Lambda N$ and $\Sigma N$ data on the LO level.
The ${\Lambda\Lambda}$ $^1S_0$ scattering length predicted by the NLO interaction
is $a_{\Lambda\Lambda} = -0.70\cdot\cdot\cdot -0.62$ fm \cite{Haidenbauer2015c}. 
These values are well within the range found in the aforementioned analyses which 
are
$a_{\Lambda\Lambda} = (-1.2 \pm 0.6)$ fm \cite{Gasparyan2011} and
${-1.92} < a_{\Lambda\Lambda} < {-0.50}$ fm \cite{Ohnishi2016}, respectively. 
The values for the $\Xi^0p$ and $\Xi^0n$ $S$-wave scattering length are likewise
small and typically in the order of $\pm 0.3 \sim \pm 0.6$ fm \cite{Haidenbauer2015c} 
and indicate that the $\Xi N$ interaction has to be relatively weak in order to be in
accordance with the available empirical constraints. Indeed, the present results obtained
in chiral EFT up to NLO imply that the published values and upper bounds for the $\Xi^- p$
elastic and inelastic cross sections \cite{Aoki1998,Ahn2005a} practically rule out
a somewhat stronger attractive $\Xi N$ force.

Also for $\Xi N$ scattering an alternative NLO $\chi$EFT potential has been presented 
recently \cite{Haidenbauer2019}. 
Here the aim is to explore the possibility to establish a $\Xi N$ interaction
that is still in line with all the experimental constraints for $\Lambda \Lambda$ 
and $\Xi N$ scattering, but at the same time is somewhat more attractive. 
Recent experimental evidence for the existence of $\Xi$-hypernuclei \cite{Nakazawa:2015}
suggests that the in-medium interaction of the $\Xi$-hyperon should be moderately 
attractive \cite{Gal2016}.

   \subsection{Hyperons in nuclear matter}  \label{subsec:YMatter}

\renewcommand{\figscale}{0.65}

Experimental investigations of nuclear many-body systems including strange baryons, for instance, the spectroscopy of hypernuclei, provide important constraints on the underlying hyperon-nucleon interaction.
The analysis of data for single \(\Lambda\)-hypernuclei over a wide range in mass number leads to the result, that the attractive \(\Lambda\) single-particle potential is about half as deep (\(\apr -28\ \mathrm{MeV}\)) as the one for nucleons \ct*{Millener1988,Yamamoto1988}.
At the same time the \(\Lambda\)-nuclear spin-orbit interaction is found to be exceptionally weak \ct*{Ajimura2001a,Akikawa2002}.
Recently, the repulsive nature of the \(\Sigma\)-nuclear potential has been experimentally established in \(\Sigma^-\)-formation reactions on heavy nuclei \ct*{Friedman2007}.
Baryon-baryon potentials derived within \cheft as presented in \sect{sec:BB} are consistent with these observations \cts*{Haidenbauer2015a,Petschauer2015}.
In this section we summarize results of hyperons in infinite homogeneous nuclear matter of \ct{Petschauer2015} obtained by employing the interaction potentials from \cheft as microscopic input.
The many-body problem is solved within first-order Brueckner theory.
A detailed introduction can be found, \eg in \cts{Day1967,Baldo1999a,Fetter2003}.

Brueckner theory is founded on the so-called \emph{Goldstone expansion}, a linked-cluster perturbation series for the ground state energy of a fermionic many-body system.
Let us consider a system of \(A\) identical fermions, described by the Hamiltonian
\begin{equation}
H = T + V \,,
\end{equation}
where \(T\) is the kinetic part and \(V\) corresponds to the two-body interaction.
The goal is to calculate the ground state energy of this interacting \(A\)-body system.
It is advantageous to introduced a so-called \emph{auxiliary potential}, or single-particle potential, \(U\).
The Hamiltonian is then split into two parts
\begin{equation} \label{eq:introU}
H
= ( T + U ) + ( V - U )
= H_0 + H_1\,,
\end{equation}
the unperturbed part \(H_0\) and the perturbed part \(H_1\).
One expects the perturbed part to be small, if the single particle potential describes well the averaged effect of the medium on the particle.
In fact, the proper introduction of the auxiliary potential is crucial for the convergence of Brueckner theory.

\begin{figure*}
\centering
\hfill
\begin{subfigure}[b]{.65\textwidth}
 \centering
 \(
 \vcenter{\hbox{\includegraphics[scale=.55]{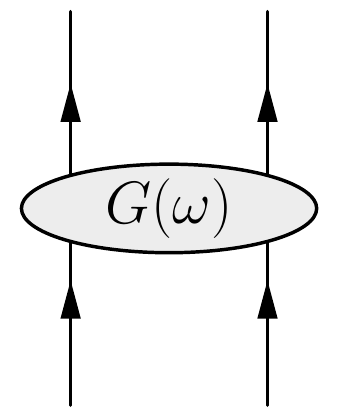}}}
 \ =\
 \vcenter{\hbox{\includegraphics[scale=.55]{files/Feynman/blob/lse2}}}
 \ +\
 \vcenter{\hbox{\includegraphics[scale=.55]{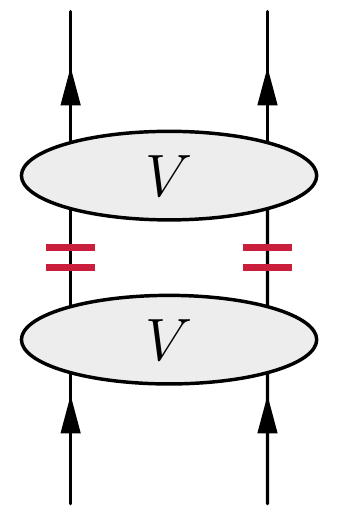}}}
 \ +\
 \vcenter{\hbox{\includegraphics[scale=.55]{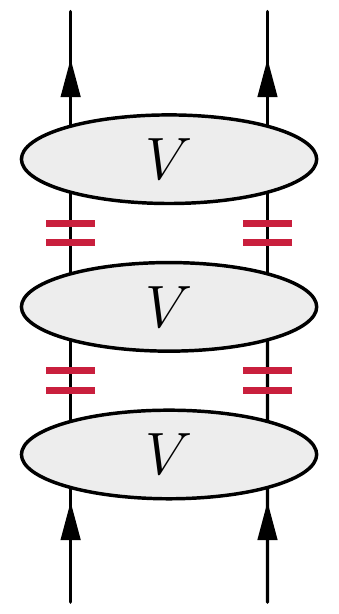}}}
 \ +
 \) \
 {\large \(\cdots\)}
 \caption{Bethe-Goldstone equation}
 \label{fig:BGE}
\end{subfigure}
\hfill
\begin{subfigure}[b]{.3\textwidth}
 \centering
 \large
 \(
 U\ = \vcenter{\hbox{\includegraphics[scale=.8]{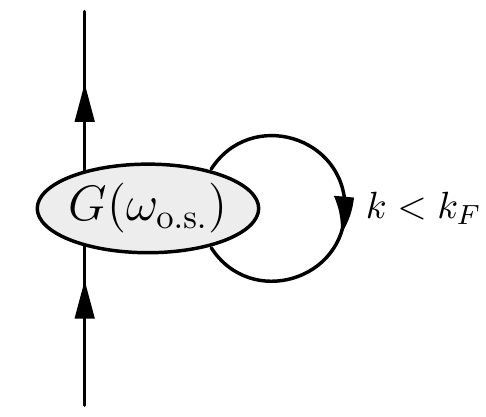}}}
 \)
 \caption{Single particle potential}
 \label{fig:UGmat}
\end{subfigure}
\hfill\mbox{}
\caption{Graphical representation of the determination of the single-particle potential from the G-matrix interaction \subref{fig:UGmat}
and of the Bethe-Goldstone equation \subref{fig:BGE}.
The symbol \(\omega_\mathrm{o.s}\) denotes the on-shell starting energy.}
\end{figure*}

Conventional nucleon-nucleon potentials exhibit a strong short-range repulsion that leads to very large matrix elements.
Hence, the Goldstone expansion in the form described above will not converge for such hard-core potentials.
One way to approach this problem is the introduction of the so-called \emph{Brueckner reaction matrix}, or \(G\)-matrix.
The idea behind it is illustrated in \fig{fig:BGE}.
Instead of only using the simple interaction, an infinite number of diagrams with increasing number of interactions is summed up.
This defines the \(G\)-matrix interaction, which is, in contrast to the bare potential, weak and of reasonable range.
In a mathematical way, the reaction matrix is defined by the \emph{Bethe-Goldstone equation}:%
\begin{equation} \label{eq:BGEop}
G(\omega) = V + V \frac Q{\omega - H_0 + \mathrm i\epsilon} G(\omega)\,,
\end{equation}
with the so-called starting energy \(\omega\).
The Pauli operator \(Q\) ensures, that the intermediate states are from outside the Fermi sea.
As shown in \fig{fig:BGE} this equation represents a resummation of the ladder diagrams to all orders.
The arising \(G\)-matrix interaction is an effective interaction of two particles in the presence of the medium.
The medium effects come in solely through the Pauli operator and the energy denominator via the single-particle potentials.
If we set the single-particle potentials to zero and omit the Pauli operator (\(Q=1\)), we recover the usual Lippmann-Schwinger equation for two-body scattering in vacuum, see also \fig{fig:LSE}.
This medium effect on the intermediate states is denoted by horizontal double lines in \fig{fig:BGE}.
An appropriate expansion using the \(G\)-matrix interaction instead of the bare potential is the so-called Brueckner-Bethe-Goldstone expansion, or \emph{hole-line expansion}.

Finally, the form of the auxiliary potential \(U\) needs to be chosen.
This choice is important for the convergence of the hole-line expansion.
Bethe, Brandow and Petschek \ct*{Bethe1963} showed for nuclear matter that important higher-order diagrams cancel each other if the auxiliary potential is taken as
\begin{equation} \label{eq:Udef}
U_m = \Re \sum_{n\leq A} \langle mn\vert G(\omega=\omega_\text{o.s.}) \vert mn\rangle_\mathcal A \,,
\end{equation}
where the Brueckner reaction matrix is evaluated \emph{on-shell}, \ie the starting energy is equal to the energy of the two particles \(m\), \(n\) in the initial state:
\begin{align*} \label{eq:omos}
 \omega_\text{o.s.} &= E_{1}(k_{1})+E_{2}(k_{2})\,, \\
 E_{B_i}(k_i) &= M_i + \frac{k_i^2}{2M_i} + \Re U_{i}(k_i) \,.
 \numberthis
\end{align*}
Pictorially \eq{eq:Udef} means, that the single-particle potential can be obtained by taking the on-shell \(G\)-matrix interaction and by closing one of the baryon lines, as illustrated in \fig{fig:UGmat}.
Note that this implies a non-trivial self-consistency problem.
On the one hand, \(U\) is calculated from the \(G\)-matrix elements via \eq{eq:Udef}, and on the other hand the starting energy of the \(G\)-matrix elements depends on \(U\) through the single-particle energies \(E_i\) in \eq{eq:omos}.

At the (leading) level of two hole-lines, called \emph{Brueckner-Hartree-Fock approximation} (BHF), the total energy is given by
\begin{align*} \label{eq:E-BHF}
E &= \sum_{n\leq A} \langle n\vert T\vert n\rangle
+\frac12 \sum_{m,n\leq A} \langle mn\vert G\vert mn\rangle_\mathcal A \\
&= \sum_{n\leq A} \langle n\vert T\vert n\rangle
+\frac12 \sum_{n\leq A} \langle n\vert U\vert n\rangle \,, \numberthis
\end{align*}
\ie the ground-state energy \(E\) can be calculated directly after the single-particle potential has been determined.

The definition of \(U\) in \eq{eq:Udef} applies only to occupied states within the Fermi sea.
For intermediate-state energies above the Fermi sea, typically two choices for the single-particle potential are employed.
In the so-called \emph{gap choice}, the single-particle potential is given by \eq{eq:Udef} for \(k\leq k_F\) and set to zero for \(k>k_F\), implying a ``gap'' (discontinuity) in the single-particle potential.
Then only the free particle energies (\(M+\vec p{}^{\,2}/2M\)) of the intermediate states appear in the energy denominator of the Bethe-Goldstone equation \eq*{eq:BGEop} since the Pauli-blocking operator is zero for momenta below the Fermi momentum.
In the so-called \emph{continuous choice} \eq{eq:Udef} is used for the \emph{whole} momentum range, hence the single-particle potentials enter also into the energy denominator.
In \ct{Song1998a} the equation of state in symmetric nuclear matter has been considered.
It has been shown, that the result including three hole-lines is almost independent of the choice of the auxiliary potential.
Furthermore the two-hole line result with the continuous choice comes out closer to the three hole-line result, than the two-hole line calculation with the gap choice.
Another advantage of the continuous choice for intermediate spectra is that it allows for a reliable determination of the single-particle potentials including their imaginary parts~\ct*{Schulze1998}.
The results presented here employ the continuous choice.

In the following we present some results of \ct{Petschauer2015} for the in-medium properties of hyperons, based on the $YN$ interaction derived from SU(3) \cheft at NLO\@.
The same potential \(V\) as in the Lippmann-Schwinger equation \eq*{eq:LSE} for free scattering is used.
However, as in \ct{Haidenbauer2015a} the contact term \(c^{8as}\) for the antisymmetric spin-orbit force in the \(YN\) interaction, allowing spin singlet-triplet transitions, has been fitted to the weak \(\Lambda\)-nuclear spin-orbit interaction \ct*{Gal:2010xn,Botta2012}.
Additionally, for the ease of comparison, the \(G\)-matrix results obtained with two phenomenological \(YN\) potentials, namely of the J\"ulich~'04 \ct*{Haidenbauer2005} and the Nijmegen NSC97f \ct*{Rijken1998} meson-exchange models, are given.
Note that, like the EFT potentials, these phenomenological \(YN\) interactions produce a bound hypertriton \ct*{Nogga2014a}.
For more details about derivation and the commonly employed approximations, we refer the reader to \cts{Reuber1994a,Rijken1998,Schulze1998,Vidana1999,Kohno1999}.

\begin{figure*}
\centerline{
\includegraphics[scale=\figscale]{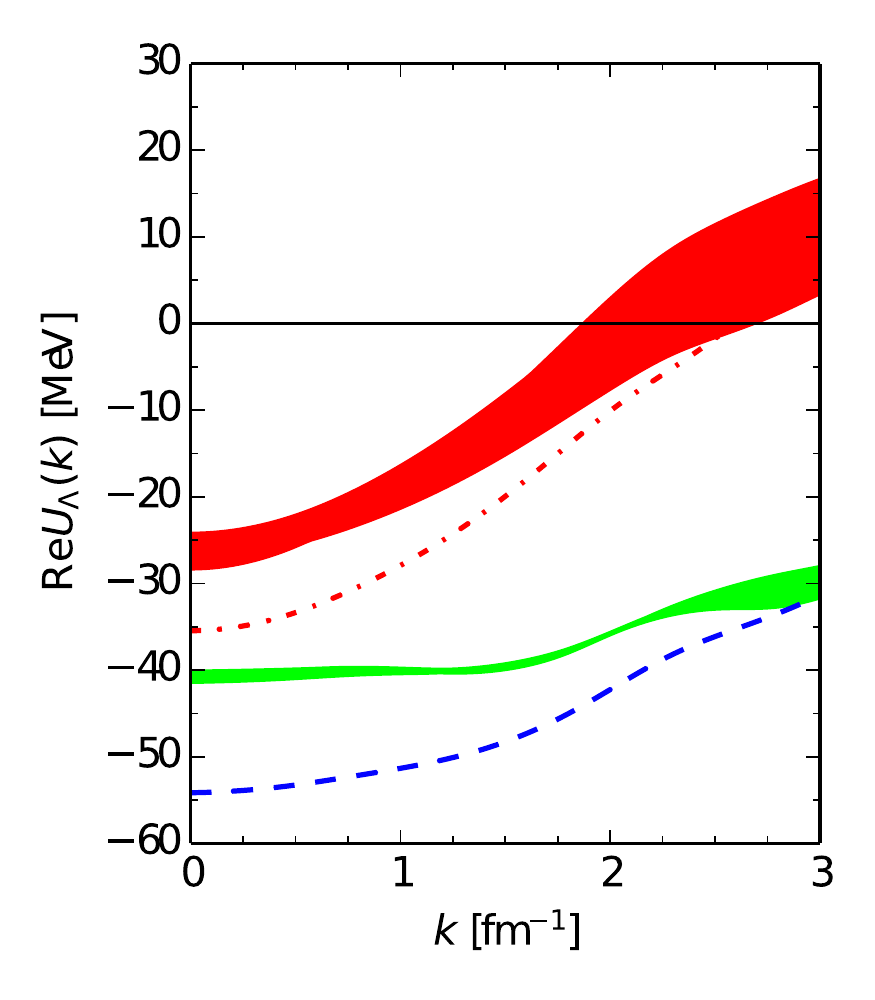}
\includegraphics[scale=\figscale]{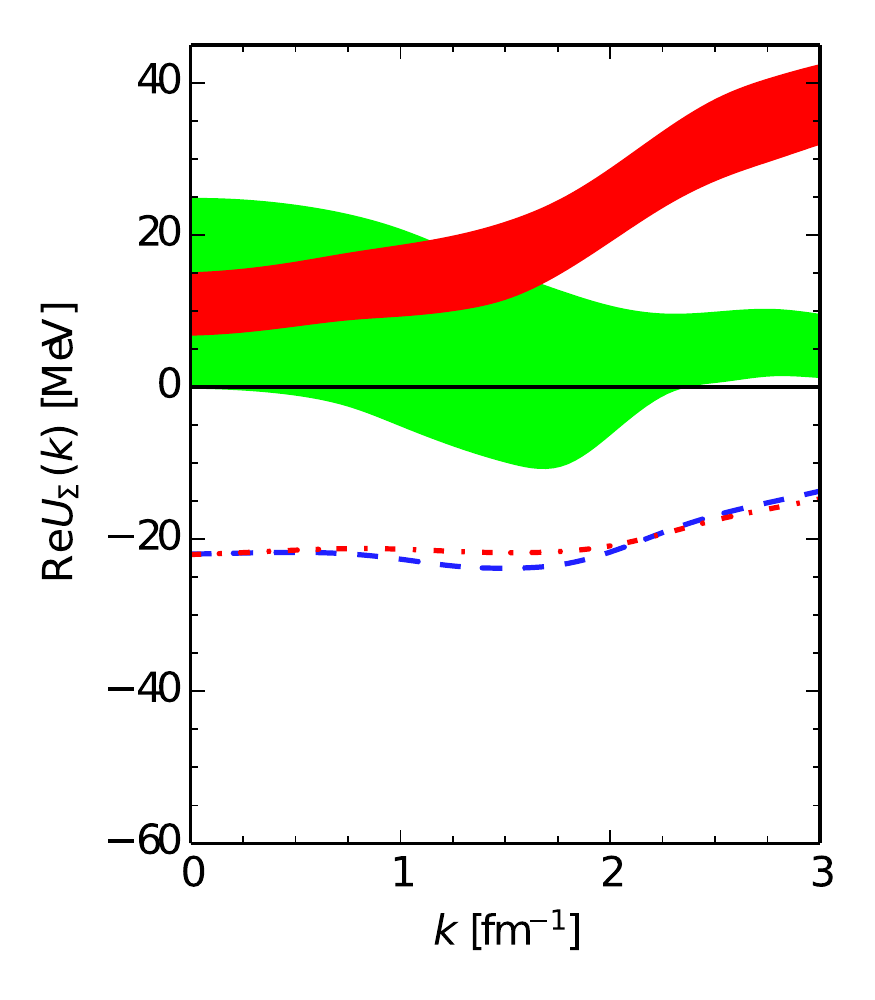}
\includegraphics[scale=\figscale]{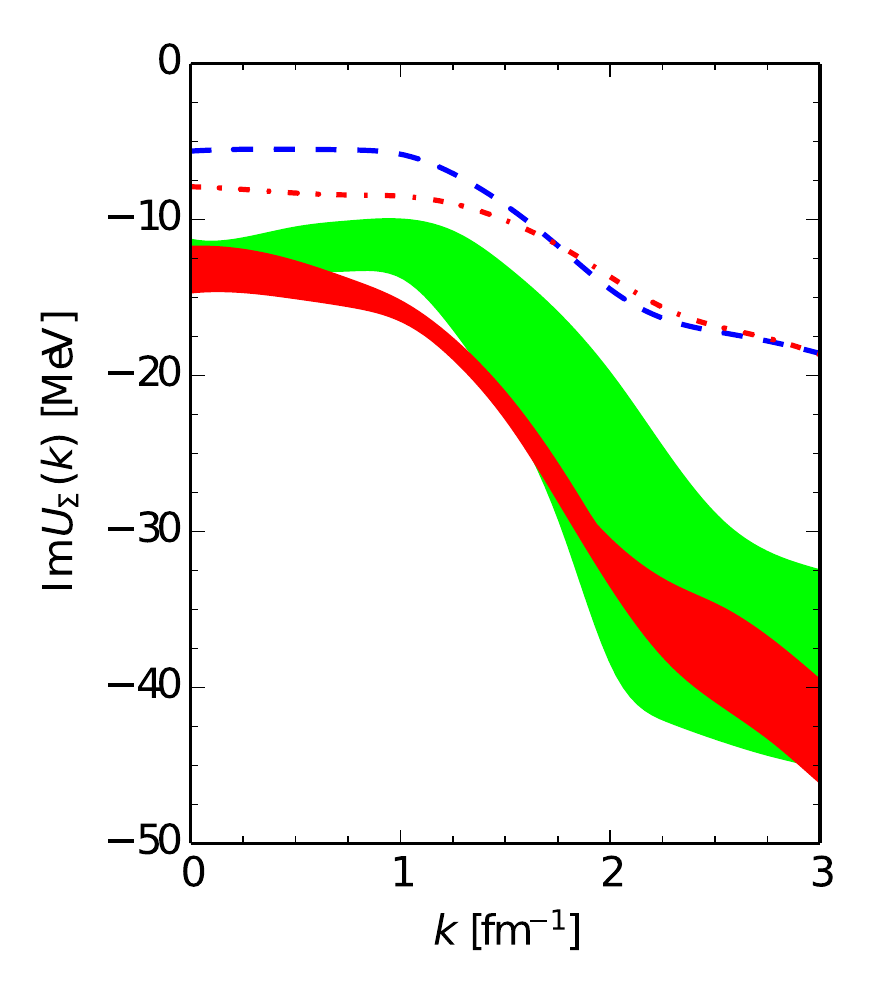}
}\vspace{-\baselineskip}
\caption{Momentum dependence of the real part of the single-particle potential of a \(\Lambda\) hyperon
and of the real and imaginary parts of the single-particle potential of a \(\Sigma\) hyperon
in isospin-symmetric nuclear matter at saturation density, \(k_F=1.35\ \mathrm{fm}^{-1}\) \ct*{Petschauer2015}.
The red band, green band, blue dashed curve and red dash-dotted curve are for \cheft NLO, \cheft LO, the J{\"u}lich '04 model and the NSC97f model, respectively.
}
\label{fig:UL_band}
\end{figure*}

Let us start with the properties of hyperons in symmetric nuclear matter.
\fig{fig:UL_band} shows the momentum dependence of the real parts of the \(\Lambda\) single-particle potential.
The values for the depth of the \(\Lambda\) single-particle potential \(U_\Lambda(k=0)\) at saturation density, \(k_F=1.35\ \mathrm{fm}^{-1}\), at NLO are between \(27.0\) and \(28.3\) MeV\@.
In the Brueckner-Hartree-Fock approximation the binding energy of a hyperon in infinite nuclear matter is given by \(B_Y(\infty)=-U_Y(k=0)\).
The results of the LO and NLO calculation are consistent with the empirical value of about \(U_\Lambda(0)\apr-28~\mathrm{MeV}\) \ct*{Millener1988,Yamamoto1988}.
The phenomenological models (J\"ulich~'04, Nijmegen NSC97f) lead to more attractive values of \(U_\Lambda(0)=(-35\ldots-50)\ \mathrm{MeV}\), where the main difference is due to the contribution in the $^3S_1$ partial wave.
In contrast to LO, at NLO the \(\Lambda\) single-particle potential at NLO turns to repulsion at fairly low momenta around \(k\approx 2\ \mathrm{fm}^{-1}\), which is also the case for the NSC97f potential.

An important quantity of the interaction of hyperons with heavy nuclei is the strength of the \(\Lambda\)-nuclear spin-orbit coupling.
It is experimentally well established \ct*{Ajimura2001a,Akikawa2002} that the \(\Lambda\)-nucleus spin-orbit force is very small.
For the \(YN\) interaction of \ct{Haidenbauer2015a} it was indeed possible to tune the strength of the antisymmetric spin-orbit contact interaction (via the constant \(c^{8as}\)),
generating a spin singlet-triplet mixing (\({}^1P_1\leftrightarrow {}^3P_1\)), in a way to achieve such a small nuclear spin-orbit potential.

Results for \(\Sigma\) hyperons in isospin-symmetric nuclear matter at saturation density are also displayed in \fig{fig:UL_band}.
Analyses of data on \((\pi^-,K^+)\) spectra related to \(\Sigma^-\) formation in heavy nuclei lead to the observation, that the \(\Sigma\)-nuclear potential in symmetric nuclear matter is moderately repulsive \ct*{Friedman2007}.
The LO as well as the NLO results are consistent with this observation.
Meson-exchange models often fail to produce such a repulsive \(\Sigma\)-nuclear potential.
The imaginary part of the \(\Sigma\)-nuclear potential at saturation density is consistent with the empirical value of \(-16\ \mathrm{MeV}\) as extracted from \(\Sigma^-\)-atom data \ct*{Dover1989}.
The imaginary potential is mainly induced by the \(\Sigma N\) to \(\Lambda N\) conversion in nuclear matter.
The bands representing the cutoff dependence of the chiral potentials, become smaller when going to higher order in the chiral expansion.

\begin{figure*}
\centering
\includegraphics[width=.9\textwidth]{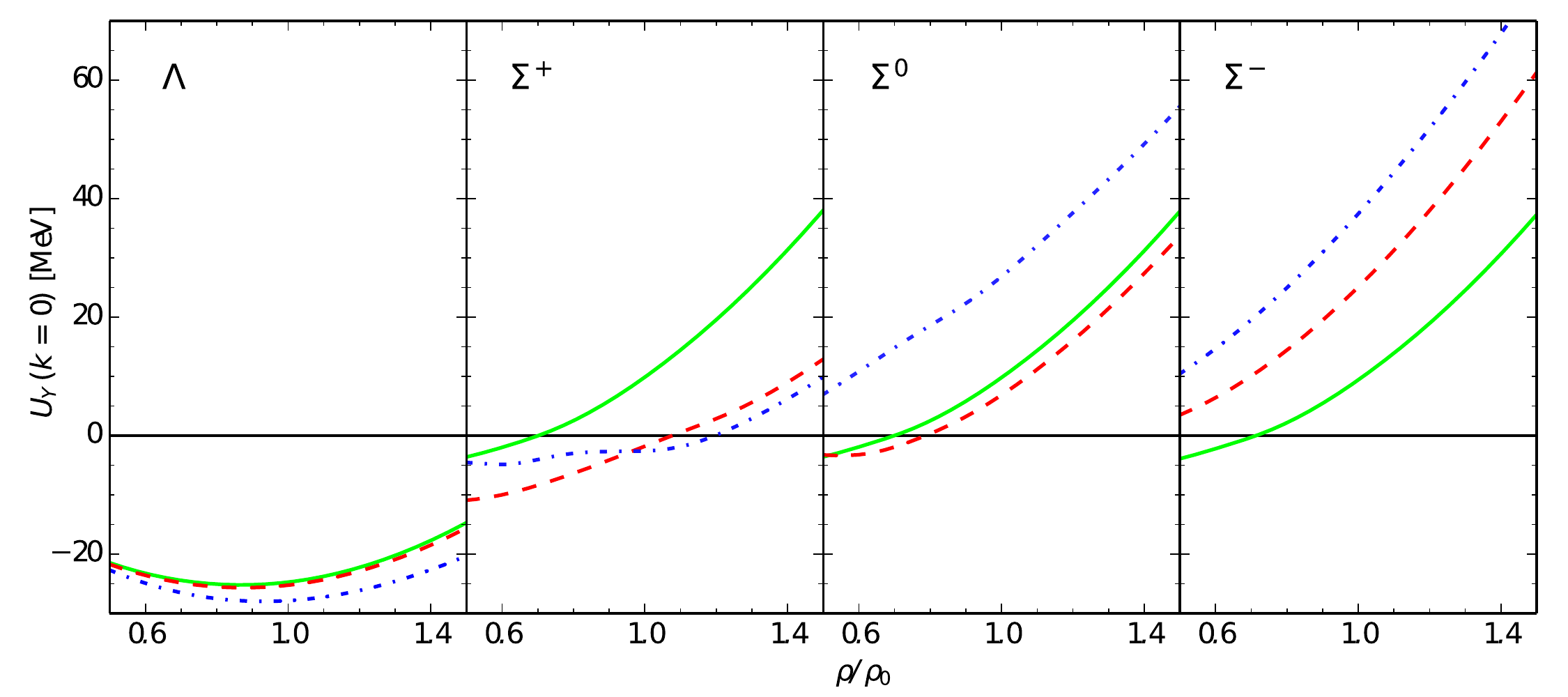}
\vspace{-.5\baselineskip}
\caption{Density dependence of the hyperon single-particle potentials at \(k=0\) with different compositions of the nuclear matter, 
calculated in \(\chi\)EFT at NLO with a cutoff $\Lambda=600$ MeV \ct*{Petschauer2015}.
The green solid, red dashed and blue dash-dotted curves are for \(\rho_p=0.5\rho\), \(\rho_p=0.25\rho\) and \(\rho_p=0\), respectively.
}
\label{fig:rho}
\end{figure*}
In \fig{fig:rho} the density dependence of the depth of the nuclear mean-field of \(\Lambda\) or \(\Sigma\) hyperons at rest (\(k=0\)).
In order to see the influence of the composition  nuclear matter on the single-particle potentials, results for isospin-symmetric nuclear matter, asymmetric nuclear matter with \(\rho_p=0.25\rho\) and pure neutron matter are shown.
The single-particle potential of the \(\Lambda\) hyperon is almost independent of the composition of the nuclear medium, because of its isosinglet nature.
Furthermore, it is attractive over the whole considered range of density \(0.5\leq \rho/\rho_0\leq 1.5\).
In symmetric nuclear matter, the three \(\Sigma\) hyperons behave almost identical (up to small differences from the mass splittings).
When introducing isospin asymmetry in the nuclear medium a splitting of the single-particle potentials occurs due to the strong isospin dependence of the \(\Sigma N\) interaction.
The splittings among the \(\Sigma^+\), \(\Sigma^0\) and \(\Sigma^-\) potentials have a non-linear dependence on the isospin asymmetry which goes beyond the usual (linear) parametrization in terms of an isovector Lane potential \ct*{Dabrowski1999}.

Recently the in-medium properties of the $\Xi$ have been investigated for $\Xi N$ potentials from
$\chi$EFT \cite{Haidenbauer2019}. For a more extensive discussion and further applications see
Ref.~\cite{Kohno:2019oyw}.

   \subsection{Hypernuclei and hyperons in neutron stars} 

\label{subsec:hypnucl}

The density-dependent single-particle potentials of hyperons interacting with nucleons in nuclear and neutron matter find their applications in several areas of high current interest: the physics of hypernuclei and the role played by hyperons in dense baryonic matter as it is realized in the core of neutron stars.

From hypernuclear spectroscopy, the deduced attractive strength of the phenomenological $\Lambda$-nuclear Woods-Saxon potential is $U_0 \simeq -30$ MeV at the nuclear center~\cite{Gal2016}.
This provides an important constraint for $U_\Lambda(k=0)$ at $\rho = \rho_0$.
The non-existence of bound $\Sigma$-hypernuclei, on the other hand, is consistent with the repulsive nature of the $\Sigma$-nuclear potential as shown in \fig{fig:rho}.
In this context effects of $YNN$ three-body forces are a key issue.
While their contributions at normal nuclear densities characteristic of hypernuclei are significant but modest, they play an increasingly important role when extrapolating to high baryon densities in neutron stars.

First calculations of hyperon-nuclear potentials based on chiral $SU(3)$ EFT and using Brueckner theory have been reported in~\ct{Haidenbauer2017, Kohno2018}.
Further investigations of (finite) $\Lambda$ hypernuclei utilizing the EFT interactions can be found in~\cite{Haidenbauer:2019thx}, based on the formalism described in \ct{Vidana2017}.
For even lighter hypernuclei, the interactions are also currently studied~\cite{Nogga:2019bwo,Le2019}.
Examples of three- and four-body results can be found in~\cts{Haidenbauer:2007ra,Nogga2012,Nogga2014a,Haidenbauer2019b}.

Here we give a brief survey of $\Lambda$-nuclear interactions for hypernuclei and extrapolations to high densities relvant to neutron stars, with special focus on the role of the (a priori unknown) contact terms of the $\Lambda NN$ three-body force.
Details can be found in~\ct{Haidenbauer2017}.
Further extended work including explicit 3-body coupled channels ($\Lambda NN\leftrightarrow\Sigma NN$) in the Brueckner-Bethe-Goldstone equation is proceeding \ct*{Gerstung:2020ktv}.

\begin{figure*}
\centerline{
\includegraphics[scale=0.35]{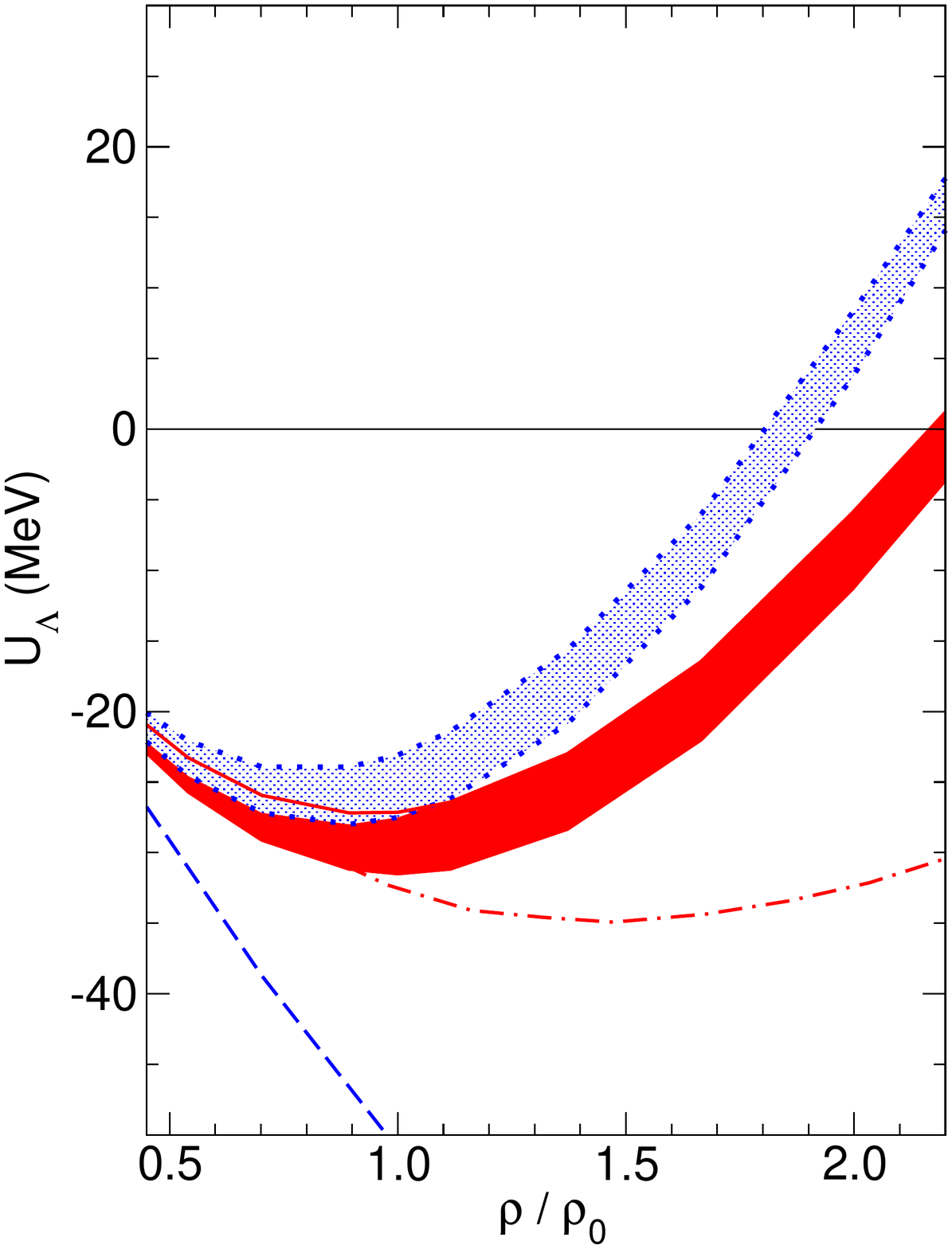} \hspace{-3ex}
\includegraphics[scale=0.35]{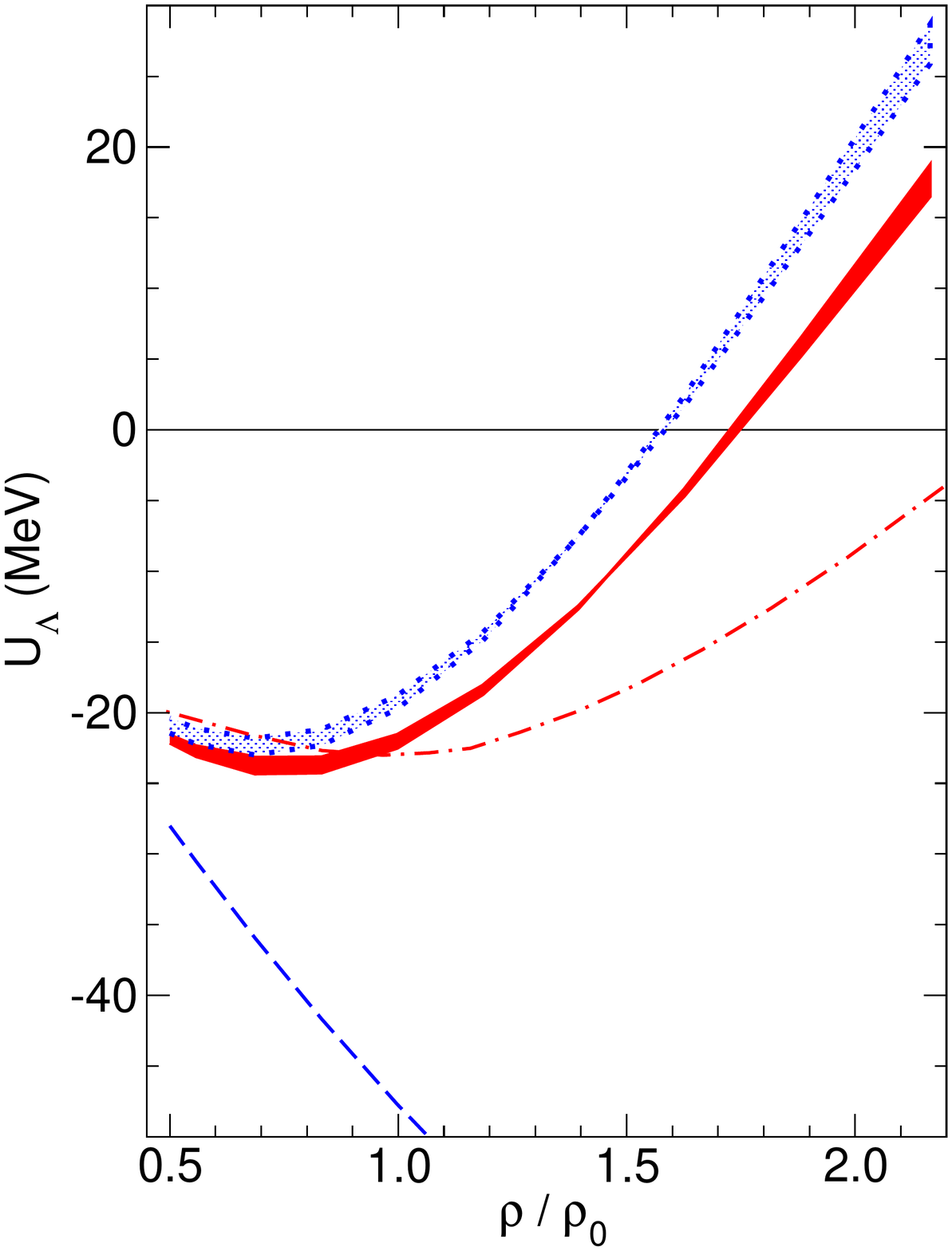}
}\vspace{-2\baselineskip}
\caption{The $\Lambda$ single-particle potential $U_\Lambda (p_\Lambda = 0,\rho)$ 
as a function of $\rho/\rho_0$ in symmetric nuclear matter (a) and in neutron matter (b).
The solid (red) band shows the chiral EFT results at NLO for cutoff variations $\Lambda =450\text{-}500~\text{MeV}$.
The dotted (blue) band includes the density-dependent $\Lambda N$-interaction derived from the $\Lambda NN$ three-body force.
The dashed curve is the result of the J{\"u}lich '04 meson-exchange model~\cite{Haidenbauer2005},
the dash-dotted curve that of the Nijmegen NSC97f potential~\cite{Rijken1998}, taken from \ct{Yamamoto:2000jh}. }
\label{fig:G}
\end{figure*}

Results for the density dependence of the $\Lambda$ single-particle potential are presented
in \fig{fig:G} for symmetric nuclear matter (a) and for neutron matter (b).
Predictions from the chiral SU(3) EFT interactions (bands) are shown in comparison with those for meson-exchange $YN$ models constructed by the J\"ulich~\cite{Haidenbauer2005}
(dashed line) and Nijmegen~\cite{Rijken1998} (dash-dotted line) groups.
One observes an onset of repulsive effects around the saturation density of nuclear matter, \ie $\rho = \rho_0$.
The repulsion increases strongly as the density increases.
Already around $\rho \approx  2\rho_0$, $U_\Lambda (0,\rho)$ turns over to net repulsion. 

Let us discuss possible implications for neutron stars.
It should be clear that it is mandatory to include the $\Lambda N$--$\Sigma N$ coupling in the pertinent calculations.
This represents a challenging task since
standard microscopic calculations without this coupling are already quite complex.
However, without the $\Lambda N$--$\Sigma N$ coupling, which has such a strong influence on the in-medium properties of hyperons, it will be difficult if not impossible to draw reliable conclusions.

The majority of $YN$-interactions employed so far in microscopic calculations of neutron stars 
have properties similar to those of the J\"ulich '04 model.
In such calculations, hyperons start appearing in the core of neutron stars typically at relatively low  densities around $(2-3)\rho_0$~\cite{Djapo2010,Lonardoni2015}.
This causes the so-called hyperon puzzle: a strong softening of the equation-of-state, such that the maximum neutron star mass falls
far below the constraint provided by the existence of several neutron stars with masses around $2 M_\odot$.
Assume now that nature favors a scenario with a weak diagonal $\Lambda N$-interaction 
and a strong $\Lambda N$--$\Sigma N$ coupling as predicted by SU(3) chiral EFT. The present study demonstrates that, 
in this case, the $\Lambda$ single-particle potential $U_\Lambda(k=0,\rho)$ based on chiral EFT two-body interactions is already repulsive at densities $\rho \sim (2-3)\rho_0$.
The one of the $\Sigma$-hyperon is likewise repulsive~\cite{Haidenbauer2015}.
We thus expect that the appearance of hyperons in neutron stars will be shifted to much higher densities.
In addition there is a repulsive density-dependent
effective $\Lambda N$-interaction that arises within the same chiral EFT framework from the leading chiral $YNN$ three-baryon forces.
This enhances the aforementioned repulsive effect further. It makes the appearance of $\Lambda$-hyperons in neutron star matter energetically unfavorable.
In summary, all these aspects taken together may well point to a solution of the hyperon puzzle in neutron stars without resorting to exotic mechanisms.

\section{Conclusions}  \label{sec:concl}

In this review we have presented the basics to derive the forces between octet baryons (\(N,\Lambda,\Sigma,\Xi\)) at next-to-leading order in SU(3) chiral effective field theory.
The connection of SU(3) \cheft to quantum chromodynamics via the chiral symmetry and its symmetry breaking patterns, and the change of the degrees of freedom has been shown.
The construction principles of the chiral effective Lagrangian and the external-field method have been presented and the Weinberg power-counting scheme has been introduced.

Within SU(3) \cheft the baryon-baryon interaction potentials have been considered at NLO\@.
The effective baryon-baryon potentials include contributions from pure four-baryon contact terms, one-meson-exchange diagrams, and two-meson-exchange diagrams at one-loop level.
The leading three-baryon forces, which formally start to contribute at NNLO\@,
consist of a three-baryon contact interaction, a one-meson exchange and a two-meson exchange component.
We have presented explicitly potentials for the \(\Lambda NN\) interaction in the spin and isospin basis.
The emerging low-energy constants can be estimated via decuplet saturation,
which leads to a promotion of some parts of the three-baryon forces to NLO\@.
The expressions of the corresponding effective two-body potential in the nuclear medium has been presented.

In the second part of this review we have presented selected applications of these potentials.
An excellent description of the available \(YN\) data has been achieved with \cheft, comparable to the most advanced phenomenological models.
Furthermore, in studies of the properties of hyperons in isospin symmetric and asymmetric infinite nuclear matter, the chiral baryon-baryon potentials at NLO are consistent with the empirical knowledge about hyperon-nuclear single-particle potentials.
The exceptionally weak \(\Lambda\)-nuclear spin-orbit force is found to be related to the contact term responsible for an antisymmetric spin-orbit interaction.
Concerning hypernuclei and neutron stars promising results have been obtained and could point to a solution of the hyperon puzzle in neutron stars.

In summary, \cheft is an appropriate tool for constructing the interaction among baryons in a systematic way.
It sets the framework for many promising applications in strangeness-nuclear physics.

\begin{acknowledgments}
This work is supported in part by the Deutsche Forschungsgemeinschaft (DFG) through the funds provided to the Sino-German Collaborative
Research Center ``Symmetries and the Emergence of Structure in QCD"  (Grant No.~TRR110), by the CAS President's International Fellowship
Initiative (PIFI) (Grant No.~2018DM0034), and by the VolkswagenStiftung (Grant No.~93562).
\end{acknowledgments}

%


\end{document}